\documentclass{statsoc}

\usepackage[a4paper]{geometry}
\usepackage{graphicx}
\usepackage{array}
\usepackage{multirow}
\usepackage[textwidth=8em,textsize=small]{todonotes}
\usepackage{amsmath}
\usepackage{hyperref}
\usepackage{amssymb}
\usepackage{stmaryrd}
\usepackage{natbib}

\numberwithin{figure}{section}
\numberwithin{table}{section}
\numberwithin{equation}{section}

\newtheorem{remark}{Remark}[section]
\newtheorem{theorem}{Theorem}[section]
\newtheorem{proposition}{Proposition}[section]

\title[Parameter estimation in NLME-ODEs via optimal control theory]{Parameter estimation in nonlinear mixed effect models based on ordinary differential equations: an optimal control approach}
\author[Author 1 {\it et al.}]{Quentin Clairon}
\address{University of Bordeaux, Inria Bordeaux Sud-Ouest, Inserm, Bordeaux Population Health Research Center, SISTM Team, UMR1219, F-33000 Bordeaux, France}
\address{Vaccine Research Institute, F-94000 Créteil, France}
\email{quentin.clairon@u-bordeaux.fr}
\author{Chlo\'e Pasin}
\address{Institute of Medical Virology, University of Zurich, Zurich, Switzerland.}
\address{Department of Infectious Diseases and Hospital Epidemiology, University Hospital,
Zurich, Switzerland.}
\author{Irene Balelli}
\address{Universit\'e Côte d'Azur, INRIA Sophia Antipolis, EPIONE Research Project, Valbonne, France.}
\author{Rodolphe Thi\'ebaut}
\address{University of Bordeaux, Inria Bordeaux Sud-Ouest, Inserm, Bordeaux Population Health Research Center, SISTM Team, UMR1219, F-33000 Bordeaux, France}
\address{CHU Bordeaux, F-33000 Bordeaux, France}
\address{Vaccine Research Institute, F-94000 Créteil, France}
\author{M\'elanie Prague}
\address{University of Bordeaux, Inria Bordeaux Sud-Ouest, Inserm, Bordeaux Population Health Research Center, SISTM Team, UMR1219, F-33000 Bordeaux, France}
\address{Vaccine Research Institute, F-94000 Créteil, France}

\begin{document}
\begin{abstract}
We present a parameter estimation method for nonlinear mixed effect
models based on ordinary differential equations (NLME-ODEs). The method
presented here aims at regularizing the estimation problem in presence
of model misspecifications, practical identifiability issues and unknown
initial conditions. For doing so, we define our estimator as the minimizer
of a cost function which incorporates a possible gap between the assumed
model at the population level and the specific individual dynamic.
The cost function computation leads to formulate and solve optimal
control problems at the subject level. This control theory approach
allows to bypass the need to know or estimate initial conditions for
each subject and it regularizes the estimation problem in presence
of poorly identifiable parameters. Comparing to maximum likelihood,
we show on simulation examples that our method improves estimation
accuracy in possibly partially observed systems with unknown initial
conditions or poorly identifiable parameters with or without model
error. We conclude this work with a real application on antibody concentration
data after vaccination against Ebola virus coming from phase 1 trials.
We use the estimated model discrepancy at the subject level to analyze
the presence of model misspecification.
\end{abstract}

\section{Introduction}

ODE models are standard in population dynamics, epidemiology, virology,
pharmacokinetics, or genetic regulation networks analysis due to their
ability to describe the main mechanisms of interaction between different
biological components of complex systems, their evolution in time
and to provide reasonable approximations of stochastic dynamics \cite{Perelson1996,Lavielle2007,Wakefield1995,Andraud2012,Pasin2019,Lavielle2011,Le2015,Engl2009,Wu2014}.
Evidence of the relevance of ODEs resides for example in their joint
use with control theory methods for the purpose of optimal treatment
design \cite{Guo2012,Agusto2014,Zhang2016,Pasin2018,villain2019adaptive}.
In cases of experimental designs involving a large number of subjects
and limited number of individual measurements, non-linear mixed-effect
models may be more relevant than subject-by-subject model to gather
information from the whole population while allowing between-individual
variability. For example, clinical trials and pharmacokinetics/pharmacodynamics
studies often fall into this category \cite{Lavielle2011,Guedj2007,Huang2008,Wang2014,Pragues2013}.
Formally, we are interested in  a population where the dynamics of
the compartments of each subject $i\in\left\llbracket 1,\,n\right\rrbracket $
is modeled by the $d$-dimensional ODE: 
\begin{equation}
\left\{ \begin{array}{l}
\dot{x}_{i}(t)=f_{\theta,b_{i}}(t,x_{i}(t),z_{i}(t))\\
x_{i}(0)=x_{i,0}
\end{array}\right.\label{eq:ODEmodel-gen}
\end{equation}
where $f$ is a $d-$dimensional vector field, $\theta$ is a $p-$dimensional
parameter, $b_{i}\sim N(0,\,\Psi)$ is a $q-$dimensional random effect
where $\Psi$ is a variance-covariance matrix, $x_{i,0}$ is the initial
condition for subject $i$ belonging to $\mathbb{R}^{d}$ and $z_{i}$
is a covariate function. We denote $X_{\theta,b_{i},x_{i,0}}$ the
solution of (\ref{eq:ODEmodel-gen}) for a given set $\left(\theta,b_{i},x_{i,0}\right)$.

Our goal is to estimate the true population parameters $\left(\theta^{*},\,\Psi^{*}\right)$
as well as the true subject specific realizations $\left\{ b_{i}^{*}\right\} _{i\in\left\llbracket 1,\,n\right\rrbracket }$
from partial and noisy observations coming from $n$ subjects and
described by the following observational model:
\[
y_{ij}=CX_{\theta^{*},b_{i}^{*},x_{i,0}^{*}}(t_{ij})+\epsilon_{ij}
\]
where $t_{ij}$ is the $j$-th measurement time-point for the $i-$th
subject on the observation interval $\left[0,\,T\right]$. Here $C$
is a $d^{o}\times d$ sized observation matrix emphasizing the potentially
partially observed nature of the process and $\epsilon_{ij}\sim\sigma^{*}\times N(0,I_{d^{o}})$
is the measurement error. We also assume only a subset of the true
initial condition $x_{i,0}^{*}$, denoted $x_{i,0}^{k*}$, is known,
the other ones, denoted $x_{i,0}^{u*}$, being unknown. For the sake
of clarity, we order the state variables as follows: $x_{i,0}=\left(\left(x_{i,0}^{u}\right)^{T},\left(x_{i,0}^{k}\right)^{T}\right)^{T}.$
We denote $n_{i}$ the number of observations for the $i$-th subject,
$\mathbf{y_{i}}=\left\{ y_{ij}\right\} _{j\in\left\llbracket 1,\,n_{i}\right\rrbracket }$
its corresponding set of observations and $\mathbf{y}=\left\{ \mathbf{y_{i}}\right\} _{i\in\left\llbracket 1,\,n\right\rrbracket }$
the set of all observations in the population.

Our problem belongs to the class parameter estimation problem in nonlinear
mixed effect models. In this context, frequentist methods based on
likelihood maximization (via different numerical procedures: Laplace
approximation \cite{Pinheiro1994}, Gaussian quadrature \cite{Pinheiro1994,Lindstrom1990,Guedj2007}
or SAEM \cite{Lavielle2005,Lavielle2007,Comets2017}) and Bayesian
ones aiming to reconstruct the a posteriori distribution or to derive
the maximum a posteriori estimator (via MCMC algorithms \cite{Lunn2000,Huang2011,Huang2010},
importance sampling \cite{Raftery2010}, approximation of the asymptotic
posterior distribution \cite{Pragues2013}) have been proposed. In
particular, dedicated methods/softwares using the structure of ODE
models have been implemented to increase numerical stability and speed
up convergence rate \cite{Tornoe2004}, to reduce the computational
time \cite{Donnet2006} or to avoid the repeated model integration
and estimation of initial conditions \cite{Wang2014}. However, all
the preceding methods face similar pitfalls due to specific features
of population models based on ODEs (with the exception of \cite{Wang2014}):
\begin{enumerate}
\item They do not account for model misspecification presence, a common
feature in ODE models used in biology. Indeed, the ODE modeling process
suffers from model inadequacy, understood as the discrepancy between
the mean model response and real world process, and residual variability
issues, that is subject specific stochastic perturbations or missed
elements which disappear by averaging over the whole population \cite{Kennedy2001}.
As examples of model inadequacy causes, one can think of ODE models
used in epidemiology and virology which are derived by approximations
where for instance, interactions are modeled by pairwise products
while higher order terms and/or the influence of unknown/unmeasured
external factors are neglected \cite{Stein2013}. Regarding residual
variability, let us remind that biological processes are often stochastic
\cite{Bowsher2012,Komorowski2013} and the justification of deterministic
modeling comes from the approximation of stochastic processes \cite{Kurtz1978,Gillespie2000,Kampen1992}.
Moreover, in the context of population models, new sources of model
uncertainties emerge. Firstly, error measurement in covariates $z_{i}$
which is not often considered leads to use a proxy function $\widehat{z_{i}}$
instead of $z_{i}$ \cite{Huang2011}. Secondly, the sequential nature
of most inference methods leads to estimate $\left\{ b_{i}^{*}\right\} _{i\in\left\llbracket 1,\,n\right\rrbracket }$
based on an approximation $\widehat{\theta}$ instead of the true
population parameter value $\theta^{*}$. Thus, the structure of mixed-effect
models spread measurement uncertainty into the mechanistic model structure
during the estimation. It turns classical statistical uncertainties
into model error causes. Estimation of $\theta^{*}$, $\Psi^{*}$
and $\left\{ b_{i}^{*}\right\} _{i\in\left\llbracket 1,\,n\right\rrbracket }$
has to be done with model misspecification presence although it is
known to dramatically impair the accuracy of methods which do not
take into account potential modeling error \cite{Brynjarsdottir2014,kirk2016reverse}. 
\item They have to estimate or make assumptions on $x_{i,0}^{u*}$ values.
In ODE models, the initial conditions are generally nuisance parameters
in the sense that knowing their values does not bring answers to the
scientific questions which motivate the model construction but the
estimation of the relevant parameters requires $x_{i,0}^{*}$ inference
as well. For example partially observed compartmental models used
in pharmacokinetics/pharmacodynamics often involve unknown initial
conditions which needs to be inferred to estimate the transmission
rates between compartments which are the true parameters of interest.
Unknown initial conditions imply either: assumptions on their values
\cite{Lavielle2011,Guedj2007,Thiebaut2014}, another potential cause
of model misspecifications, or the need to estimate them \cite{Huang2006b,Huang2008}
which increases the optimization problem dimension and degrades estimation
accuracy due to covariance effect between $\left(\theta^{*},\Psi^{*}\right)$
and $x_{i,0}^{u*}$ estimate. 
\item They can face accuracy degradation when the inverse problem of parameter
estimation is ill-posed \cite{Engl2009,Stuart2010} due to practical
identifiability issues. Ill-posedness in ODE models is often due to
the geometry induced by the mapping $\left(\theta,b_{i},x_{i,0}\right)\longmapsto CX_{\theta,b_{i},x_{i,0}}$,
where there can be a small number of relevant directions of variation
skewed from the original parameter axes \cite{Gutenkunst2007,Transtrum2011,Transtrum2015,OLeary2015}.
This problem, called sloppiness, often appears in ODE models used
in biology \cite{Gutenkunst2007,OLeary2015} and leads to an ill-conditioned
Fisher Information Matrix. For maximum likelihood estimators this
is a cause for high variance due to the Cram\'er-Rao bound. For Bayesian
inference, it leads to a nearly singular asymptotic a posteriori distribution
because of Bernstein--von Mises theorem (see \cite{campbell2007}
for the computational induced problems). Despite this problem is in
part mitigated by the population approach which merges different subjects
for estimating $\left(\theta^{*},\Psi^{*}\right)$ and uses distribution
of $b_{i}\mid\Psi$ as prior at the subject level \cite{Lavielle2015},
estimation accuracy can benefit from the use of regularization techniques
for the inverse problem. 
\end{enumerate}
These specific features of ODE-based population models limit the amount
of information classic approaches can extract for estimation purposes
from observations no matter their qualities or abundances. This advocates
for the development of new estimation procedures. Approximate methods
\cite{Varah1982,Ramsay2007,BrunelClairon_Pontryagin2017}
have already proven to be useful for ODE models facing these issues
with observations coming from one subject. These approaches rely on
an approximation of the solution of the original ODE (\ref{eq:ODEmodel-gen})
which is expected to have a smoother dependence with respect to the
parameters and to relax the constraint imposed by the model during
the estimation process. The interest of such approximations is twofold.
Firstly they produce estimators with a better conditioned variance
matrix comparing to classic likelihood based approaches and they reduce
the effect of model error on estimator accuracy. Secondly, some of
these approximations bypass the need to estimate initial conditions
\cite{Ramsay2007,Clairon2019}. In this work, we generalize one of
these approaches to population models by developing a new estimation
method specific to NLME-ODEs aiming to integrate such approximations
to mitigate the effect of model misspecification and poorly identifiable
parameter on estimation accuracy, while avoiding the need to estimate
$x_{i,0}^{u*}$. We propose here a nested estimation procedure where
population parameters $\left(\theta^{*},\Psi^{*},\sigma^{*}\right)$
are estimated through the maximization of an outer criterion. This
requires in turn an estimator for the $\left\{ b_{i}^{*}\right\} _{i\in\left\llbracket 1,\,n\right\rrbracket }$
obtained through the repeated optimization of inner criteria. We consider
that the actual dynamic for each subject is described by a perturbed
version of the ODE (\ref{eq:ODEmodel-gen}) where the added perturbation
captures different sources of errors at the subject level \cite{Brynjarsdottir2014,TuoWu2015}.
We control the magnitude of the acceptable perturbations by defining
the inner criteria through a cost function balancing the two contrary
objectives of fidelity to the observations and to the original model:
to this end,  we introduce a model discrepancy penalization term.
The practical computation of the $\left\{ b_{i}^{*}\right\} _{i\in\left\llbracket 1,\,n\right\rrbracket }$
estimators require to solve optimal control problems \cite{clarke2013variationalcalculus,Kirk1998optimalcontrol,Sontag1998}
known as tracking problems. This is done using a method inspired by
\cite{Cimen2004,CimenBanks2004} based on pseudo-linear representation
and Linear-Quadratic theory. In addition, our method does not need
to estimate $x_{i,0}^{u*}$. Nevertheless, it can provide an estimator
of $x_{i,0}^{u*}$ as a direct byproduct of structural parameters
estimation with no additional computational costs. 

In section \ref{sec:Inner_Outer_crit}, we present the estimation
method and derive the inner and outer criteria as well as an estimator
of the asymptotic Variance-Covariance matrix for the estimators of
$\left(\theta^{*},\Psi^{*}\right)$. In section \ref{sec:OCA_method_presentation},
we present the optimal control problems related to the different criteria
as well as the algorithms used to solve them. In section \ref{sec:Simulation},
we compare our approach with classic maximum likelihood in simulations.
We then proceed to the real data analysis coming from clinical studies
and a model of the antibody concentration dynamics following immunization
with an Ebola vaccine in East African participants \cite{Pasin2019}.
Section \ref{sec:Conclusion} concludes and discuss further applications
and extensions of the method. 

\section{\label{sec:Inner_Outer_crit}Construction of the estimator: definition
of the inner and outer criteria}

From now on, we use the following Choleski decomposition $\sigma^{2}\Psi^{-1}=\triangle^{T}\triangle$
(or equivalently $\Psi=\sigma^{2}\left(\triangle^{T}\triangle\right)^{-1}$)
and the parametrization $\left(\theta,\Delta,\sigma\right)$ instead
of $\left(\theta,\Psi,\sigma\right)$. This parametrization will allows
us to enforce positiveness and symmetry of $\Psi$ and to derive an
explicit estimator of $\sigma$ given a value for $\left(\theta,\Delta\right)$.
The norm $\left\Vert .\right\Vert _{2}$ will denote the classic Euclidean
one defined by $\left\Vert b\right\Vert _{2}=\sqrt{b^{T}b}.$ Similarly
as in the Expectation-Maximization (EM) algorithm, we estimate the
population and individual parameters via a nested procedure:
\begin{itemize}
\item Estimation of $\widehat{b_{i}}:=\widehat{b_{i}}(\theta,\Delta)$ for
each subject $i$ by minimization of the \textbf{inner criterion}
$g_{i}$, a modified version of the log joint-likelihood function
of the data and the random effects.
\item Estimation of $\left(\theta,\Delta,\sigma\right)$ via the maximization
of an \textbf{outer criterion} defined as an approximation of the
profiled joint distribution of $\left(\theta,\Delta,\sigma,b\right)$
with respect to $b$ and denoted $G(\theta,\Delta,\sigma).$ 
\end{itemize}

\subsection{Inner criteria}

In this section, we describe the procedure used to estimate the $q-$dimensional
random effects $\left\{ b_{i}^{*}\right\} _{i\in\left\llbracket 1,\,n\right\rrbracket }$
for a given $(\theta,\Delta,\sigma)$ value. A straightforward approach
would be to look for the minimum of the log joint-likelihood function
of the data and $\left\{ b_{i},x_{0,i}^{u}\right\} .$ However, we
want to: 
\begin{enumerate}
\item avoid estimation of unknown initial conditions,
\item allow for each subject an acceptable departure from the assumed model
at the population level to take into account possible model misspecifications. 
\end{enumerate}
To solve the first point, we define our estimator as the maximizer
of the joint conditional likelihood $\mathbb{P}(\mathbf{y_{i}},b_{i}\mid x_{0,i}^{u},\theta,\Delta,\sigma)$
profiled on the unknown initial condition. Since
\[
\begin{array}{lll}
\mathbb{P}(\mathbf{y_{i}},b_{i}\mid x_{0,i}^{u},\theta,\Delta,\sigma) & = & \mathbb{P}(\mathbf{y_{i}}\mid b_{i},x_{0,i}^{u},\theta,\Delta,\sigma)\mathbb{P}(b_{i}\mid\theta,\Delta,\sigma)\\
 & = & \left(2\pi\right)^{-\left(d^{o}n_{i}+q\right)/2}\sigma^{-\left(d^{o}n_{i}+q\right)}\left|\triangle\right|e^{-0.5\left(\sum_{j}\left\Vert CX_{\theta,b_{i},x_{0,i}}(t_{ij})-y_{ij}\right\Vert _{2}^{2}+b_{i}^{T}\left(\triangle^{T}\triangle\right)b_{i}\right)/\sigma^{2}}
\end{array}
\]
by using $\mathbb{P}(\mathbf{y_{i}}\mid b_{i},\theta,\Delta,\sigma)=\prod_{j}\mathbb{P}(y_{ij}\mid b_{i},\theta,\Delta,\sigma)=\prod_{j}\left(2\pi\right)^{-d^{o}/2}\sigma^{-d^{o}}e^{-0.5\left\Vert CX_{\theta,b_{i},x_{0,i}}(t_{ij})-y_{ij}\right\Vert _{2}^{2}/\sigma^{2}}$,
$\mathbb{P}(b_{i}\mid\theta,\Delta,\sigma)=\left(2\pi\right)^{-q/2}\left|\Psi\right|^{-1/2}e^{-0.5b_{i}^{T}\Psi^{-1}b_{i}}$
and $\sigma^{2q}\left|\Psi\right|^{-1}=\left|\triangle\right|^{2}$,
a straightforward mixed-effect estimator would be $\widehat{b_{i}}=\arg\min_{b_{i}}\min_{x_{0,i}^{u}}\left\{ \sum_{j}\left\Vert CX_{\theta,b_{i},x_{0,i}}(t_{ij})-y_{ij}\right\Vert _{2}^{2}+\left\Vert \Delta b_{i}\right\Vert _{2}^{2}\right\} $
that is, the classic maximum likelihood criteria profiled on $x_{0,i}^{u}.$
Concerning the second point, we allow perturbations comparing to the
original model, by assuming that the dynamic of each subject $i$
follows a perturbed version of ODE (\ref{eq:ODEmodel-gen}):
\begin{equation}
\left\{ \begin{array}{l}
\dot{x}_{i}(t)=f_{\theta,b_{i}}(t,x_{i}(t),z_{i}(t))+Bu_{i}(t)\\
x_{i}(0)=x_{i,0}
\end{array}\right.\label{eq:ControlledODEmodel}
\end{equation}
with the addition of the forcing term $t\mapsto Bu_{i}(t)$ with $B$
a $d\times d_{u}$ matrix and $u_{i}$ a function in $L^{2}\left(\left[0,T\right],\mathbb{R}^{d_{u}}\right)$.
We denote $X_{\theta,b_{i},x_{i,0},u_{i}}$ the solution of this new
ODE (\ref{eq:ControlledODEmodel}). However, to ensure the possible
perturbation remains small, we replace the data fitting criterion
$\sum_{j}\left\Vert CX_{\theta,b_{i},x_{0,i}}(t_{ij})-y_{ij}\right\Vert _{2}^{2}$
by $\min_{u_{i}}\mathcal{C}_{i}(b_{i},x_{i,0},u_{i}\mid\theta,U)$
where $\mathcal{C}_{i}(b_{i},x_{i,0},u_{i}\mid\theta,U)=\sum_{j}\left\Vert CX_{\theta,b_{i},x_{0,i},u_{i}}(t_{ij})-y_{ij}\right\Vert _{2}^{2}+\left\Vert u_{i}\right\Vert _{U,L^{2}}^{2}$
and $\left\Vert u_{i}\right\Vert _{U,L^{2}}^{2}=\int_{0}^{T}u_{i}(t)^{T}Uu_{i}(t)dt$
is the weighted Euclidean norm. Therefore the magnitude of the allowed
perturbations is controlled by a positive definite and symmetric weighting
matrix $U.$ Finally, we obtain:
\begin{equation}
\begin{array}{lll}
\widehat{b_{i}}\left(\theta,\Delta\right) & := & \arg\min_{b_{i}}g_{i}(b_{i}\mid\theta,\Delta,U)\end{array}\label{eq:inner_estimator}
\end{equation}
where: 
\[
g_{i}(b_{i}\mid\theta,\Delta,U)=\min_{x_{0,i}^{u}}\left\{ \min_{u_{i}}\mathcal{C}_{i}(b_{i},x_{i,0},u_{i}\mid\theta,U)+\left\Vert \Delta b_{i}\right\Vert _{2}^{2}\right\} .
\]
This requires to solve the infinite dimensional optimization problem
$\min_{u_{i}}\mathcal{C}_{i}(b_{i},x_{i,0},u_{i}\mid\theta,U)$ in
$L^{2}\left(\left[0,T\right],\mathbb{R}^{d_{u}}\right)$. This problem
belongs to the field of optimal control theory for which dedicated
approaches have been developed to solve them \cite{Sontag1998,Aliyu2011,clarke2013variationalcalculus}.
Here we use the same method as in \cite{Clairon2019}, which ensures
the existence and uniqueness of the solution and provides a computationally
efficient way to find it for linear ODEs. This method can be extended
to non-linear ODEs through an iterative procedure where the original
problem is replaced by a sequence of problems involving only linear
ODEs. In addition, the methods from \cite{Clairon2019} presents the
advantage of formulating $\min_{u_{i}}\mathcal{C}_{i}(b_{i},x_{i,0},u_{i}\mid\theta,U)$
as a quadratic form (or a sequence of quadratic forms) with respect
to $x_{0,i}^{u}$. Thus, the computation of $\min_{x_{0,i}^{u}}\left\{ \min_{u_{i}}\mathcal{C}_{i}(b_{i},x_{i,0},u_{i}\mid\theta,U)\right\} $
does not add any computational complexity comparing to $\min_{u_{i}}\mathcal{C}_{i}(b_{i},x_{i,0},u_{i}\mid\theta,U)$. 

The control corresponding to the solution of $\min_{x_{0,i}^{u}}\left\{ \min_{u_{i}}\mathcal{C}_{i}(b_{i},x_{i,0},u_{i}\mid\theta,U)\right\} $
is named optimal control and denoted $\overline{u}_{i,\theta,b_{i}}$.
The corresponding solution of (\ref{eq:ControlledODEmodel}) is denoted
$\overline{X}_{\theta,b_{i}}$ and named optimal trajectory. In particular,
$\overline{X}_{\theta,b_{i}}$ and $\overline{u}_{i,\theta,b_{i}}$
are respectively the subject specific state variable and perturbation
such that:
\begin{equation}
g_{i}(b_{i}\mid\theta,\Delta,U)=\sum_{j}\left\Vert C\overline{X}_{\theta,b_{i}}(t_{ij})-y_{ij}\right\Vert _{2}^{2}+\left\Vert \overline{u}_{i,\theta,b_{i}}\right\Vert _{U,L^{2}}^{2}+\left\Vert \Delta b_{i}\right\Vert _{2}^{2}.\label{eq:inner_criteria}
\end{equation}
To incorporate possible model errors in the estimation process, e.g.
due to subject specific exogenous perturbations, $\overline{X}_{\theta,b_{i}}$
is now assumed to be the subject specific regression function, defined
as the state-variable which needs the smallest perturbation in order
to get close to the observations. The numerical procedure to derive
$\overline{X}_{\theta,b_{i}}$ and $g_{i}$ is presented in section
\ref{sec:OCA_method_presentation}. 
\begin{remark}
The definition of the optimal control $\overline{u}_{i,\theta,b_{i}}$
has an interpretation in terms of Bayesian inference in an infinite
dimensional space. According to \cite{Dashti2013} (theorem 3.5 and
Corollary 3.10), $\overline{u}_{i,\theta,b_{i}}$ is a maximum a posteriori
estimator where the chosen prior measure is a centered Gaussian random
field with the covariance operator determined by $U$. This link can
be fruitful to import tools coming from deterministic control theory
to solve statistical problem formalized in functional spaces. 
\end{remark}

\subsection{Outer criteria definition}

We focus in this section on population parameter estimation. Classic
approaches rely on maximum a posteriori distribution or the likelihood
of the observations in which they get rid of the unknown subject specific
parameters by taking the mean value of $\mathbb{P}\left[\theta,\Delta,\sigma,b\mid\mathbf{y}\right]$
or $\mathbb{P}\left[\mathbf{y}\mid\theta,\Delta,\sigma,b\right]$,
$\mathbb{E}_{b}\left[\mathbb{P}\left[\theta,\Delta,\sigma,b\mid\mathbf{y}\right]\right]$
or $\mathbb{E}_{b}\left[\mathbb{P}\left[\mathbf{y}\mid\theta,\Delta,\sigma,b\right]\right]$
respectively, as outer criteria. This generally requires the numerical
approximation of integrals of possibly high dimensions (the same as
$b$), a source of approximation and computational issues \cite{Pinheiro1994}.
To avoid this, we consider the random effects as nuisance parameters
and rely on a classic profiling approach for $\left(\theta^{*},\triangle^{*}\right)$
estimation \cite{Murphy2000}. Instead of taking the mean, we rely
on the maximal value of the joint distribution with respect to $b$.
We consider the cost function $\max_{b}\mathbb{P}\left[\theta,\Delta,\sigma,b\mid\mathbf{y}\right]$
(or equivalently $\max_{b}\ln\mathbb{P}\left[\theta,\Delta,\sigma,b\mid\mathbf{y}\right]$).
Bayes formula gives us $\mathbb{P}\left[\theta,\Delta,\sigma,b\mid\mathbf{y}\right]\propto\mathbb{P}\left[\mathbf{y}\mid\theta,\Delta,\sigma,b\right]\mathbb{P}\left[\theta,\Delta,\sigma,b\right].$
Since $\mathbb{P}\left[\theta,\Delta,\sigma,b\right]=\mathbb{P}\left[b\mid\theta,\Delta,\sigma\right]\mathbb{P}\left[\theta,\Delta\right]$,
we get $\mathbb{P}\left[\theta,\Delta,\sigma,b\mid\mathbf{y}\right]\propto\left(\prod_{i}\mathbb{P}\left[\mathbf{y_{i}}\mid\theta,\Delta,\sigma,b_{i}\right]\mathbb{P}\left[b_{i}\mid\theta,\Delta,\sigma\right]\right)\mathbb{P}\left[\theta,\Delta\right]$
by conditional independence of subject by subject observations and
subject specific parameters. It follows that $\max_{b}\ln\mathbb{P}\left[\theta,\Delta,\sigma,b\mid\mathbf{y}\right]\propto\sum_{i}\max_{b_{i}}\left(\ln\mathbb{P}\left[\mathbf{y_{i}}\mid\theta,\Delta,\sigma,b_{i}\right]+\ln\mathbb{P}\left[b_{i}\mid\theta,\Delta,\sigma\right]\right)+\ln\mathbb{P}\left[\theta,\Delta\right].$
From now on we will use the estimate (\ref{eq:inner_estimator}) of
the previous section to construct a suitable approximation of 
\[
\overline{G}^{(1)}(\theta,\Delta,\sigma\mid\mathbf{y})=\sum_{i}\max_{b_{i}}\left(\ln\mathbb{P}\left[\mathbf{y_{i}}\mid\theta,\Delta,\sigma,b_{i}\right]+\ln\mathbb{P}\left[b_{i}\mid\theta,\Delta,\sigma\right]\right)+\ln\mathbb{P}\left[\theta,\Delta\right]
\]
as our criteria to estimate population parameters. As said in the
previous section, we define the optimal trajectory $\overline{X}_{\theta,b_{i}}$
as the regression function for each subject. Therefore, we approximate
$\mathbb{P}\left[\mathbf{y_{i}}\mid\theta,\Delta,\sigma,b_{i}\right]$
by $\widetilde{\mathbb{P}}\left[\mathbf{y_{i}}\mid\theta,\Delta,\sigma,b_{i}\right]\simeq\prod_{j}\left(2\pi\right)^{-d^{o}/2}\sigma^{-d^{o}}e^{-0.5\left\Vert C\overline{X}_{\theta,b_{i}}(t_{ij})-y_{ij}\right\Vert _{2}^{2}/\sigma^{2}}.$
By using the previous section computations, we get $\arg\max_{b_{i}}\left(\ln\widetilde{\mathbb{P}}\left[\mathbf{y_{i}}\mid\theta,\Delta,\sigma,b_{i}\right]+\ln\mathbb{P}\left[b_{i}\mid\theta,\Delta,\sigma\right]\right) =\arg\max_{b_{i}}\left(\sum_{j}\left\Vert C\overline{X}_{\theta,b_{i}}(t_{ij})-y_{ij}\right\Vert _{2}^{2}+\left\Vert \Delta b_{i}\right\Vert _{2}^{2}\right).$
We regularize this estimation problem by approximating it via the
addition of the Tikhonov penalization term on perturbation magnitude
$\left\Vert \overline{u}_{i,\theta,b_{i}}\right\Vert _{U,L^{2}}^{2}$,
thus $\arg\max_{b_{i}}\left(\ln\widetilde{\mathbb{P}}\left[\mathbf{y_{i}}\mid\theta,\Delta,\sigma,b_{i}\right]+\ln\mathbb{P}\left[b_{i}\mid\theta,\Delta,\sigma\right]\right)\simeq\arg\max_{b_{i}}g_{i}(b_{i}\mid\theta,\Delta,U)=\widehat{b_{i}}\left(\theta,\Delta\right)$
by using definition (\ref{eq:inner_criteria}). Also, we use 
\[
\overline{G}^{(2)}\left[\theta,\Delta,\sigma\mid\mathbf{y}\right]=\sum_{i}\left(\ln\widetilde{\mathbb{P}}\left[\mathbf{y_{i}}\mid\theta,\Delta,\sigma,\widehat{b_{i}}\left(\theta,\Delta\right)\right]+\ln\mathbb{P}\left[\widehat{b_{i}}\left(\theta,\Delta\right)\mid\theta,\Delta,\sigma\right]\right)+\ln\mathbb{P}\left[\theta,\Delta\right]
\]
as an approximation of $\overline{G}^{(1)}.$ By replacing $\widetilde{\mathbb{P}}\left[\mathbf{y_{i}}\mid\theta,\Delta,\sigma,b_{i}\right]$
and $\mathbb{P}\left[b_{i}\mid\theta,\Delta,\sigma\right]$ by their
values, we notice that $\arg\max_{\left(\theta,\Delta\right)}\left\{ \overline{G}^{(2)}\left[\theta,\Delta,\sigma\mid\mathbf{y}\right]\right\} =\arg\max_{\left(\theta,\Delta\right)}\left\{ \overline{G}^{(3)}\left[\theta,\Delta,\sigma\mid\mathbf{y}\right]\right\} $
for every $\sigma>0$ where 
\[
\begin{array}{lll}
\overline{G}^{(3)}\left[\theta,\Delta,\sigma\mid\mathbf{y}\right] & = & -\frac{1}{2\sigma^{2}}\sum_{i}\left(\sum_{j}\left\Vert C\overline{X}_{\theta,\widehat{b_{i}}\left(\theta,\Delta\right)}(t_{ij})-y_{ij}\right\Vert _{2}^{2}+\left\Vert \triangle\widehat{b_{i}}\left(\theta,\Delta\right)\right\Vert _{2}^{2}\right)\\
 & - & 0.5\left(d^{o}\sum_{i}n_{i}+qn\right)\ln\left(\sigma^{2}\right)+0.5n\ln\left(\left|\triangle^{T}\triangle\right|\right)+\ln\mathbb{P}\left[\theta,\Delta\right].
\end{array}
\]
Moreover, for each $\left(\theta,\Delta\right)$, the maximizer in
$\sigma^{2}$ of $\overline{G}^{(3)}$ has a closed form expression: 
\begin{equation}
\sigma^{2}\left(\theta,\Delta\right)=\frac{1}{\left(d^{o}\sum_{i}n_{i}+qn\right)}\sum_{i}\left(\sum_{j}\left\Vert C\overline{X}_{\theta,\widehat{b_{i}}\left(\theta,\Delta\right)}(t_{ij})-y_{ij}\right\Vert _{2}^{2}+\left\Vert \triangle\widehat{b_{i}}\left(\theta,\Delta\right)\right\Vert _{2}^{2}\right).\label{eq:sigma_est}
\end{equation}
By using the expression of $\sigma^{2}\left(\theta,\Delta\right)$
given by equation (\ref{eq:sigma_est}), we get that $\arg\max_{\left(\theta,\Delta\right)}\max_{\sigma^{2}}\overline{G}^{(3)}(\theta,\Delta,\sigma\mid\mathbf{y})=\arg\max_{\left(\theta,\Delta\right)}\left\{ G\left[\theta,\Delta\mid\mathbf{y}\right]\right\} $
where: 
\[
G\left[\theta,\Delta\mid\mathbf{y}\right]=-0.5\left(d^{o}\sum_{i}n_{i}+qn\right)\ln\left(\sigma^{2}\left(\theta,\Delta\right)\right)+n\ln\left|\triangle\right|+\ln\mathbb{P}\left[\theta,\Delta\right].
\]
Thus we can profile $\overline{G}^{(3)}$ on sigma $\sigma^{2}$ and
define our estimator as:
\begin{equation}
\begin{array}{lll}
\left(\widehat{\theta},\widehat{\Delta}\right) & = & \arg\max_{\left(\theta,\Delta\right)}\left\{ G\left[\theta,\Delta\mid\mathbf{y}\right]\right\} \end{array}\label{eq:pop_param_est}
\end{equation}
 to reduce the optimization problem dimension and focus on the structural
parameters. An estimator of $\sigma^{*}$ is obtained from there by
computing $\sigma^{2}\left(\widehat{\theta},\widehat{\Delta}\right)$
given by equation (\ref{eq:sigma_est}). The details of the outer
criteria derivation are left in appendix A.

\subsection{An asymptotic Variance-Covariance matrix estimator of population
parameters}

In this section, we derive an estimator of the asymptotic variance
of $\left(\widehat{\theta},\widehat{\Delta}\right).$ We highlight
that in practice the matrix $\Delta$ is parametrized by a vector
$\delta$ of dimension $q'$, i.e $\triangle:=\triangle(\delta)$.
We give here a variance estimator of $\left(\widehat{\theta},\widehat{\delta}\right).$
The variance of $\widehat{\Delta}$ can be obtained using classic
delta-methods (see \cite{Vaart1998} chapter 3). First of all, we
drop the vector field dependence in $z$ and we introduce the function:
\[
h(b_{i},\theta,\Delta,\mathbf{y_{i}})=\left\Vert \Delta b_{i}\right\Vert _{2}^{2}+\sum_{j}\left\Vert C\overline{X}_{\theta,b_{i}}(t_{ij})-y_{ij}\right\Vert _{2}^{2}
\]
in order to present sufficient conditions ensuring our estimator is
asymptotically normal:
\begin{enumerate}
\item the function $\widetilde{G}\left[\theta,\Delta(\delta)\right]=-0.5\left(d^{o}\mathbb{E}\left[n_{1}\right]+q\right)\ln\left(\frac{\lim_{n}\frac{1}{n}\sum_{i}^{n}\mathbb{E}\left[h(\widehat{b}(\theta,\Delta(\delta)),\theta,\Delta(\delta),y_{i})\right]}{d^{o}\mathbb{E}\left[n_{1}\right]+q}\right)+\ln\left|\Delta(\delta)\right|$
has a well separated minimum $\left(\overline{\theta},\overline{\delta}\right)$
belonging to the interior of a compact $\Theta\times\Omega\subset\mathbb{R}^{d\times q'}$
\item the true initial conditions $\left\{ x_{0,i}^{*}\right\} _{i\in\left\llbracket 1,n\right\rrbracket }\in\left\llbracket 1,\,n\right\rrbracket $
have finite variance and either
\begin{enumerate}
\item they are i.i.d,
\item for $\nu=0$ and $\nu=1$:
\[
\lim_{n\longrightarrow\infty}\frac{1}{\left(V^{(\nu)}\right)^{2}}\mathbb{E}\left[\sum_{i=1}^{n}\left(\overline{h}^{(\nu)}(\mathbf{y_{i}})-\mathbb{E}\left[\overline{h}^{(\nu)}(\mathbf{y_{i}})\right]\right)^{2}1_{\left\{ \overline{h}(\mathbf{y_{i}})-\mathbb{E}\left[\overline{h}(\mathbf{y_{i}})\right]>\varepsilon\sqrt{V^{(\nu)}}\right\} }\right]
\]
where $\overline{h}^{(\nu)}(\mathbf{y_{i}})=\frac{d^{(\nu)}h}{d^{(\nu)}\left(\theta,\delta\right)}(\widehat{b}_{i}(\overline{\theta},\Delta(\overline{\delta})),\overline{\theta},\Delta(\overline{\delta}),\mathbf{y_{i}})$
and $V^{(\nu)}=\sqrt{\sum_{i}Var(\overline{h}^{(\nu)}(\mathbf{y_{i}}))^{2}}$,
\end{enumerate}
\item the subject specific number of observations $\left\{ n_{i}\right\} _{i\in\left\llbracket 1,n\right\rrbracket }$
are i.i.d and uniformly bounded,
\item for all possibles values $\left(\theta,b_{i}\right)$, the solution
$X_{\theta,b_{i},x_{0,i}^{*}}$ belongs to a compact $\chi$ of $\mathbb{R}^{d}$,
and for all $(t,\theta,x)$, the mapping $b_{i}\longmapsto f_{\theta,b_{i}}(t,x)$
has a compact support $\Theta_{b}$,
\item $\left(\theta,b_{i},t,x\right)\longmapsto f_{\theta,b_{i}}(t,x)$
belongs to $C^{1}(\Theta\times\Theta_{b}\times\left[0,\,T\right]\times\chi,\mathbb{R}^{d})$,
\item the matrices $\frac{\partial^{2}}{\partial^{2}b_{i}}g_{i}(\widehat{b_{i}}\left(\overline{\theta},\Delta(\overline{\delta})\right)\mid\overline{\theta},\Delta(\overline{\delta}),U)$
and $\frac{\partial^{2}\mathcal{C}_{i}}{\partial^{2}x_{0,i}}(\widehat{b_{i}}\left(\overline{\theta},\Delta(\overline{\delta})\right),\overline{X}_{\overline{\theta},\widehat{b}_{i}(\overline{\theta},\Delta(\overline{\delta}))}(0),\overline{u}_{\overline{\theta},\widehat{b}_{i}(\overline{\theta},\Delta(\overline{\delta}))}\mid\overline{\theta},U)$
are of full rank almost surely for every sequence $\mathbf{y_{i}}$,
\item there is a neighborhood $\Theta_{\overline{\theta}}$ of $\overline{\theta}$
such that $\left(\theta,b_{i},t,x\right)\longmapsto f_{\theta,b_{i}}(t,x)\in C^{5}(\Theta_{\overline{\theta}}\times\Theta_{b}\times\left[0,\,T\right]\times\chi,\mathbb{R}^{d})$.
\end{enumerate}
Conditions 1-4 are used to derive the consistency of our estimator
toward $\left(\overline{\theta},\overline{\delta}\right)$ by following
classic steps for M-estimator by proving 1/the uniform convergence
of our stochastic cost function to a deterministic one, 2/the existence
of a well-separated minimum for this deterministic function (\cite{Vaart1998}
chapter 5). Conditions 6-7 ensures that our cost function is asymptotically
smooth enough in the vicinity of $\left(\overline{\theta},\overline{\delta}\right)$
to proceed to a Taylor expansion and transfer the regularity of the
cost function to the asymptotic behavior of $\sqrt{n}(\widehat{\theta}-\overline{\theta},\widehat{\delta}-\overline{\delta})$.
Less restrictive conditions can be established under which our estimator
is still asymptotically normal, in particular regarding $f_{\theta,b_{i}}$
regularity with respect to $t$. Also, we emphasize that the second
assumption does not require to know the distribution of the $x_{0,i}^{*}.$
\begin{theorem}
\label{thm:Asymptotic_normality}Under conditions 1-7, there is a
model dependent lower bound $\lambda$ such that if $\left\Vert U\right\Vert _{2}>\lambda$
then the estimator $\left(\widehat{\theta},\widehat{\delta}\right)$
is asymptotically normal and:
\[
\sqrt{n}(\widehat{\theta}-\overline{\theta},\widehat{\delta}-\overline{\delta})\rightsquigarrow N\left(0,A(\overline{\theta},\overline{\delta})^{-1}B(\overline{\theta},\overline{\delta})\left(A(\overline{\theta},\overline{\delta})^{-1}\right)^{T}\right)
\]
where $A(\overline{\theta},\overline{\delta})=\lim_{n}\frac{1}{n}\sum_{i=1}^{n}\left[\frac{\partial\widetilde{J}(\overline{\theta},\overline{\delta},\mathbf{y_{i}})}{\partial(\theta,\delta)}\right]$,
$B(\overline{\theta},\overline{\delta})=\lim_{n}\frac{1}{n}\left[\sum_{i}\widetilde{J}(\overline{\theta},\overline{\delta},\mathbf{y_{i}})\widetilde{J}(\overline{\theta},\overline{\delta},\mathbf{y_{i}})^{T}\right]$
and the vector valued function $\widetilde{J}(\theta,\delta,\mathbf{y_{i}})=\left(\begin{array}{l}
\widetilde{J}_{\theta}(\theta,\delta,\mathbf{y_{i}})\\
\widetilde{J}_{\delta}(\theta,\delta,\mathbf{y_{i}})
\end{array}\right)$ is given by:
\[
\begin{array}{l}
\widetilde{J}_{\theta}(\theta,\delta,\mathbf{y_{i}})=\frac{d}{d\theta}h(\widehat{b}(\theta,\Delta(\delta)),\theta,\Delta(\delta),y_{i})\\
\widetilde{J}_{\delta}(\theta,\delta,\mathbf{y_{i}})=\frac{d}{d\delta}h(\widehat{b}_{i}(\theta,\Delta(\delta)),\theta,\Delta(\delta),y_{i})-\frac{2}{d^{o}\mathbb{E}\left[n_{1}\right]+q}Tr\left(\triangle(\delta)^{-1}\frac{\partial\triangle(\delta)}{\partial\delta_{k}}\right)h(\widehat{b}_{i}(\theta,\Delta(\delta)),\theta,\Delta(\delta),y_{i}).
\end{array}
\]
\end{theorem}

The proof is left in appendix C. The practical interest of this theorem
is to give an estimator of Variance-Covariance:
\[
V(\widehat{\theta},\widehat{\delta})\simeq\widehat{A}(\widehat{\theta},\widehat{\delta})^{-1}\widehat{B}(\widehat{\theta},\widehat{\delta})\left(\widehat{A}(\widehat{\theta},\widehat{\delta})^{-1}\right)^{T}/n.
\]
In the last equation the matrices $\widehat{A}$ and $\widehat{B}$
are defined by:
\[
\left\{ \begin{array}{l}
\widehat{A}(\widehat{\theta},\widehat{\delta})=-\frac{1}{n}\sum_{i=1}^{n}\frac{\partial J(\widehat{\theta},\widehat{\delta},\mathbf{y_{i}})}{\partial(\theta,\delta)}\\
\widehat{B}(\widehat{\theta},\widehat{\delta})=\frac{1}{n}\sum_{i=1}^{n}J(\widehat{\theta},\widehat{\delta},\mathbf{y_{i}})J(\widehat{\theta},\widehat{\delta},\mathbf{y_{i}})^{T}
\end{array}\right.
\]
where the $(p+q)$ components of the vector valued function $J$ for
$1\leq k\leq p$ are given by 
\[
J_{k}(\theta,\delta,\mathbf{y_{i}})=\frac{d}{d\theta_{k}}h(\widehat{b}(\theta,\Delta(\delta)),\theta,\Delta(\delta),\mathbf{y_{i}})
\]
 and for $p+1\leq k\leq p+q$ by
\[
J_{k}(\theta,\delta,\mathbf{y_{i}})=\frac{d}{d\delta_{k}}h(\widehat{b}_{i}(\theta,\Delta(\delta)),\theta,\Delta(\delta),\mathbf{y_{i}})-\frac{2n}{d^{o}\sum_{i}n_{i}+qn}Tr\left(\triangle(\delta)^{-1}\frac{\partial\triangle(\delta)}{\partial\delta_{k}}\right)h(\widehat{b}_{i}(\theta,\Delta(\delta)),\theta,\Delta(\delta),\mathbf{y_{i}}).
\]
Now that we have proven the existence of the variance matrix $V(\theta^{*},\delta^{*})$
such that $\widehat{\delta}-\delta^{*}\rightsquigarrow N\left(0,V(\theta^{*},\delta^{*})\right)$,
we can use the Delta method to derive the asymptotic normality of
the original matrix $\Psi\left(\widehat{\delta}\right)=\sigma^{2}\left(\Delta(\widehat{\delta})^{T}\Delta(\widehat{\delta})\right)^{-1}$
as well as an estimator of its asymptotic variance. In the case of
a diagonal matrix $\Psi$, composed of the elements $\left(\Psi_{1}^{2},\ldots\Psi_{q}^{2}\right)$
and of the parametrization $\triangle(\delta)=\left(\begin{array}{ccc}
e^{\delta_{1}} & 0 & 0\\
0 & \ddots & 0\\
0 & 0 & e^{\delta_{q}}
\end{array}\right)$ used in section \ref{sec:Simulation}, we derive:
\[
\left(\begin{array}{c}
\Psi_{1}(\widehat{\delta})\\
\vdots\\
\Psi_{q}(\widehat{\delta})
\end{array}\right)-\left(\begin{array}{c}
\Psi_{1}(\delta^{*})\\
\vdots\\
\Psi_{q}(\delta^{*})
\end{array}\right)\rightsquigarrow N\left(0,\sigma^{2}\left(\begin{array}{ccc}
e^{-\delta_{1}^{*}} & 0 & 0\\
0 & \ddots & 0\\
0 & 0 & e^{-\delta_{q}^{*}}
\end{array}\right)V(\theta^{*},\delta^{*})\left(\begin{array}{ccc}
e^{-\delta_{1}^{*}} & 0 & 0\\
0 & \ddots & 0\\
0 & 0 & e^{-\delta_{q}^{*}}
\end{array}\right)\right).
\]
\begin{remark}
The previous theorem \ref{thm:Asymptotic_normality} states that we
retrieve a parametric convergence rate despite a number of nuisance
parameter increasing with the number of subjects. We avoid the pitfall
described in \cite{Sartori2003} for profiled methods, thanks to the
i.i.d structure of the nuisance parameters. This allows us to prevent
bias accumulation for score functions among subjects by using the
central limit theorem. Our estimator shares similarities with conditional
maximum likelihood ones and our proof for asymptotic normality follows
similar steps as in \cite{Andersen1970} since the $\left\{ b_{i}\right\} _{i\in\left\llbracket 1,\,n\right\rrbracket }$
are i.i.d.
\end{remark}

\section{\label{sec:OCA_method_presentation}Numerical procedure for $\overline{u}_{i,\theta,b_{i}},$
$\overline{X}_{\theta,b_{i}}$ and $g_{i}$ computation}

In this section we explain how to get numerical approximations for
$\min_{x_{0,i}^{u}}\left\{ \min_{u_{i}}\mathcal{C}_{i}(b_{i},x_{i,0},u_{i}\mid\theta,U)\right\} $
and $\overline{u}_{i,\theta,b_{i}}$ which are then used to evaluate
$\overline{X}_{\theta,b_{i}}$ and $g_{i}$ defined by equation (\ref{eq:inner_criteria}).
Firstly we approximate $g_{i}$ with a special type of discrete time
optimal control problem, known as 'tracking problem'. Secondly we
adapt the method proposed by \cite{CimenBanks2004,Cimen2004} to solve
it. 

\subsection{\label{sub:OCA_to_solve}$g_{i}$ expression as an optimal control
problem}

We introduce a pseudo-linear version of model (\ref{eq:ControlledODEmodel}):

\begin{equation}
\left\{ \begin{array}{l}
\dot{x}_{i}(t)=A_{\theta,b_{i}}\left(t,x_{i}(t),z_{i}(t)\right)x_{i}(t)+r_{\theta,b_{i}}(t,z_{i}(t))+Bu_{i}(t)\\
x_{i}(0)=x_{i,0}
\end{array}\right.\label{eq:pseudoLinearControlledODEmodel}
\end{equation}
where $A_{\theta,b_{i}}$ (resp. $r_{\theta,b_{i}}$) is a $d\times d$
sized matrix (resp. $d$ dimensional vector) valued function, linked
to the original model by the relation $A_{\theta,b_{i}}\left(t,x_{i}(t),z_{i}(t)\right)x_{i}(t)+r_{\theta,b_{i}}(t,z_{i}(t))=f_{\theta,b_{i}}(t,x_{i}(t),z_{i}(t))$.
This formulation is crucial for solving the optimal control problem
defining our estimators in a computationally efficient way. Linear
models already fit in this formalism with $A_{\theta,b_{i}}\left(t,z_{i}(t)\right)\,:=A_{\theta,b_{i}}\left(t,x_{i}(t),z_{i}(t)\right)$.
For nonlinear models, the pseudo-linear representation is not unique
but always exists \cite{Cimen2004} (in order to exploit this non-uniqueness
as an additional degree of freedom, see \cite{Cimen2008} section
6). 

We consider a discretized version of the perturbed ODE (\ref{eq:ControlledODEmodel})
to proceed to parametric estimation: 
\begin{equation}
\left\{ \begin{array}{l}
x_{i}(t_{k+1}^{d})=\left(I_{d}+\Delta_{k}A_{\theta,b_{i}}(t_{k}^{d},x_{i}(t_{k}^{d}),z_{i}(t_{k}^{d}))\right)x_{i}(t_{k}^{d})+\Delta_{k}r_{\theta,b_{i}}(t_{k}^{d},z_{i}(t_{k}^{d}))+B\Delta_{k}u_{i}(t_{k}^{d})\\
x_{i}(0)=x_{i,0}
\end{array}\right.\label{eq:discretized_equation}
\end{equation}
where the discretization is made at $K_{i}+1$ time points $\left\{ t_{k}^{d}\right\} _{0\leq k\leq K_{i}}$
with $t_{0}^{d}=0$ and $t_{K_{i}}^{d}=t_{in_{i}}$. This set contains
the observations time points i.e. $\left\{ t_{ij}\right\} _{0\leq j\leq n_{i}}\subset\left\{ t_{k}^{d}\right\} _{0\leq k\leq K_{i}}$,
but can be bigger and patient specific, allowing to accurately approximate
$X_{\theta,b_{i},x_{i,0}}$ even when the observations are sparse
on $\left[0,\,T\right]$. We define: 
\begin{itemize}
\item $\Delta_{k}=t_{k+1}^{d}-t_{k}^{d}$, the mesh size between two discretization
time-points,
\item $u_{i}^{d}$ the set of discrete values taken by the control at each
time step i.e $u_{i}^{d}=\left(u(t_{k}^{d}),\ldots,u(t_{K_{i}-1}^{d})\right)$,
\item $w_{k}=1_{\left\{ \exists t_{ij}\,\mid\,t_{ij}=t_{k}^{d}\right\} }/(t_{k+1}^{d}-t_{k}^{d})$
i.e. $w_{k}$ is equal to $1/(t_{k+1}^{d}-t_{k}^{d})$ if $t_{k}^{d}$
corresponds to an observation time $t_{ij}$, otherwise $w_{k}=0$,
\item $y_{k}^{d}$=$y_{ij}$ if $t_{k}^{d}=t_{ij}$, $0$ otherwise,
\item $X_{\theta,b_{i},x_{i,0},u_{i}^{d}}^{d}$ the solution of (\ref{eq:discretized_equation}).
\end{itemize}
The weights $w_{k}$ and the set of extended data $\left\{ y_{k}^{d}\right\} $
are introduced to have a vector of observations with the same length
as $\left\{ t_{k}^{d}\right\} _{0\leq k\leq K_{i}}$. We now introduce
the discretized version of the cost $\mathcal{C}_{i}$ to be minimized:
\begin{equation}
\begin{array}{lll}
\mathcal{C}_{i}^{d}(b_{i},x_{i,0},u_{i}^{d}\mid\theta,U) & = & \sum_{j=0}^{n_{i}}\left\Vert CX_{\theta,b_{i},x_{i,0},u_{i}^{d}}^{d}(t_{ij})-y_{ij}\right\Vert _{2}^{2}+\sum_{k=0}^{K_{i}-1}\triangle_{k}u_{i}(t_{k})^{T}Uu_{i}(t_{k})\\
 & = & \left\Vert CX_{\theta,b_{i},x_{i,0},u_{i}^{d}}^{d}(t_{in_{i}})-y_{in_{i}}\right\Vert _{2}^{2}\\
 & + & \sum_{k=0}^{K_{i}-1}\triangle_{k}\left(\left\Vert CX_{\theta,b_{i},x_{i,0},u_{i}^{d}}^{d}(t_{k}^{d})-y_{k}^{d}\right\Vert _{2}^{2}w_{k}+u_{i}(t_{k})^{T}Uu_{i}(t_{k})\right).
\end{array}\label{eq:discrete_general_cost_function}
\end{equation}
such that our inner criteria $g_{i}$ can be approximated by:
\[
g_{i}(b_{i}\mid\theta,\Delta,U)\simeq\min_{x_{0,i}^{u}}\min_{u_{i}^{d}}\mathcal{C}_{i}^{d}(b_{i},x_{i,0},u_{i}^{d}\mid\theta,U)+\left\Vert \Delta b_{i}\right\Vert _{2}^{2}.
\]
The solution of this discrete control problem will be denoted $\overline{u}_{i,\theta,b_{i}}^{d}$,
and the related optimal trajectory $\overline{X}_{\theta,b_{i}}^{d}$:
they will be used as numerical approximations of $\overline{u}_{i,\theta,b_{i}}$
and $\overline{X}_{\theta,b_{i}}$ respectively.

\subsection{\label{sec:S_tractable_form}Numerical methods for solving the tracking
problem}

We present how to numerically obtain $\min_{x_{0,i}^{u}}\min_{u_{i}^{d}}\mathcal{C}_{i}^{d}(b_{i},x_{i,0},u_{i}^{d}\mid\theta,U)$
as well as the corresponding minimizer $\overline{u}_{i,\theta,b_{i}}^{d}.$
We start with linear ODE models (section \ref{subsec:Linear-models}),
then we consider nonlinear models  (section \ref{subsec:Non-linear-models}).

\subsubsection{\label{subsec:Linear-models}Linear models}

Here, we suppose $A_{\theta,b_{i}}(t,z_{i}(t))\,:=A_{\theta,b_{i}}(t,x,z_{i}(t))$
in model (\ref{eq:ODEmodel-gen}), for the sake of clarity we drop the dependence of $A_{\theta,b_{i}}$ and $r_{\theta,b_{i}}$ in $z_{i}$. For a given set $\left(\theta,b_{i},x_{i,0}\right)$,
Linear-Quadratic theory ensures the existence and uniqueness of the
optimal control $\overline{u}_{i,\theta,b_{i}}^{d}$ and that $\min_{x_{0,i}^{u}}\min_{u_{i}^{d}}\mathcal{C}_{i}^{d}(b_{i},x_{i,0},u_{i}^{d}\mid\theta,U)$
can be computed by solving a discrete final value problem, called
the Riccati equation (e.g. \cite{Sontag1998}). 
\begin{proposition}
\label{prop:discrete_Riccati_accurate_representation}Let us introduce
$(R_{\theta,b_{i},k},\,h_{\theta,b_{i},k})$ for $1\leq k\leq K_{i}$,
the solution of the discrete Riccati equation: 
\begin{equation}
\left\{ \begin{array}{lll}
R_{\theta,b_{i},k} & = & R_{\theta,b_{i},k+1}+\triangle_{k}w_{k}C^{T}C+\Delta_{k}\left(R_{\theta,b_{i},k+1}A_{\theta,b_{i}}(t_{k}^{d})+A_{\theta,b_{i}}(t_{k}^{d})^{T}R_{\theta,b_{i},k+1}\right)\\
 & + & \triangle_{k}^{2}A_{\theta,b_{i}}(t_{k}^{d})^{T}R_{\theta,b_{i},k+1}A_{\theta,b_{i}}(t_{k}^{d})\\
 & - & \triangle_{k}(I_{d}+\triangle_{k}A_{\theta,b_{i}}(t_{k}^{d})^{T})R_{\theta,b_{i},k+1}BG(R_{\theta,b_{i},k+1})B^{T}R_{\theta,b_{i},k+1}(I_{d}+\triangle_{k}A_{\theta,b_{i}}(t_{k}^{d}))\\
h_{\theta,b_{i},k} & = & h_{\theta,b_{i},k+1}-\triangle_{k}w_{k}C^{T}y_{k}^{d}+\triangle_{k}A_{\theta,b_{i}}(t_{k}^{d})^{T}h_{\theta,b_{i},k+1}\\
 & + & \Delta_{k}\left(I_{d}+\Delta_{k}A_{\theta,b_{i}}(t_{k}^{d})\right)^{T}R_{\theta,b_{i},k+1}r_{\theta,b_{i}}(t_{k}^{d})\\
 & - & \Delta_{k}(I_{d}+\Delta_{k}A_{\theta,b_{i}}(t_{k}^{d}))^{T}R_{\theta,b_{i},k+1}BG(R_{\theta,b_{i},k+1})B^{T}\left(h_{\theta,b_{i},k+1}+\Delta_{k}R_{\theta,b_{i},k+1}r_{\theta,b_{i}}(t_{k}^{d})\right)
\end{array}\right.\label{eq:discrete_accurate_Riccati_equation_lincase}
\end{equation}
with final condition $(R_{\theta,b_{i},K_{i}},\,h_{\theta,b_{i},K_{i}})=(C^{T}C,\,-C^{T}y_{in_{i}})$
and $G(R_{\theta,b_{i},k+1})\,=\left[U+\triangle_{k}B^{T}R_{\theta,b_{i},k+1}B\right]^{-1}.$
Hence we get:
\begin{equation}
\begin{array}{l}
g_{i}(b_{i}\mid\theta,\Delta,U)=\left\Vert \Delta b_{i}\right\Vert _{2}^{2}+y_{in_{i}}^{T}y_{in_{i}}\\
-\left(R_{\theta,b_{i},0}^{uk}x_{0,i}^{k}+h_{\theta,b_{i},0}^{u}\right)^{T}\left(R_{\theta,b_{i},0}^{u}\right)^{-1}\left(R_{\theta,b_{i},0}^{uk}x_{0,i}^{k}+h_{\theta,b_{i},0}^{u}\right)+\left(x_{0,i}^{k}\right)^{T}R_{\theta,b_{i},0}^{k}x_{0,i}^{k}+2\left(h_{\theta,b_{i},0}^{k}\right)^{T}x_{0,i}^{k}\\
+\sum_{k=0}^{K_{m}-1}\triangle_{k}\left(w_{k}\left(y_{k}^{d}\right)^{T}y_{k}^{d}+\left(2\left(h_{\theta,b_{i},k+1}\right)^{T}+\Delta_{k}r_{\theta,b_{i}}(t_{k}^{d})^{T}R_{\theta,b_{i},k+1}\right)r_{\theta,b_{i}}(t_{k}^{d})\right)\\
-\sum_{k=0}^{K_{m}-1}\triangle_{k}\left(h_{\theta,b_{i},k+1}+\Delta_{k}R_{\theta,b_{i},k+1}r_{\theta,b_{i}}(t_{k}^{d})\right)^{T}BG(R_{\theta,b_{i},k+1})B^{T}\left(h_{\theta,b_{i},k+1}+\Delta_{k}R_{\theta,b_{i},k+1}r_{\theta,b_{i}}(t_{k}^{d})\right)
\end{array}\label{eq:S_n_CI_linear_case-1}
\end{equation}
where $R_{\theta,b_{i},0}^{u},R_{\theta,b_{i},0}^{uk}$, $R_{\theta,b_{i},0}^{k}$,
$h_{\theta,b_{i},0}^{u}$ and $h_{\theta,b_{i},0}^{k}$ are given
by the following decomposition $R_{\theta,b_{i},0}:=\left(\begin{array}{cc}
R_{\theta,b_{i},0}^{u} & R_{\theta,b_{i},0}^{uk}\\
\left(R_{\theta,b_{i},0}^{uk}\right)^{T} & R_{\theta,b_{i},0}^{k}
\end{array}\right)$ and $h_{\theta,b_{i},0}:=\left(\begin{array}{cc}
h_{\theta,b_{i},0}^{u} & h_{\theta,b_{i},0}^{k}\end{array}\right).$ Moreover, the control $\overline{u}_{i,\theta,b_{i}}^{d}$ which
minimizes the cost (\ref{eq:discrete_general_cost_function}) is unique
and equal to:
\begin{equation}
\overline{u}_{i,\theta,b_{i}}^{d}(t_{k}^{d})=-G(R_{\theta,b_{i},k+1})B^{T}\left(R_{\theta,b_{i},k+1}\left(\left(I_{d}+\triangle_{k}A_{\theta,b_{i}}(t_{k}^{d})\right)\overline{X}_{\theta,b_{i}}^{d}(t_{k}^{d})+\Delta_{k}r_{\theta,b_{i}}(t_{k}^{d})\right)+h_{\theta,b_{i},k+1}\right)\label{eq:opt_control_expression}
\end{equation}
where $\overline{X}_{\theta,b_{i}}^{d}$ is the optimal trajectory,
i.e. the solution of the initial value problem:
\begin{equation}
\left\{ \begin{array}{lll}
\overline{X}_{\theta,b_{i}}^{d}(t_{k+1}^{d}) & = & \left(I_{d}+\triangle_{k}A_{\theta,b_{i}}(t_{k}^{d})\right)\overline{X}_{\theta,b_{i}}^{d}(t_{k}^{d})+\Delta_{k}r_{\theta,b_{i}}(t_{k}^{d})\\
 & - & \triangle_{k}BG(R_{\theta,b_{i},k+1})B^{T}R_{\theta,b_{i},k+1}\left(\left(I_{d}+\triangle_{k}A_{\theta,b_{i}}(t_{k}^{d})\right)\overline{X}_{\theta,b_{i}}^{d}(t_{k})+\Delta_{k}r_{\theta,b_{i}}(t_{k}^{d})\right)\\
 & - & \triangle_{k}BG(R_{\theta,b_{i},k+1})B^{T}h_{\theta,b_{i},k+1}
\end{array}\right.\label{eq:opt_trajectory_expression}
\end{equation}
with estimator $\widehat{x_{i,0}^{u}}=-\left(R_{\theta,b_{i},0}^{u}\right)^{-1}\left(R_{\theta,b_{i},0}^{uk}x_{0}^{k}+h_{\theta,b_{i},0}^{u}\right)$
for $x_{i,0}^{u}$.
\end{proposition}
\begin{remark}
The theoretical basis for replacing $\mathcal{C}_{i}^{d}$ and the
perturbed ODE (\ref{eq:ControlledODEmodel}) by their discretized
counterparts can be found in \cite{Clairon2019} where, under mild
regularity conditions on $A_{\theta,b_{i}}$ and $r_{\theta,b_{i}}$,
$\overline{X}_{\theta,b_{i}}^{d}$ and $\overline{u}_{i,\theta,b_{i}}^{d}$
converge to the solution of the continuous optimal control problem.
\end{remark}

\subsubsection{\label{subsec:Non-linear-models}Non-linear models}

We adapt the method proposed by \cite{Cimen2004} to solve tracking
problem for discrete time models. The outline of the method is the
following: we replace the original problem (\ref{eq:discrete_general_cost_function})
by a recursive sequence of problems, where the $l$-th one is defined
by:
\begin{equation}
\begin{array}{l}
\begin{array}{lll}
\min_{u_{i}^{d}}\mathcal{C}_{i}^{d,l}(b_{i},x_{i,0},u_{i}^{d}\mid\theta,U) & := & \left\Vert CX_{\theta,b_{i},x_{i,0},u_{i}^{d}}^{d,l}(t_{in_{i}})-y_{in_{i}}\right\Vert _{2}^{2}\\
 & + & \sum_{k=0}^{K_{i}-1}\triangle_{k}\left(\left\Vert CX_{\theta,b_{i},x_{i,0},u_{i}^{d}}^{d,l}(t_{k}^{d})-y_{k}^{d}\right\Vert _{2}^{2}w_{k}+u_{i}(t_{k}){}^{T}Uu_{i}(t_{k})\right)
\end{array}\\
\textrm{such that }\left\{ \begin{array}{l}
x_{i}(t_{k+1}^{d})=\left(I_{d}+\Delta_{k}A_{\theta,b_{i}}(t_{k}^{d},\overline{X}_{\theta,b_{i}}^{d,l-1}(t_{k}^{d}),z_{i}(t_{k}^{d}))\right)x_{i}(t_{k}^{d})+\Delta_{k}r_{\theta,b_{i}}(t_{k}^{d},z_{i}(t_{k}^{d}))+B\Delta_{k}u_{i}(t_{k})\\
x_{i}(0)=x_{i,0}.
\end{array}\right.
\end{array}\label{eq:Sequence_LQ_problem}
\end{equation}
where $\overline{X}_{\theta,b_{i}}^{d,l-1}$ is the solution of problem
(\ref{eq:Sequence_LQ_problem}) at iteration $l-1$. Thus, for each
$l$ the matrix $A_{\theta,b_{i}}(t_{k}^{d},\overline{X}_{\theta,b_{i}}^{d,l-1}(t_{k}^{d}),z_{i}(t_{k}^{d}))$
does not depend on $x_{i}$ and the problem (\ref{eq:Sequence_LQ_problem})
is a Linear-Quadratic one. We use the results of section \ref{subsec:Linear-models}
to construct the following algorithm:
\begin{enumerate}
\item Initialization phase: $\overline{X}_{\theta,b_{i}}^{u,d,0}(t_{k}^{d})=x_{i,0}^{u,r}$
for all $k\in\left\llbracket 0,\,n_{i}\right\rrbracket $ where $x_{i,0}^{u,r}$
is an arbitrary starting point for the unknown initial condition and
$\overline{X}_{\theta,b_{i}}^{k,d,0}(t_{k}^{d})=x_{i,0}^{k}.$
\item At iteration $l$: use proposition \ref{prop:discrete_Riccati_accurate_representation}
to obtain $(R_{\theta,b_{i}}^{l},\,h_{\theta,b_{i}}^{l}),$ $\overline{u}_{i,\theta,b_{i}}^{d,l}$,$\overline{X}_{\theta,b_{i}}^{d,l}$
and $g_{i}^{l}(b_{i}\mid\theta,\Delta,U)$. 
\item If $\sum_{k=1}^{K_{i}}\left\Vert \overline{X_{\theta,b_{i}}^{d,l}}(t_{k}^{d})-\overline{X_{\theta,b_{i}}^{d,l-1}}(t_{k}^{d})\right\Vert _{2}^{2}<\varepsilon_{1}$
and $\left|g_{i}^{l}(b_{i}\mid\theta,\Delta,U)-g_{i}^{l-1}(b_{i}\mid\theta,\Delta,U)\right|<\varepsilon_{2}$,
then step 4; otherwise get back to step 2.
\item Set $(R_{\theta,b_{i}},\,h_{\theta,b_{i}})\,=(R_{\theta,b_{i}}^{l},\,h_{\theta,b_{i}}^{l})$
, $\overline{u}_{i,\theta,b_{i}}^{d}\,=\overline{u}_{i,\theta,b_{i}}^{d,l}$,
$\overline{X}_{\theta,b_{i}}^{d}\,=\overline{X}_{\theta,b_{i}}^{d,l}$
and $g_{i}(b_{i}\mid\theta,\Delta,U)\,=g_{i}^{l}(b_{i}\mid\theta,\Delta,U)$. 
\end{enumerate}

\section{\label{sec:Simulation}Results on simulated data}

We compare the accuracy of our approach with maximum likelihood (ML)
in different models and experimental designs reflecting the problems
exposed in introduction, that is estimation in 1/presence of model
error, 2/partially observed framework with unknown initial conditions
and 3/presence of poorly identifiable parameters. For the fairness
of comparison with ML where no prior is specified, we choose a non-informative
one i.e. $\ln\mathbb{P}\left[\theta,\Delta\right]=0$ for our method
throughout this section. If the differential equation (\ref{eq:ODEmodel-gen})
has an analytical solution, the ML estimator is computed via SAEM
algorithm (SAEMIX package \cite{Comets2017}). Otherwise it is done
via a restricted likelihood method dedicated to ODE models implemented
in the nlmeODE package \cite{Tornoe2004}. For both our method and
the ML, we proceed to Monte-Carlo simulations based on $N_{MC}=100$
runs. At each run, we generate $n_{i}$ observations coming from $n$
subjects on an observation interval $\left[0,\,T\right]$ with Gaussian
measurement noise of standard deviation $\sigma^{*}$. From these
data, we estimate the true population parameters $\theta^{*}$, $\Psi^{*}$
as well as the subject parameter realizations $b_{i}^{*}\sim N(0,\Psi^{*})$
with both estimation methods. We quantify the accuracy of each entry
$\widehat{\psi}_{p}$ of the population parameters estimate $\widehat{\psi}=\left(\widehat{\theta},\widehat{\Psi}\right)$
via Monte-Carlo computation of the bias $b(\widehat{\psi}_{p})=\mathbb{E}\left[\widehat{\psi}_{p}\right]-\psi_{p}^{*}$,
the empirical variance $V_{emp}(\widehat{\psi}_{p})=\mathbb{E}\left[\left(\mathbb{E}\left[\widehat{\psi}_{p}\right]-\psi_{p}^{*}\right)^{2}\right]$,
the mean square error $MSE(\widehat{\psi}_{p})=b(\widehat{\psi}_{p})^{2}+V_{emp}(\widehat{\psi}_{p})$,
the estimated variance $\widehat{V}\left(\widehat{\psi}_{p}\right)$
as well as the coverage rate of the 95\%-confidence interval derived
from it, it corresponds to the frequency at which the interval $\left[\widehat{\psi}_{p}\pm z_{0.975}\sqrt{\widehat{V}\left(\widehat{\psi}_{p}\right)}\right]$
contains $\psi_{p}^{*}$ with $z_{0.975}$ the $0.975-$quantile of
the centered Gaussian law. We compute the previous quantities for
the normalized values $\widehat{\psi}_{p}^{norm}:=\frac{\widehat{\psi}_{p}}{\psi_{p}^{*}}$
to make relevant comparisons among parameters with different order
of magnitude. For the subject specific parameter, we estimate the
mean square error $MSE(\widehat{b}_{i})=\mathbb{E}\left[\left\Vert b_{i}^{*}-\widehat{b_{i}}\right\Vert _{2}^{2}\right]$.
For each subsequent examples, we give the results for $n=50$ and
present in appendix B the case $n=20$ to analyze the evolution of
each estimator accuracy with respect to the sparsity of the available
observations.

For our method, we need to select $U$ the matrix appearing in the
inner criteria definition (\ref{eq:inner_criteria}) balancing model
and data fidelity. We use for this the forward cross-validation method
presented in \cite{Hooker2011}. Let us denote $\widehat{\theta_{U}}$,
$\left\{ \widehat{b_{i,U}}\right\} _{i\in\left\llbracket 1,\,n\right\rrbracket }$
the estimators obtained for a given matrix $U$. For each subject
$i$, we split $\left[0,T\right]$ into $H=2$ sub-intervals $\left[t_{h},\,t_{h+1}\right]$,
such that $t_{1}=0$ and $t_{H}=T$. We denote $X_{\theta,b_{i}}(.,t_{h},x_{h})$
the solution of $\dot{x}(t)=f_{\theta,b_{i}}\left(t,x_{i}(t),z_{i}(t)\right)$
defined on the interval $\left[t_{h},\,t_{h+1}\right]$ with initial
condition $X_{\theta,b_{i}}(t_{h},t_{h},x_{h})=x_{h}.$The forward
cross-validation uses the causal relation imposed to the data by the
ODE to quantify the prediction error:
\[
\textrm{EP}(i,U)=\sum_{h=1}^{H}\sum_{\left\{ t_{ij}\in\left[t_{h},\,t_{h+1}\right]\right\} }\left\Vert y_{ij}-CX_{\widehat{\theta_{U}},\widehat{b_{i,U}}}(t_{ij},t_{h},\overline{X}_{\widehat{\theta_{U}},\widehat{b_{i,U}}}(t_{h}))\right\Vert _{2}^{2}.
\]
The rationale of this selection method is the following: if $U$ is
too small, $C\overline{X}_{\widehat{\theta_{U}},\widehat{b_{i,U}}}(t_{h})$
will be close to $y_{h}$ but not to the actual ODE solution, and
$t\longmapsto CX_{\widehat{\theta_{U}},\widehat{b_{i,U}}}(t,t_{h},\overline{X}_{\widehat{\theta_{U}},\widehat{b_{i,U}}}(t_{h}))$
will diverge from the observations on $\left[t_{h},\,t_{h+1}\right]$.
If $U$ is too large, $\overline{X}_{\widehat{\theta_{U}},\widehat{b_{i,U}}}(t_{h})$
will be close to the ODE solution but far from $y_{h}$ and it will
lead to a large value for $\textrm{EP}(i,U)$. Thus, a proper value
for $U$ which minimizes $\textrm{EP}(i,U)$ will be chosen between
these two extreme cases. The global prediction error for the whole
population is computed with $\textrm{EP}(U)=\sum_{i}\textrm{EP}(i,U)$.
We retain the matrix $U$ which minimizes EP among a trial of tested
values and we denote $\widehat{\theta},\widehat{\Psi},\,\left\{ \widehat{b_{i,}}\right\} _{i\in\left\llbracket 1,\,n\right\rrbracket }$
the corresponding estimator. In the following, we use the subscript
$ML$ to denote the ML estimator.

For solving the optimization problems required for computing our inner
and outer criteria, we use the Nelder-Mead algorithm implemented in
the optimr package \cite{nash2016using}. All optimization algorithms
used by the estimation methods require a starting guess value. We
start from the true parameter value for each of them. By doing so,
we aim to do not mix two distinct problems: 1)the numerical stability
of the estimation procedures, 2)the intrinsic accuracy of the different
estimators. These two problems are correlated, but we aim
to adress only the latter which corresponds to the issues raised in
introduction. Still, we check on preliminary analysis that local minima
presence was not an issue in the vicinity of $\left(\theta^{*},\triangle^{*}\right)$
by testing different starting points for all methods. No problem appears
for our method and SAEMIX. A negligible number of non convergence
cases appear for nlmODE which have been discarded thanks to the convergence
criteria embedded in the package.

\subsection{Partially observed linear model}

\begin{figure}
\includegraphics[scale=0.35]{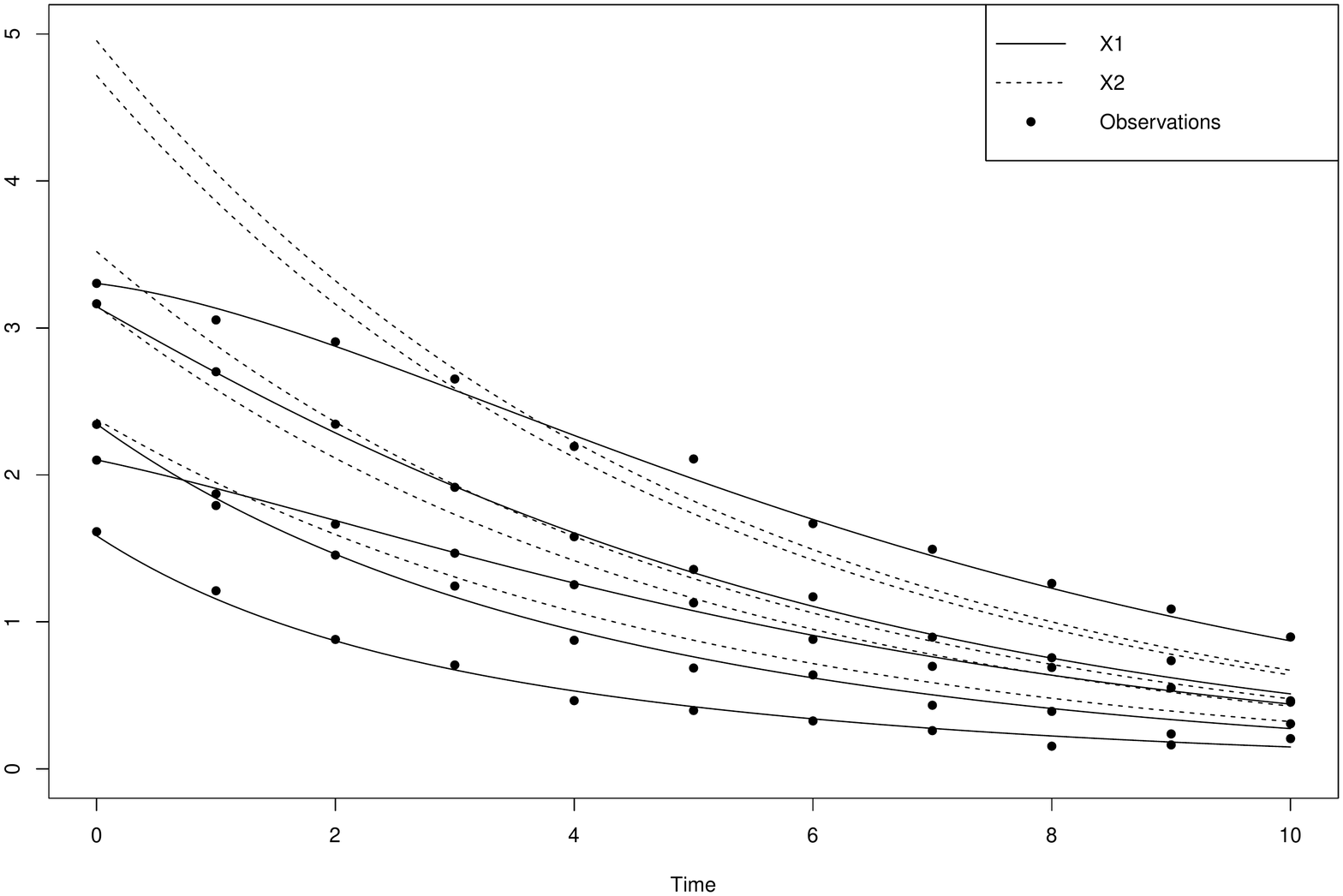}\caption{\label{fig:Lin2dim_exampletraj}Examples of (\ref{eq:lin_model_2dim})
solutions and corresponding observations.}
\end{figure}

We consider the population model where each subject $i$ follows the
ODE: 
\begin{equation}
\left\{ \begin{array}{l}
\dot{X}_{1,i}=\phi_{2,i}X_{2,i}-\phi_{1,i}X_{1,i}\\
\dot{X}_{2,i}=-\phi_{2,i}X_{2,i}\\
\left(X_{1,i}(0),X_{2,i}(0)\right)=\left(x_{1,0},x_{2,0,i}\right)
\end{array}\right.\label{eq:lin_model_2dim}
\end{equation}
with the following parametrization:
\[
\left\{ \begin{array}{l}
\log(\phi_{1,i})=\theta_{1}+b_{i}\\
\log(\phi_{2,i})=\theta_{2}
\end{array}\right.
\]
where $b_{i}\sim N(0,\Psi)$. The true population parameter values
are $\theta^{*}=(\theta_{1}^{*},\theta_{2}^{*})=\left(\log\left(0.5\right),\,\log\left(2\right)\right)$,
and $\Psi^{*}=0.5^{2}$ and we are in a partially observed framework
where only $X_{1,i}$ is accessible. The true initial conditions are subject specifics and normally distributed with $x_{1,0,i}^{*}\sim N(2,\,0.5)$
and $x_{2,0,i}^{*}\sim N(3,\,1).$ ODE (\ref{eq:lin_model_2dim})
has an analytic solution given by $X_{1,i}(t)=e^{-\phi_{1,i}t}(x_{1,0}+\frac{x_{2,0}\phi_{2,i}}{\phi_{1,i}-\phi_{2,i}}(e^{\left(\phi_{1,i}-\phi_{2,i}\right)t}-1))$
for its first component which will be used for parameter estimation
with the SAEMIX package. We generate $n_{i}=11$ observations per
subject on $\left[0,\,T\right]=\left[0,\,10\right]$ with Gaussian
measurement noise of standard deviation $\sigma=0.05$. An example
of observations and corresponding solution is plotted in figure \ref{fig:Lin2dim_exampletraj}.

We want to investigate the impact of initial condition, especially
the unobserved one $x_{2,0,i}^{*}$, on the ML estimator accuracy.
Indeed, our method does not need to estimate $x_{2,0,i}^{*}$ and
thus no additional difficulties appear in this partially observed
framework. For the ML, however, it is nuisance subject-specific parameter
that should be estimated and for which no observations are available.
For this, we compute $\widehat{\theta}_{MLx_{0,1},x_{0,2}}$, $\widehat{\theta}_{ML,x_{0,2}}$
and $\widehat{\theta}_{ML}$ the ML estimator respectively when: 1)
both initial conditions are perfectly known, 2) $x_{1,0,i}^{*}$ is
replaced by the measured value, 3)in addition $x_{2,0,i}^{*}$ has
to be estimated.

\subsubsection{Correct model case}

We present the estimation results in table \ref{tab:lin2dim_Bias_Var_pop_param}.
For ML, the results are goods in terms of accuracy and consistent
in terms of asymptotic confidence interval coverage rate when both
initial conditions are known: 95\% for $\theta_{1}$ and $\theta_{2}$
in accordance with theoretical results. However, there is a significant
drop in accuracy when $x_{2,0,i}^{*}$ has to be estimated, especially
for $\theta_{2}$. In particular, the coverage rate drops to 86\%
and 80\% for $\theta_{1}$ and $\theta_{2}$ respectively. Interestingly,
ML inaccuracy is driven by bias and under-estimated variance when
initial conditions are not known. In this case our method provides
a relevant alternative: it gives accurate estimations with a good
coverage rate for all parameters while avoiding the estimation of
the unobserved initial conditions. Estimation of individual random
effects is also more accurate with our method, with a decrease of
more than 90\% of MSE for $b_{i}$ comparing to ML.

\begin{table}
\caption{\label{tab:lin2dim_Bias_Var_pop_param}Results of estimation for model
(\ref{eq:lin_model_2dim}). The different subscripts stand for the
following estimation scenarios: 1)$\left(x_{0,1},x_{0,2}\right)$
when both initial conditions are set to $\left(x_{0,1}^{*},x_{0,2}^{*}\right)$,
2)$x_{0,2}$ when $x_{0,i}$ is set to $y_{i,0}$ and $x_{0,2}$ to
$x_{0,2}^{*}$, 3/absence of subscript when $x_{0,i}$ is set to $y_{i,0}$
and $x_{0,2}$ is estimated. Results from our method are in bold.}
{\footnotesize{}}%
\begin{tabular}{c|lcccccc}
\hline 
\multicolumn{8}{l}{Well-specified model}\tabularnewline
\hline 
\hline 
\multicolumn{1}{c}{} &  & {\footnotesize{}MSE} & {\footnotesize{}Bias} & {\footnotesize{}Emp. Var} & {\footnotesize{}Est. Var} & {\footnotesize{}Cov. Rate} & {\footnotesize{}MSE $b_{i}$}\tabularnewline
\hline 
\hline 
\multirow{4}{*}{{\footnotesize{}$\theta_{1}$}} & {\footnotesize{}$\widehat{\theta}_{ML,x_{0,1},x_{0,2}}$} & 0.01 & 0.01 & 0.01 & 0.01 & 0.95 & \tabularnewline
 & {\footnotesize{}$\widehat{\theta}_{ML,x_{0,2}}$} & 0.01 & 0.01 & 0.01 & 0.01 & 0.94 & \tabularnewline
 & {\footnotesize{}$\widehat{\theta}_{ML}$} & 0.04 & -0.04 & 0.04 & 0.01 & 0.86 & \tabularnewline
 & \textbf{\footnotesize{}$\widehat{\theta}$} & \textbf{5e-3} & \textbf{8e-3} & \textbf{8e-3} & \textbf{1e-2} & \textbf{0.97} & \tabularnewline
\hline 
\multirow{4}{*}{{\footnotesize{}$\theta_{2}$}} & {\footnotesize{}$\widehat{\theta}_{ML,x_{0,1},x_{0,2}}$} & 4e-5 & 1e-3 & 4e-5 & 4e-5 & 0.95 & \tabularnewline
 & {\footnotesize{}$\widehat{\theta}_{ML,x_{0,2}}$} & 6e-5 & 1e-3 & 6e-5 & 8e-5 & 0.94 & \tabularnewline
 & {\footnotesize{}$\widehat{\theta}_{ML}$} & 4e-3 & -0.01 & 3e-3 & 1e-4 & 0.80 & \tabularnewline
 & \textbf{\footnotesize{}$\widehat{\theta}$} & \textbf{5e-5} & \textbf{2e-3} & \textbf{4e-5} & \textbf{4e-5} & \textbf{0.93} & \tabularnewline
\hline 
\multirow{4}{*}{{\footnotesize{}$\Psi$}} & {\footnotesize{}$\widehat{\theta}_{ML,x_{0,1},x_{0,2}}$} & 0.01 & -0.03 & 0.01 & 7e-3 & 1 & 5e-3\tabularnewline
 & {\footnotesize{}$\widehat{\theta}_{ML,x_{0,2}}$} & 0.02 & -0.03 & 0.01 & 7e-3 & 1 & 5e-3\tabularnewline
 & {\footnotesize{}$\widehat{\theta}_{ML}$} & 0.05 & 0.17 & 0.02 & 0.02 & 1 & 0.10\tabularnewline
 & \textbf{\footnotesize{}$\widehat{\theta}$} & \textbf{0.01} & \textbf{-0.01} & \textbf{0.01} & \textbf{0.01} & \textbf{0.92} & \textbf{0.01}\tabularnewline
\hline 
\multicolumn{8}{l}{Misspecified model}\tabularnewline
\hline 
\hline 
\multicolumn{1}{c}{} &  & {\footnotesize{}MSE} & {\footnotesize{}Bias} & {\footnotesize{}Emp. Var} & {\footnotesize{}Est. Var} & {\footnotesize{}Cov. Rate} & {\footnotesize{}MSE $b_{i}$}\tabularnewline
\hline 
\hline 
\multirow{4}{*}{{\footnotesize{}$\theta_{1}$}} & {\footnotesize{}$\widehat{\theta}_{ML,x_{0,1},x_{0,2}}$} & 0.01 & 4e-4 & 0.01 & 0.01 & 0.91 & \tabularnewline
 & {\footnotesize{}$\widehat{\theta}_{ML,x_{0,2}}$} & 0.01 & -3e-4 & 0.01 & 1e-4 & 0.89 & \tabularnewline
 & {\footnotesize{}$\widehat{\theta}_{ML}$} & 0.05 & 0.02 & 0.05 & 0.01 & 0.81 & \tabularnewline
 & \textbf{\footnotesize{}$\widehat{\theta}$} & \textbf{0.01} & \textbf{-8e-3} & \textbf{7e-3} & \textbf{0.05} & \textbf{0.97} & \tabularnewline
\hline 
\multirow{4}{*}{{\footnotesize{}$\theta_{2}$}} & {\footnotesize{}$\widehat{\theta}_{ML,x_{0,1},x_{0,2}}$} & 1e-4 & -1e-3 & 1e-4 & 1e-4 & 0.83 & \tabularnewline
 & {\footnotesize{}$\widehat{\theta}_{ML,x_{0,2}}$} & 1e-4 & -1e-3 & 2e-4 & 0.01 & 0.82 & \tabularnewline
 & {\footnotesize{}$\widehat{\theta}_{ML}$} & 4e-3 & -2e-3 & 4e-3 & 2e-4 & 0.63 & \tabularnewline
 & \textbf{\footnotesize{}$\widehat{\theta}$} & \textbf{1e-4} & \textbf{2e-5} & \textbf{1e-4} & \textbf{1e-4} & \textbf{0.92} & \tabularnewline
\hline 
\multirow{4}{*}{{\footnotesize{}$\Psi$}} & {\footnotesize{}$\widehat{\theta}_{ML,x_{0,1},x_{0,2}}$} & 0.01 & -0.003 & 0.01 & 0.01 & 1 & 0.01\tabularnewline
 & {\footnotesize{}$\widehat{\theta}_{ML,x_{0,2}}$} & 0.01 & -0.005 & 0.01 & 0.01 & 1 & 0.01\tabularnewline
 & {\footnotesize{}$\widehat{\theta}_{ML}$} & 0.09 & 0.21 & 0.04 & 0.03 & 1 & 0.12\tabularnewline
 & \textbf{\footnotesize{}$\widehat{\theta}$} & \textbf{0.02} & \textbf{-0.02} & \textbf{0.02} & \textbf{0.01} & \textbf{0.90} & \textbf{0.01}\tabularnewline
\hline 
\end{tabular}
{\footnotesize\par}

\end{table}
\subsubsection{Estimation in presence of model error at the subject level}

\begin{figure}
\includegraphics[scale=0.35]{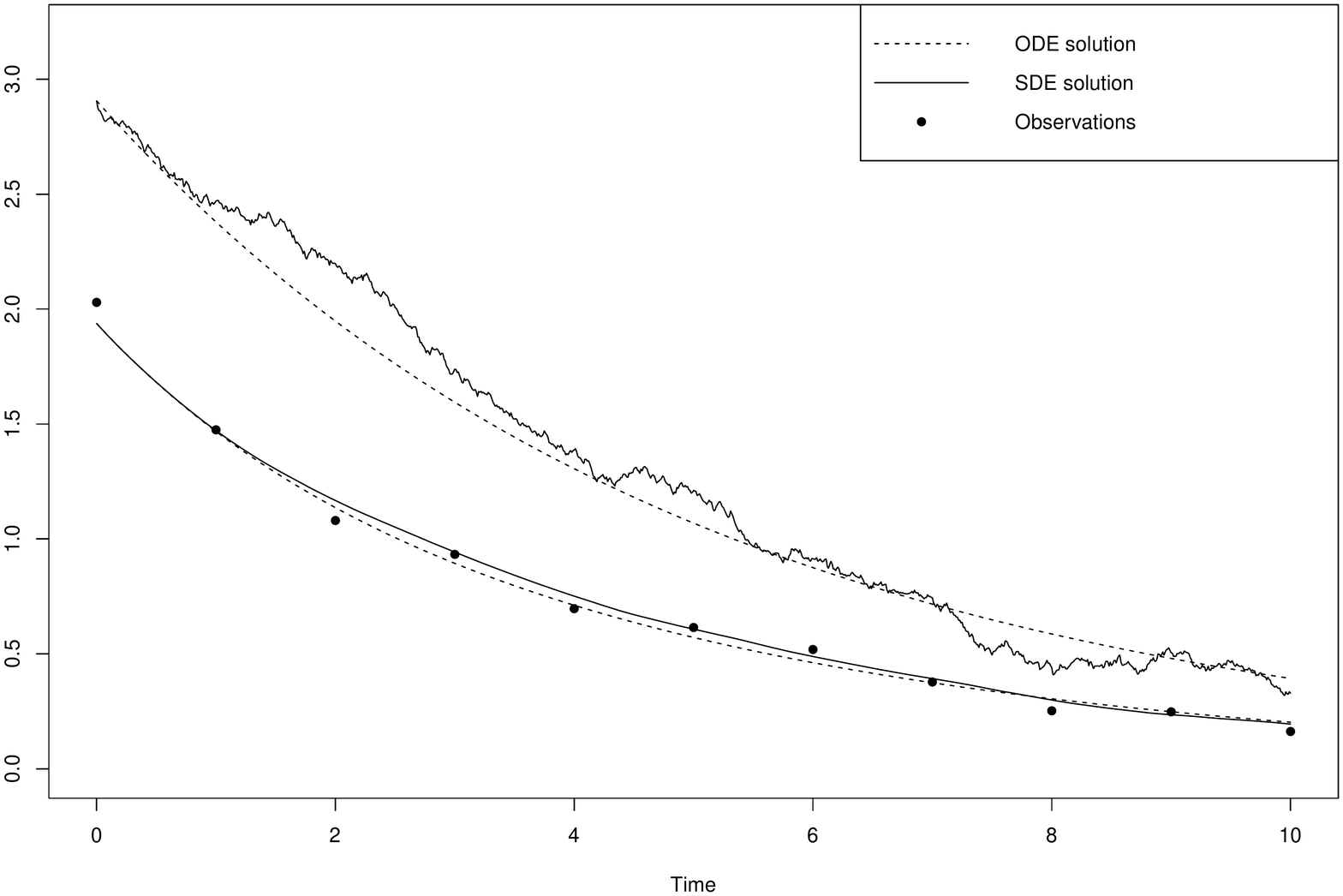}\caption{\label{fig:lin2dim_pert_ODE_rec}Solution of (\ref{eq:lin_model_2dim})
and a realization of (\ref{eq:stoch_lin_model_2dim}) for the same
parameter values.}
\end{figure}

To mimic misspecification presence, we now generate the observations
from the hypoelliptic stochastic model:
\begin{equation}
\left\{ \begin{array}{l}
dX_{1,i}=\phi_{2,i}X_{2,i}dt-\phi_{1,i}X_{1,i}dt\\
dX_{2,i}=-\phi_{2,i}X_{2,i}dt+\alpha dB_{t}\\
\left(X_{1,i}(0),X_{2,i}(0)\right)=\left(x_{1,0},x_{2,0,i}\right)
\end{array}\right.\label{eq:stoch_lin_model_2dim}
\end{equation}
with $B_{t}$ a Wiener process and $\alpha=0.1$ the diffusion coefficient.
For the sake of comparison, a solution of (\ref{eq:lin_model_2dim})
and a realization of its perturbed counterpart given by (\ref{eq:stoch_lin_model_2dim})
are plotted in figure \ref{fig:lin2dim_pert_ODE_rec}. This framework
where stochasticity only affects the unmeasured compartment is known
to be problematic for parameter estimation and inference procedures
are yet to be developed for sparse sampling case. From figure \ref{fig:lin2dim_pert_ODE_rec}
it is easy to see the diffusion $\alpha$ will be hard to estimate
when we only have observations for $X_{1,i}$. Thus, we still estimate
the parameters from the model (\ref{eq:lin_model_2dim}) which is
now seen as a deterministic approximation of the true stochastic process.
Still, it is expected that our method will mitigate the effect of
stochasticity on the estimation accuracy by taking into account model
error presence. Results are presented in table \ref{tab:lin2dim_Bias_Var_pop_param}.
The differences between the two methods are similar to the previous
well-specified case with an additional loss of accuracy coming from
model error for both estimators. However, the misspecification effect
for SAEM is more pronounced than for our method which manages to limit
the damages done. This confirms the benefits of taking into account
model uncertainty for the regularization of the inverse problem, in
particular when model error occurs in the unobserved compartment,
a situation in which classic statistical criteria for model assessment
based on a data fitting criterion are difficult to use. 

\subsection{Partially observed nonlinear model}

We consider a simplified version of the model used in \cite{Tornoe2004}
for the analysis of glucose and insulin regulation:
\begin{equation}
\left\{ \begin{array}{l}
\dot{G_{i}}=S_{G}(G_{B}-G_{i})-X_{i}G_{i}\\
\dot{I_{i}}=\gamma t(G_{i}-h)-n_{i}(I_{i}-I_{B})\\
\dot{X_{i}}=-p_{2}(X_{i}+S_{I}(I_{i}-I_{B})).
\end{array}\right.\label{eq:Insulin_model}
\end{equation}
We are in a partially observed framework where only the glucose ($G_{i}$)
and insulin ($I_{i}$) concentration are measured. The values of parameters
$\left(p_{2},\gamma,h,G_{B},I_{B}\right)$ are fixed to $\left(-4.93,-6.85,4.14,100,100\right)$
and we aim to estimate $\theta=\left(\theta_{S_{G}},\theta_{S_{I}},\theta_{n}\right)$,
linked to the original model via the parametrization:
\[
\left\{ \begin{array}{l}
\log(S_{G})=\theta_{S_{G}}\\
\log(S_{I})=\theta_{S_{I}}\\
\log(n_{i})=\theta_{n}+b_{i}
\end{array}\right.
\]
where $b_{i}\sim N(0,\Psi)$. The true population parameter values
are $\theta^{*}=\left(-3.89,\,-7.09,\,-1.81\right)$ and $\Psi^{*}=0.26^{2}$.
The true initial conditions $x_{i,0}^{*}=\left(G_{0,i}^{*},I_{0,i}^{*},X_{0,i}^{*}\right)$
are subject-specific and distributed according to
$\ln(x_{i,0}^{*})\sim N(l_{x_{0}^{*}},\Psi_{l_{x_{0}^{*}}})$ with
$l_{x_{0}^{*}}=(5.52,4.88,-7)$ and $\Psi_{l_{x_{0}^{*}}}=\left(0.17^{2},\,0.1^{2},10^{-4}\right).$
We generate $n_{i}=5$ observations on $\left[0,\,T\right]=\left[0,\,180\right]$
with Gaussian measurement noise of standard deviation $\sigma^{*}=3$.
As in the previous example, we investigate the impact of unknown initial
conditions on estimators accuracy. We are particularly interested
in the joint estimation of $\theta_{S_{I}}$, which appears only in
the equation ruling the unobserved state variable $X_{i},$ and $X_{0,i}^{*}$
required for each subject by the maximum likelihood based method.
For this, we distinguish two cases, 1)when $\theta_{S_{I}}$ is known,
2)when $\theta_{S_{I}}$ has to be estimated here and we respectively denote
$\widehat{\theta_{S_{i}}}$ and $\widehat{\theta}$ the corresponding
estimators. Finally, since the model is nonlinear we have to specify
a pseudo-linear representation to use the algorithm presented in section
3.2.2:
\[
A_{\theta,b_{i}}\left(t,G_{i},I_{i},X_{i}\right)=\left(\begin{array}{ccc}
-S_{G} & 0 & -G_{i}\\
\gamma t & -n_{i} & 0\\
0 & -p_{2}S_{I} & -p_{2}
\end{array}\right),\,r_{\theta,b_{i}}\left(t\right)=\left(\begin{array}{c}
S_{G}G_{B}\\
-\gamma th+n_{i}I_{B}\\
-p_{2}S_{I}I_{B}
\end{array}\right).
\]

\subsubsection{Correct model case}

We present the estimation results in table \ref{tab:Insulin_model_Bias_Var_pop_param}.
Our method obtains smaller MSE than ML and escapes the drop in coverage
rate of the confidence interval in the case of $\theta_{S_{I}}^{*}$
estimation. The difference between the two estimators behavior is
explained by the fact that they are defined through the construction
of two different optimization problems. At the population level our
approach leads to minimize a cost function depending on a 4-dimensional
parameter whereas ML, due to its need to estimate $x_{i,0}^{*}$,
considers a 7-dimensional one. Thus, the topology of the parameter
spaces explored by each method to look for the minimum are very different.

\begin{table}
\caption{\label{tab:Insulin_model_Bias_Var_pop_param}Results of estimation
for model (\ref{eq:Insulin_model}). The different subscripts stand
for the following estimation scenarios: 1)$S_{i}$ when $S_{i}$ is
set to $S_{i}^{*}$, 2)absence of subscript when $S_{i}$ is estimated.
Results from our method are in bold.}
{\footnotesize{}}%
\begin{tabular}{c|lcccccc}
\hline 
\multicolumn{8}{l}{Well-specified model}\tabularnewline
\hline 
\multicolumn{1}{c}{} &  & {\footnotesize{}MSE} & {\footnotesize{}Bias} & {\footnotesize{}Emp. Var} & {\footnotesize{}Est. Var} & {\footnotesize{}Cov. Rate} & {\footnotesize{}MSE $b_{i}$}\tabularnewline
\hline 
\hline 
\multirow{4}{*}{{\footnotesize{}$\theta_{S_{G}}$}} & {\footnotesize{}$\widehat{\theta}_{ML,S_{i}}$} & 4.6e-5 & 2.2e-3 & 4.1e-5 & 8.8e-6 & 0.95 & \tabularnewline
 & {\footnotesize{}$\widehat{\theta}_{ML}$} & 2.0e-3 & 0.03 & 1.0e-3 & 7.6e-5 & 0.85 & \tabularnewline
 & {\footnotesize{}$\widehat{\theta}_{S_{i}}$} & \textbf{1.0e-5} & \textbf{3.8e-4} & \textbf{1.0e-5} & \textbf{8.2e-6} & \textbf{0.95} & \tabularnewline
 & {\footnotesize{}$\widehat{\theta}$} & \textbf{1.8e-4} & \textbf{-5.5e-4} & \textbf{1.8e-4} & \textbf{1.5e-4} & \textbf{0.96} & \tabularnewline
\hline 
\multirow{4}{*}{{\footnotesize{}$\theta_{S_{I}}$}} & {\footnotesize{}$\widehat{\theta}_{ML,S_{i}}$} & known &  &  &  &  & \tabularnewline
 & {\footnotesize{}$\widehat{\theta}_{ML}$} & 2.2e-3 & 0.03 & 1.2e-3 & 6.4e-5 & 0.90 & \tabularnewline
 & {\footnotesize{}$\widehat{\theta}_{S_{i}}$} & \textbf{known} &  &  &  &  & \tabularnewline
 & {\footnotesize{}$\widehat{\theta}$} & \textbf{1.3e-4} & \textbf{-7.1e-4} & \textbf{1.3e-4} & \textbf{1.1e-4} & \textbf{0.96} & \tabularnewline
\hline 
\multirow{4}{*}{{\footnotesize{}$\theta_{n}$}} & {\footnotesize{}$\widehat{\theta}_{ML,S_{i}}$} & 7.0e-4 & 2.8e-3 & 6.0e-4 & 5.0e-4 & 0.94 & \tabularnewline
 & {\footnotesize{}$\widehat{\theta}_{ML}$} & 8.5e-4 & 8.0e-3 & 8.4e-4 & 5.0e-4 & 0.86 & \tabularnewline
 & {\footnotesize{}$\widehat{\theta}_{S_{i}}$} & \textbf{5.2e-4} & \textbf{5.7e-3} & \textbf{5.1e-4} & \textbf{5.0e-4} & \textbf{0.95} & \tabularnewline
 & {\footnotesize{}$\widehat{\theta}$} & \textbf{5.6e-4} & \textbf{5.6e-3} & \textbf{5.2e-4} & \textbf{5.1e-4} & \textbf{0.95} & \tabularnewline
\hline 
\multirow{4}{*}{{\footnotesize{}$\Psi$}} & {\footnotesize{}$\widehat{\theta}_{ML,S_{i}}$} & 0.02 & 6.5e-4 & 0.02 & 0.02 & 0.95 & 0.02\tabularnewline
 & {\footnotesize{}$\widehat{\theta}_{ML}$} & 0.04 & -0.09 & 0.03 & 0.02 & 0.88 & 0.02\tabularnewline
 & {\footnotesize{}$\widehat{\theta}_{S_{i}}$} & \textbf{0.01} & \textbf{-2.0e-3} & \textbf{0.01} & \textbf{0.01} & \textbf{0.95} & \textbf{0.01}\tabularnewline
 & {\footnotesize{}$\widehat{\theta}$} & \textbf{0.01} & \textbf{2.7e-3} & \textbf{0.01} & \textbf{0.01} & \textbf{0.94} & \textbf{0.01}\tabularnewline
\hline 
\multicolumn{8}{l}{Misspecified model}\tabularnewline
\hline 
\multicolumn{1}{c}{} &  & {\footnotesize{}MSE} & {\footnotesize{}Bias} & {\footnotesize{}Emp. Var} & {\footnotesize{}Est. Var} & {\footnotesize{}Cov. Rate} & {\footnotesize{}MSE $b_{i}$}\tabularnewline
\hline 
\multirow{4}{*}{{\footnotesize{}$\theta_{S_{G}}$}} & {\footnotesize{}$\widehat{\theta}_{ML,S_{i}}$} & 6.4e-5 & 2.6e-3 & 5.7e-5 & 2.2e-5 & 0.85 & \tabularnewline
 & {\footnotesize{}$\widehat{\theta}_{ML}$} & 2.3e-3 & 3.1e-3 & 1.4e-3 & 1.9e-4 & 0.54 & \tabularnewline
 & {\footnotesize{}$\widehat{\theta}_{S_{i}}$} & \textbf{2.1e-5} & \textbf{-2.2e-5} & \textbf{2.1e-5} & \textbf{2.0e-5} & \textbf{0.93} & \tabularnewline
 & {\footnotesize{}$\widehat{\theta}$} & \textbf{3.2e-4} & \textbf{-1.2e-3} & \textbf{3.2e-4} & \textbf{4.0e-4} & \textbf{0.93} & \tabularnewline
\hline 
\multirow{4}{*}{{\footnotesize{}$\theta_{S_{I}}$}} & {\footnotesize{}$\widehat{\theta}_{ML,S_{i}}$} & known &  &  &  &  & \tabularnewline
 & {\footnotesize{}$\widehat{\theta}_{ML}$} & 0.01 & 0.04 & 0.01 & 1.2e-3 & 0.55 & \tabularnewline
 & {\footnotesize{}$\widehat{\theta}_{S_{i}}$} & \textbf{known} &  &  &  &  & \tabularnewline
 & {\footnotesize{}$\widehat{\theta}$} & \textbf{3.2e-4} & \textbf{-1.0e-3} & \textbf{3.2e-4} & \textbf{2.9e-4} & \textbf{0.92} & \tabularnewline
\hline 
\multirow{4}{*}{{\footnotesize{}$\theta_{n}$}} & {\footnotesize{}$\widehat{\theta}_{ML,S_{i}}$} & 7.7e-4 & -3.4e-3 & 7.6e-4 & 5.4e-4 & 0.89 & \tabularnewline
 & {\footnotesize{}$\widehat{\theta}_{ML}$} & 4.8e-3 & -4.8e-3 & 4.8e-3 & 5.3e-4 & 0.88 & \tabularnewline
 & {\footnotesize{}$\widehat{\theta}_{S_{i}}$} & \textbf{4.2e-4} & \textbf{6.9e-4} & \textbf{4.2e-4} & \textbf{5.1e-4} & \textbf{0.95} & \tabularnewline
 & {\footnotesize{}$\widehat{\theta}$} & \textbf{4.2e-4} & \textbf{5.8e-4} & \textbf{4.2e-4} & \textbf{4.9e-4} & \textbf{0.96} & \tabularnewline
\hline 
\multirow{4}{*}{{\footnotesize{}$\Psi$}} & {\footnotesize{}$\widehat{\theta}_{ML,S_{i}}$} & 0.03 & -3.4e-3 & 0.03 & 0.02 & 0.93 & 0.03\tabularnewline
 & {\footnotesize{}$\widehat{\theta}_{ML}$} & 0.03 & -8.1e-3 & 0.02 & 0.02 & 0.87 & 0.03\tabularnewline
 & {\footnotesize{}$\widehat{\theta}_{S_{i}}$} & \textbf{0.01} & \textbf{-3.9e-3} & \textbf{0.01} & \textbf{0.02} & \textbf{0.94} & \textbf{0.01}\tabularnewline
 & {\footnotesize{}$\widehat{\theta}$} & \textbf{0.02} & \textbf{-7.1e-3} & \textbf{0.02} & \textbf{0.02} & \textbf{0.94} & \textbf{0.02}\tabularnewline
\end{tabular}

\end{table}

\subsubsection{Estimation in presence of model error at the subject level}

To mimic misspecification presence, we generate the observations from
the stochastic model:
\begin{equation}
\left\{ \begin{array}{l}
dG_{i}=\left(S_{G}(G_{B}-G_{i})-X_{i}G_{i}\right)dt+\alpha_{1}dB_{1,t}\\
dI_{i}=\left(\gamma t(G_{i}-h)-n_{i}(I_{i}-I_{B}\right))dt+\alpha_{2}dB_{2,t}\\
dX_{i}=\left(-p_{2}(X_{i}+S_{I}(I_{i}-I_{B}))\right)dt+\alpha_{3}dB_{3,t}
\end{array}\right.\label{eq:Misspe_Insulin_Model}
\end{equation}
where the $B_{i,t}$ are Wiener processes and $\left(\alpha_{1},\alpha_{2},\alpha_{3}\right)=\left(2,2,2\times10^{-4}\right)$
their diffusion coefficients. We present the estimation results in
table \ref{tab:Insulin_model_Bias_Var_pop_param}.{\footnotesize{}
}For ML, the drop in coverage rate for $\theta_{S_{G}}^{*}$ and $\theta_{S_{I}}^{*}$
is even more striking when $\theta_{S_{I}}^{*}$ needs to be estimated.
This is explained by the effect of model misspecification which increases
bias and the fact that ML does not take into account this new source
of uncertainty leading here to under-estimation of variance and too narrow
confidence intervals.

\subsection{Antibody concentration evolution model}

We consider the model presented in \cite{Pasin2019} to analyze the
antibody concentration, denoted $Ab_{i}$, generated by two populations
of antibody secreting cells: the short lived, denoted $S_{i}$, and
the long-lived, denoted $L_{i}$: 
\begin{equation}
\left\{ \begin{array}{l}
\dot{S}_{i}=-\delta_{S}S_{i}\\
\dot{L}_{i}=-\delta_{L}L_{i}\\
\dot{Ab}_{i}=\theta_{S,i}S_{i}+\theta_{L,i}L_{i}-\delta_{Ab,i}Ab_{i}\\
\left(S_{i}(0),L_{i}(0),Ab_{i}(0)\right)=\left(S_{0,i},L_{0,i},Ab_{0,i}\right).
\end{array}\right.\label{eq:Chloe_original_antibody_ODE}
\end{equation}
This model is used to quantify the humoral response on different populations
after an Ebola vaccine injection with a 2 doses regimen seven days
after the second injection when the antibody secreting cells enter
in a decreasing phase. These cells being unobserved, the preceding
equation can be simplified to focus on antibody concentration evolution:
\begin{equation}
\dot{Ab}_{i}=\phi_{S,i}e^{-\delta_{S}t}+\phi_{L,i}e^{-\delta_{L}t}-\delta_{Ab,i}Ab_{i}\label{eq:Chloe_antibody_ODE}
\end{equation}
with $\phi_{S,i}:=\theta_{S,i}S_{0,i}$ and $\phi_{L,i}:=\theta_{L,i}L_{0,i}$.
This equation has an analytic solution which will be used for maximum
likelihood estimation with SAEMIX. We consider the following parametrization:
\[
\left\{ \begin{array}{l}
\log(\delta_{S})=\psi_{\delta_{S}}\\
\log(\phi_{S,i})=\psi_{\phi_{S}}+b_{\phi_{S},i}\\
\log(\phi_{L,i})=\psi_{\phi_{L}}+b_{\phi_{L},i}\\
\log(\delta_{Ab,i})=\psi_{\delta_{Ab}}+b_{\delta_{Ab},i}.
\end{array}\right.
\]
The true parameter values are presented in table \ref{tab:chlod_model_Param_recap}.{\footnotesize{}
}
\begin{table}
\caption{\label{tab:chlod_model_Param_recap} Biological interpretation and parameter values}
{\footnotesize{}}%
\begin{tabular}{l|lll}
\hline 
\multicolumn{2}{l}{{\footnotesize{}Parameters}} & {\footnotesize{}Biological interpretation} & Values\tabularnewline
\hline 
\hline 
\multicolumn{2}{l}{{\footnotesize{}$\delta_{L}$}} & {\footnotesize{}long-lived cells declining rate} & $\log(2)/(364\times6)$\tabularnewline
\hline 
\multirow{3}{*}{{\footnotesize{}$\theta^{*}$}} & {\footnotesize{}$\psi_{\delta_{S}}^{*}$} & {\footnotesize{}Mean log-value for $\delta_{S}$, the short-lived
cells declining rate } & $\log(\log(2)/1.2)\simeq-0.54$\tabularnewline
 & {\footnotesize{}$\psi_{\phi_{S}}^{*}$} & {\footnotesize{}Mean log-value for $\phi_{S}$, the antibodies influx
from short-lived cells } & $\log(2755)\simeq7.92$\tabularnewline
 & {\footnotesize{}$\psi_{\phi_{L}}^{*}$} & {\footnotesize{}Mean log-value for $\phi_{L}$, the antibodies influx
from long-lived cells } & $\log(16)\simeq2.78$\tabularnewline
 & {\footnotesize{}$\psi_{\delta_{Ab}}^{*}$} & {\footnotesize{}Mean log-value for $\delta_{Ab}$, the antibodies
declining rate } & $\log(\log(2)/24)\simeq-3.54$\tabularnewline
\hline 
\multirow{3}{*}{{\footnotesize{}$\Psi^{*}$}} & {\footnotesize{}$\Psi_{\phi_{S}}^{*}$} & {\footnotesize{}Inter individual variance for $\log(\phi_{S,i})$} & $0.92^{2}$\tabularnewline
 & {\footnotesize{}$\Psi_{\phi_{L}}^{*}$} & {\footnotesize{}Inter individual variance for $\log(\phi_{L,i})$} & $0.85^{2}$\tabularnewline
 & {\footnotesize{}$\Psi_{\delta_{Ab}}^{*}$} & {\footnotesize{}Inter individual variance for $\log(\delta_{Ab,i})$} & $0.3^{2}$\tabularnewline
\hline 
\end{tabular}{\footnotesize\par}
{\footnotesize\par}

\end{table}
According to \cite{Pasin2019}, the parameter $\delta_{L}$
was non-identifiable and only a lower bound has been derived for it
via profiled likelihood. So, to make fair comparisons between our
approach and maximum likelihood, we do not estimate it. Regarding
population parameters, we are particularly interested in the behavior
of estimation methods for $\psi_{\delta_{S}}$ and $\psi_{\phi_{S}}.$
Indeed a parameter sensitivity analysis shows the symmetric role of
$\psi_{\delta_{S}}$ and $\psi_{\phi_{S}}$ on the ODE solution (see
\cite{Balelli2019}). Thus, they are likely to face practical identifiability
problems. To investigate this effect, we estimate the parameters when
1) $\psi_{\delta_{S}}^{*}$ is known (the corresponding estimators
will be denoted with the subscript $\psi_{\delta_{S}}$), 2) it has
to be estimated as well. 

\subsubsection{Correct model case}

We generate $n_{i}=11$ observations on the interval $\left[0,\,T\right]=\left[0,\,364\right]$
with Gaussian measurement noise of standard deviation $\sigma^{*}=100$.
For each subject $i$, the initial condition has been generated according
to $Ab_{0,i}^{*}\sim N(\overline{Ab_{0}},\,\sigma_{\overline{Ab_{0}}}^{2})$
with $\overline{Ab_{0}}=500$ and $\sigma_{\overline{Ab_{0}}}=260$
to reflect the dispersion observed in the data presented in \cite{Pasin2019}.
We present the estimation results in table \ref{tab:Chloe_model_Bias_Var_pop_param}.

\begin{table}
\caption{\label{tab:Chloe_model_Bias_Var_pop_param} Results of estimation
for model (\ref{eq:Chloe_antibody_ODE}). The different subscripts
stand for the following estimation scenarios: 1)$\psi_{\delta_{S}}$
when $\psi_{\delta_{S}}$ is set to $\psi_{\delta_{S}}^{*}$, 2)absence
of subscript when $\psi_{\delta_{S}}$ is estimated. Results from
our method are in bold.}
{\footnotesize{}}%
\begin{tabular}{l|lllllll}
\hline 
\multicolumn{8}{l}{Well-specified model}\tabularnewline
\hline 
\multicolumn{1}{l}{} &  & {\footnotesize{}MSE} & {\footnotesize{}Bias} & {\footnotesize{}Emp. Var} & {\footnotesize{}Est. Var} & {\footnotesize{}Cov. Rate} & {\footnotesize{}MSE $b_{i}$}\tabularnewline
\hline 
\hline 
\multirow{4}{*}{{\footnotesize{}$\psi_{\delta_{S}}$}} & {\footnotesize{}$\widehat{\theta}_{ML,\psi_{\delta_{S}}}$} & \multicolumn{5}{l}{known} & \tabularnewline
 & {\footnotesize{}$\widehat{\theta}_{ML}$} & 2.13 & 0.78 & 1.51 & 70.64 & 0.92 & \tabularnewline
 & {\footnotesize{}$\widehat{\theta}_{\psi_{\delta_{S}}}$} & \multicolumn{5}{l}{\textbf{known}} & \tabularnewline
 & {\footnotesize{}$\widehat{\theta}$} & \textbf{0.62} & \textbf{-0.34} & \textbf{0.50} & \textbf{0.66} & \textbf{0.92} & \tabularnewline
\hline 
\multirow{4}{*}{{\footnotesize{}$\psi_{\phi_{S}}$}} & {\footnotesize{}$\widehat{\theta}_{ML,\psi_{\delta_{S}}}$} & 4e-4 & 0.01 & 3e-4 & 3e-4 & 0.94 & \tabularnewline
 & {\footnotesize{}$\widehat{\theta}_{ML}$} & 0.01 & -0.05 & 7e-3 & 0.40 & 0.92 & \tabularnewline
 & {\footnotesize{}$\widehat{\theta}_{\psi_{\delta_{S}}}$} & \textbf{2e-3} & \textbf{-0.05} & \textbf{2e-4} & \textbf{1e-3} & \textbf{0.94} & \tabularnewline
 & {\footnotesize{}$\widehat{\theta}$} & \textbf{2e-3} & \textbf{1e-3} & \textbf{2e-3} & \textbf{2e-3} & \textbf{0.93} & \tabularnewline
\hline 
\multirow{4}{*}{{\footnotesize{}$\psi_{\phi_{L}}$}} & {\footnotesize{}$\widehat{\theta}_{ML,\psi_{\delta_{S}}}$} & 3e-3 & 0.02 & 3e-3 & 2e-3 & 0.95 & \tabularnewline
 & {\footnotesize{}$\widehat{\theta}_{ML}$} & 4e-3 & 0.03 & 4e-3 & 3e-3 & 0.90 & \tabularnewline
 & {\footnotesize{}$\widehat{\theta}_{\psi_{\delta_{S}}}$} & \textbf{7e-4} & \textbf{-0.01} & \textbf{5e-4} & \textbf{3e-3} & \textbf{0.95} & \tabularnewline
 & {\footnotesize{}$\widehat{\theta}$} & \textbf{3e-3} & \textbf{-3e-3} & \textbf{3e-3} & \textbf{2e-3} & \textbf{0.91} & \tabularnewline
\hline 
\multirow{4}{*}{{\footnotesize{}$\psi_{\delta_{Ab}}$}} & {\footnotesize{}$\widehat{\theta}_{ML,\psi_{\delta_{S}}}$} & 7e-4 & -0.02 & 5e-4 & 3e-4 & 0.93 & \tabularnewline
 & {\footnotesize{}$\widehat{\theta}_{ML}$} & 2e-3 & -0.02 & 1e-3 & 4e-4 & 0.88 & \tabularnewline
 & {\footnotesize{}$\widehat{\theta}_{\psi_{\delta_{S}}}$} & \textbf{2e-4} & \textbf{0.01} & \textbf{1e-4} & \textbf{3e-4} & \textbf{0.95} & \tabularnewline
 & {\footnotesize{}$\widehat{\theta}$} & \textbf{4e-4} & \textbf{0.01} & \textbf{3e-4} & \textbf{2e-4} & \textbf{0.90} & \tabularnewline
\hline 
\multirow{4}{*}{{\footnotesize{}$\Psi_{\phi_{S}}$}} & {\footnotesize{}$\widehat{\theta}_{ML,\psi_{\delta_{S}}}$} & 0.04 & -1e-3 & 0.04 & 0.07 & 1 & 0.15\tabularnewline
 & {\footnotesize{}$\widehat{\theta}_{ML}$} & 0.11 & 0.01 & 0.11 & 0.05 & 1 & 0.17\tabularnewline
 & {\footnotesize{}$\widehat{\theta}_{\psi_{\delta_{S}}}$} & \textbf{0.02} & \textbf{8e-3} & \textbf{0.02} & \textbf{0.01} & \textbf{0.94} & \textbf{0.06}\tabularnewline
 & {\footnotesize{}$\widehat{\theta}$} & \textbf{0.02} & \textbf{-0.03} & \textbf{0.02} & \textbf{0.02} & \textbf{0.94} & \textbf{0.07}\tabularnewline
\hline 
\multirow{4}{*}{{\footnotesize{}$\Psi_{\phi_{L}}$}} & {\footnotesize{}$\widehat{\theta}_{ML,\psi_{\delta_{S}}}$} & 0.03 & 0.04 & 0.02 & 0.04 & 1 & 0.30\tabularnewline
 & {\footnotesize{}$\widehat{\theta}_{ML}$} & 0.03 & 0.05 & 0.02 & 0.04 & 1 & 0.60\tabularnewline
 & {\footnotesize{}$\widehat{\theta}_{\psi_{\delta_{S}}}$} & \textbf{0.02} & \textbf{-0.1} & \textbf{5e-3} & \textbf{8e-3} & \textbf{0.93} & \textbf{0.07}\tabularnewline
 & {\footnotesize{}$\widehat{\theta}$} & \textbf{0.03} & \textbf{-0.06} & \textbf{0.02} & \textbf{0.01} & \textbf{0.92} & \textbf{0.08}\tabularnewline
\hline 
\multirow{4}{*}{{\footnotesize{}$\Psi_{\delta_{Ab}}$}} & {\footnotesize{}$\widehat{\theta}_{ML,\psi_{\delta_{S}}}$} & 0.11 & 0.18 & 0.08 & 0.02 & 1 & 0.10\tabularnewline
 & {\footnotesize{}$\widehat{\theta}_{ML}$} & 0.20 & 0.29 & 0.11 & 0.02 & 1 & 0.50\tabularnewline
 & {\footnotesize{}$\widehat{\theta}_{\psi_{\delta_{S}}}$} & \textbf{0.10} & \textbf{-0.30} & \textbf{0.01} & \textbf{0.01} & \textbf{0.95} & \textbf{0.03}\tabularnewline
 & {\footnotesize{}$\widehat{\theta}$} & \textbf{0.11} & \textbf{-0.27} & \textbf{0.04} & \textbf{0.04} & \textbf{0.95} & \textbf{0.04}\tabularnewline
\end{tabular}{\footnotesize\par}

{\footnotesize\par}
\end{table}
Our method improves the estimation of $\psi_{\delta_{S}}^{*}$ facing
practical identifiability problems comparing to the ML. In particular
our method reduces its variance. As advocated in the introduction,
our approach provides an improved estimate for the $\left\{ b_{i}^{*}\right\} _{i\in\left\llbracket 1,\,n\right\rrbracket }$.
We assume that is due to the committed estimation error for $\theta^{*}$,
as it causes model error for $\left\{ b_{i}^{*}\right\} _{i\in\left\llbracket 1,\,n\right\rrbracket }$
estimation, which is not taken into account by exact methods. This
in turn explains why their variance $\Psi^{*}$ is better estimated
with our approach. In this mixed-effect context, this cause of model
error is systematically present and claims for the use of estimation
methods taking into account modeling uncertainties when subject specific
parameters are critical for the practitioner. 

\subsubsection{Estimation in presence of model error at the subject level}

The data are now generated with a stochastic perturbed version of
the original model:
\begin{equation}
dAb_{i}=\left(\phi_{S,i}e^{-\delta_{S}t}+\phi_{L,i}e^{-\delta_{L}t}-\delta_{Ab,i}Ab_{i}\right)dt+\alpha dB_{t}\label{eq:Chloe_antibody_ODE_pert}
\end{equation}
where $B_{t}$ is a Wiener process and $\alpha=10$ its diffusion
coefficient. The value for $\alpha$ has been chosen big enough to
produce significantly perturbed trajectories but small enough to ensure
that ODE (\ref{eq:Chloe_antibody_ODE}) is still a relevant approximation
for estimation purpose. We keep the same parameter values and measurement
noise level as in the previous section. The results are presented
in table \ref{tab:Chloe_model_misspe_Bias_Var_pop_param}.

\begin{table}
\caption{\label{tab:Chloe_model_misspe_Bias_Var_pop_param} Results of estimation
in presence of model error when data are generated from (\ref{eq:Chloe_antibody_ODE_pert}).
The different subscripts stand for the same estimation scenario as
in table \ref{tab:Chloe_model_Bias_Var_pop_param}. Results from our
method are in bold.}
{\footnotesize{}}%
\begin{tabular}{l|lllllll}
\hline 
\multicolumn{8}{l}{Misspecified model}\tabularnewline
\hline 
\multicolumn{1}{l}{} &  & {\footnotesize{}MSE} & {\footnotesize{}Bias} & {\footnotesize{}Emp. Var} & {\footnotesize{}Est. Var} & {\footnotesize{}Cov. Rate} & {\footnotesize{}MSE $b_{i}$}\tabularnewline
\hline 
\hline 
\multirow{4}{*}{{\footnotesize{}$\psi_{\delta_{S}}$}} & {\footnotesize{}$\widehat{\theta}_{ML,\psi_{\delta_{S}}}$} & \multicolumn{5}{l}{known} & \tabularnewline
 & {\footnotesize{}$\widehat{\theta}_{ML}$} & 3.88 & 1.48 & 1.68 & 4.10 & 0.80 & \tabularnewline
 & {\footnotesize{}$\widehat{\theta}_{\psi_{\delta_{S}}}$} & \multicolumn{5}{l}{\textbf{known}} & \tabularnewline
 & {\footnotesize{}$\widehat{\theta}$} & \textbf{0.93} & \textbf{-0.40} & \textbf{0.77} & \textbf{0.62} & \textbf{0.90} & \tabularnewline
\hline 
\multirow{4}{*}{{\footnotesize{}$\psi_{\phi_{S}}$}} & {\footnotesize{}$\widehat{\theta}_{ML,\psi_{\delta_{S}}}$} & 1e-3 & 0.02 & 1e-3 & 5e-4 & 0.91 & \tabularnewline
 & {\footnotesize{}$\widehat{\theta}_{ML}$} & 0.02 & -0.10 & 0.01 & 0.02 & 0.88 & \tabularnewline
 & {\footnotesize{}$\widehat{\theta}_{\psi_{\delta_{S}}}$} & \textbf{7e-4} & \textbf{-0.02} & \textbf{3e-4} & \textbf{1e-3} & \textbf{0.92} & \tabularnewline
 & {\footnotesize{}$\widehat{\theta}$} & \textbf{4e-3} & \textbf{-6e-3} & \textbf{3e-3} & \textbf{0.01} & \textbf{0.90} & \tabularnewline
\hline 
\multirow{4}{*}{{\footnotesize{}$\psi_{\phi_{L}}$}} & {\footnotesize{}$\widehat{\theta}_{ML,\psi_{\delta_{S}}}$} & 5e-3 & 0.03 & 4e-3 & 3e-3 & 0.93 & \tabularnewline
 & {\footnotesize{}$\widehat{\theta}_{ML}$} & 9e-3 & 0.05 & 7e-3 & 4e-3 & 0.90 & \tabularnewline
 & {\footnotesize{}$\widehat{\theta}_{\psi_{\delta_{S}}}$} & \textbf{2e-3} & \textbf{-0.02} & \textbf{3e-3} & \textbf{2e-3} & \textbf{0.97} & \tabularnewline
 & {\footnotesize{}$\widehat{\theta}$} & \textbf{6e-3} & \textbf{-8e-3} & \textbf{6e-3} & \textbf{7e-3} & \textbf{0.90} & \tabularnewline
\hline 
\multirow{4}{*}{{\footnotesize{}$\psi_{\delta_{Ab}}$}} & {\footnotesize{}$\widehat{\theta}_{ML,\psi_{\delta_{S}}}$} & 2e-3 & -0.03 & 1e-3 & 1e-3 & 0.92 & \tabularnewline
 & {\footnotesize{}$\widehat{\theta}_{ML}$} & 4e-3 & -0.04 & 3e-3 & 7e-4 & 0.88 & \tabularnewline
 & {\footnotesize{}$\widehat{\theta}_{\psi_{\delta_{S}}}$} & \textbf{3e-4} & \textbf{2e-3} & \textbf{3e-4} & \textbf{3e-4} & \textbf{0.96} & \tabularnewline
 & {\footnotesize{}$\widehat{\theta}$} & \textbf{3e-4} & \textbf{8e-3} & \textbf{3e-4} & \textbf{2e-3} & \textbf{0.89} & \tabularnewline
\hline 
\multirow{4}{*}{{\footnotesize{}$\Psi_{\phi_{S}}$}} & {\footnotesize{}$\widehat{\theta}_{ML,\psi_{\delta_{S}}}$} & 0.05 & 0.03 & 0.05 & 0.08 & 1 & 0.17\tabularnewline
 & {\footnotesize{}$\widehat{\theta}_{ML}$} & 0.13 & 0.01 & 0.13 & 0.25 & 1 & 0.21\tabularnewline
 & {\footnotesize{}$\widehat{\theta}_{\psi_{\delta_{S}}}$} & \textbf{0.02} & \textbf{2e-3} & \textbf{0.02} & \textbf{0.02} & \textbf{0.94} & \textbf{0.11}\tabularnewline
 & {\footnotesize{}$\widehat{\theta}$} & \textbf{0.02} & \textbf{-0.05} & \textbf{0.02} & \textbf{0.03} & \textbf{0.92} & \textbf{0.08}\tabularnewline
\hline 
\multirow{4}{*}{{\footnotesize{}$\Psi_{\phi_{L}}$}} & {\footnotesize{}$\widehat{\theta}_{ML,\psi_{\delta_{S}}}$} & 0.05 & 0.03 & 0.05 & 0.06 & 1 & 0.73\tabularnewline
 & {\footnotesize{}$\widehat{\theta}_{ML}$} & 0.03 & 0.05 & 0.02 & 0.07 & 1 & 0.74\tabularnewline
 & {\footnotesize{}$\widehat{\theta}_{\psi_{\delta_{S}}}$} & \textbf{0.02} & \textbf{-0.10} & \textbf{0.01} & \textbf{0.02} & \textbf{0.91} & \textbf{0.10}\tabularnewline
 & {\footnotesize{}$\widehat{\theta}$} & \textbf{0.03} & \textbf{-0.06} & \textbf{0.02} & \textbf{0.03} & \textbf{0.87} & \textbf{0.12}\tabularnewline
\hline 
\multirow{4}{*}{{\footnotesize{}$\Psi_{\delta_{Ab}}$}} & {\footnotesize{}$\widehat{\theta}_{ML,\psi_{\delta_{S}}}$} & 0.33 & 0.41 & 0.17 & 0.05 & 1 & 0.56\tabularnewline
 & {\footnotesize{}$\widehat{\theta}_{ML}$} & 0.30 & 0.34 & 0.19 & 0.05 & 1 & 0.69\tabularnewline
 & {\footnotesize{}$\widehat{\theta}_{\psi_{\delta_{S}}}$} & \textbf{0.10} & \textbf{-0.16} & \textbf{0.08} & \textbf{0.06} & \textbf{0.91} & \textbf{0.04}\tabularnewline
 & {\footnotesize{}$\widehat{\theta}$} & \textbf{0.15} & \textbf{-0.29} & \textbf{0.06} & \textbf{0.10} & \textbf{0.88} & \textbf{0.06}\tabularnewline
\end{tabular}{\footnotesize\par}

\end{table}
 Our method still outperforms the maximum likelihood for $\psi_{\delta_{S}}^{*}$
as well as the $\left\{ b_{i}^{*}\right\} _{i\in\left[1,\,n\right]}$
estimation and their variances. In addition, we mitigate the effect
of model error on estimation accuracy.

\section{\label{sec:Real-data-analysis}Real data analysis}

\begin{figure}
\includegraphics[scale=0.45]{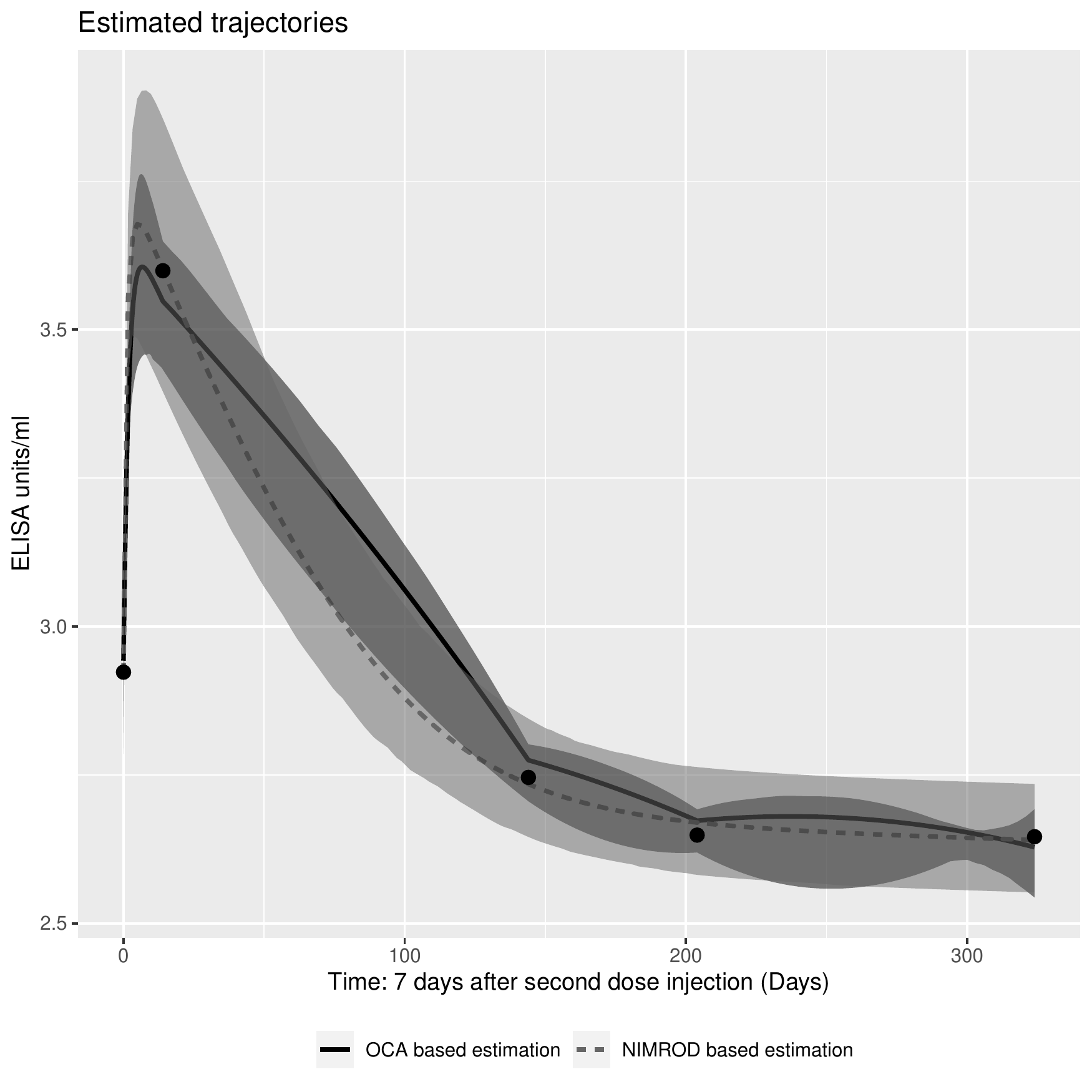}\includegraphics[scale=0.45]{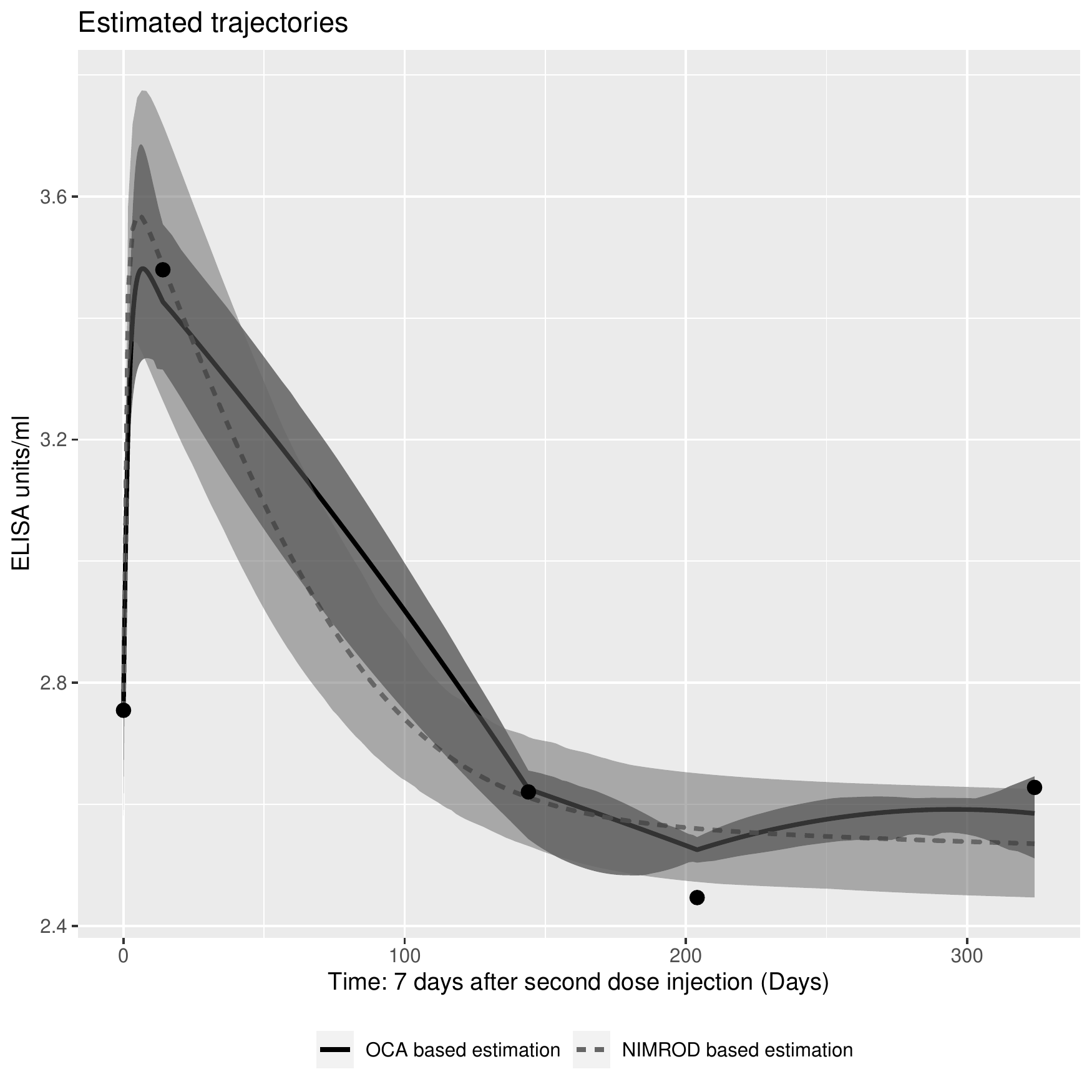}

\includegraphics[scale=0.45]{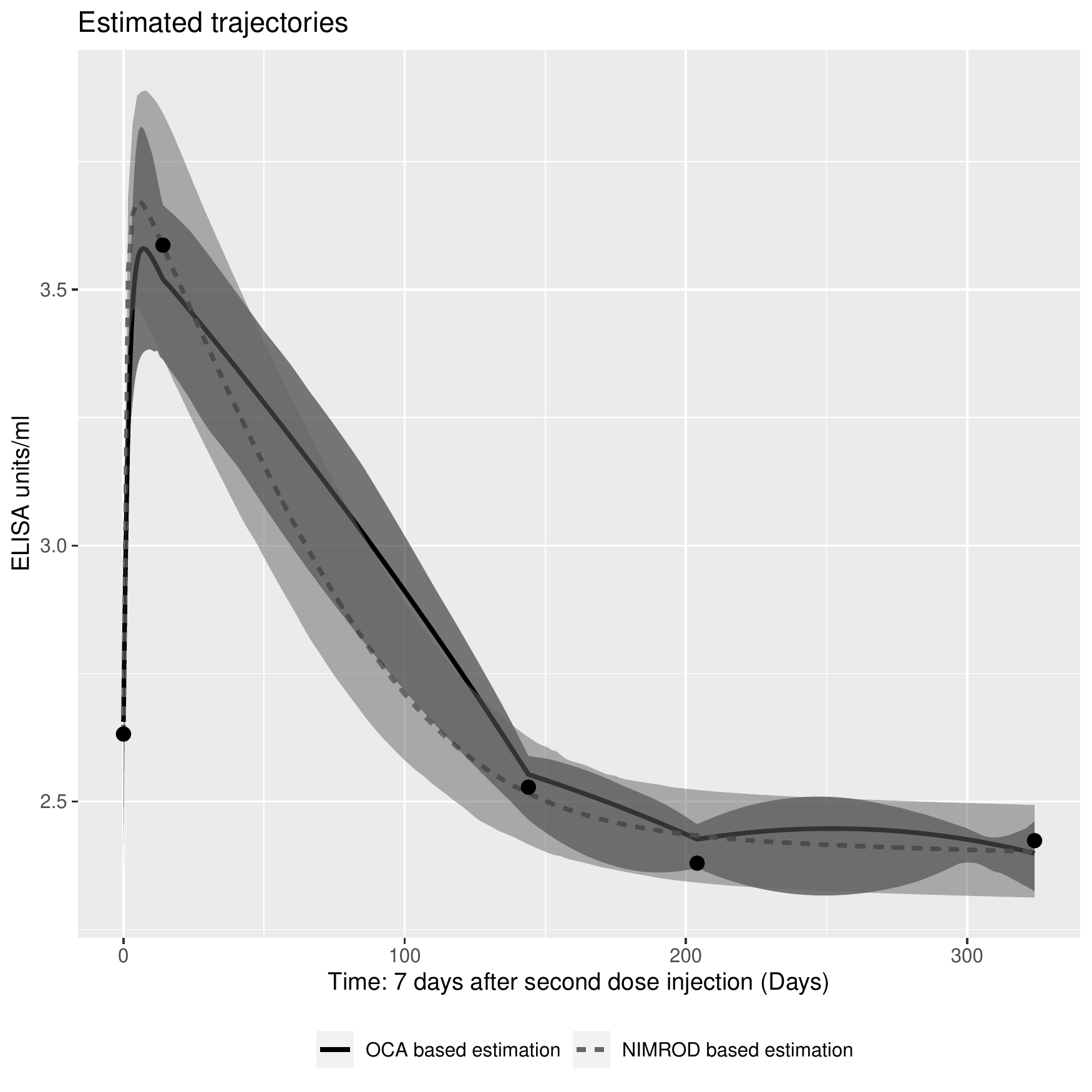}\includegraphics[scale=0.45]{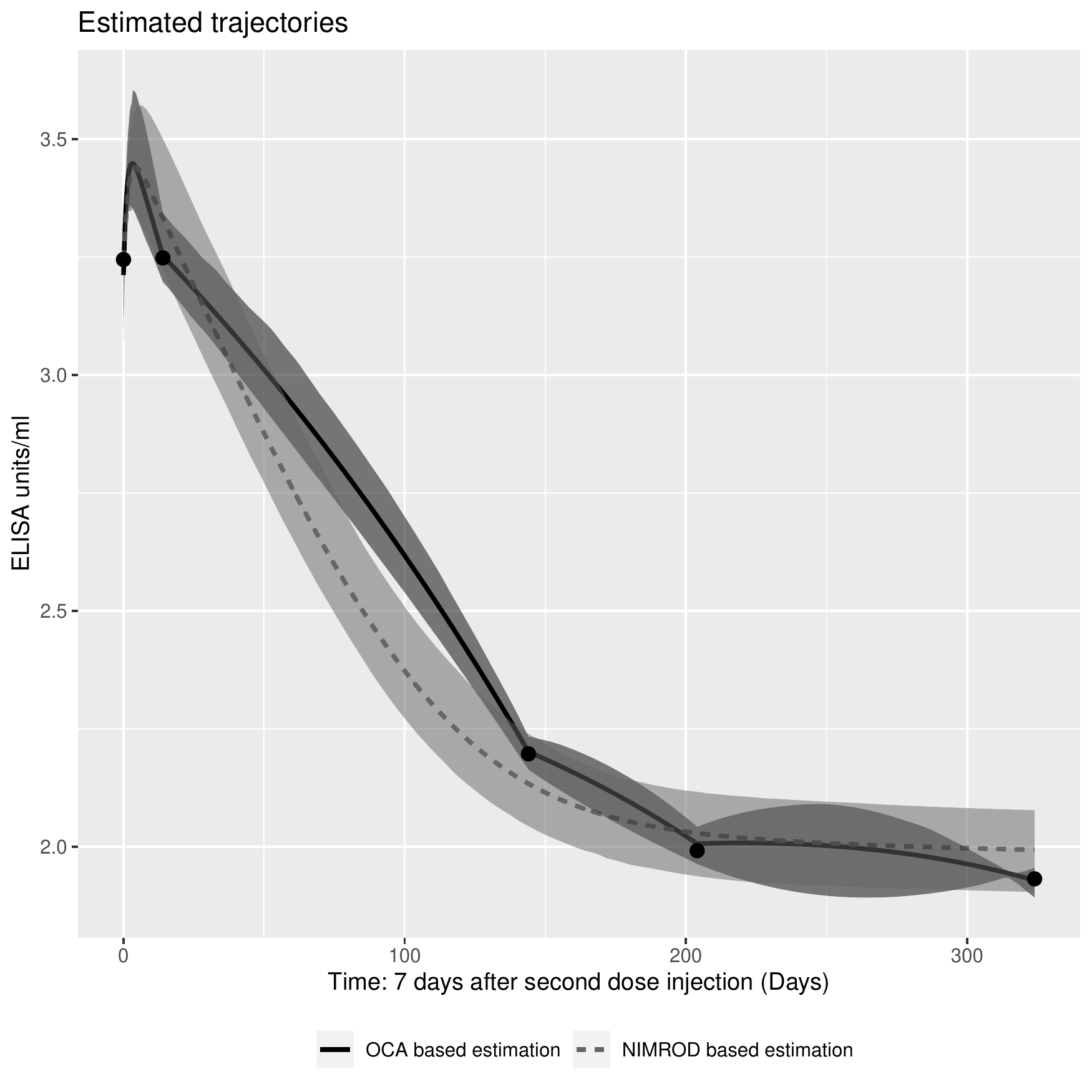}\caption{{\label{fig:Compare_trajectories_Nimrod_OCA} Examples of fitted trajectories
for both methods for different subjects. Here Time=0
is the 7th day post-second dose. Dashed lines: fitted ODE solutions
(\ref{eq:log10_transformed_Ab_model}) with $\left(\widehat{\theta}^{P},\widehat{b_{i}}^{P}\right)$.
Solid line: optimal trajectories $\overline{X}_{\widehat{\theta},\widehat{b_{i}}}.$
Shaded area are the 95\% confidence intervals.}}
\end{figure}

We now proceed to the estimation using real data presented
in \cite{Pasin2019} from which the parameter values given in table
\ref{tab:chlod_model_Param_recap} come from. In \cite{Pasin2019}
, the estimation is made from cohorts coming from three phase I trials
performed in African and European countries. Each subject was vaccinated
with two doses, Ad26.ZEBOV (Janssen Vaccines and Prevention) and MVA-BN-Filo
(Bavarian Nordic). In these cohorts, both the effect of injection
order, either Ad26.ZEBOV first and MVA-BN-Filo second, or MVA-BN-Filo
first and Ad26.ZEBOV second, and the delay between, 28 or 56 days,
were evaluated. In this study, we focus on an east African subpopulation
where Ad26.ZEBOV was injected first and then MVA-BN-Filo with a delay
of 28 days between the two doses. As in \cite{Pasin2019} and the
simulation section, to stay in the temporal domain of validity of
the model we use measurements made seven days after the second dose
injection. It leaves us with 5 measurements of antibody concentration
between days 7 up to days 330 per subject. The estimation in the original
work has been done using the NIMROD software \cite{Pragues2013} and
log-transformed antibody concentration measurement. We now estimate
the parameters with our method with the aim to compare our results
with the existing one. We used the same prior distribution $\pi(\theta)\sim N\left(\left(\begin{array}{c}
-1\\
0\\
0\\
-4.1
\end{array}\right),\left(\begin{array}{cccc}
25 & 0 & 0 & 0\\
0 & 100 & 0 & 0\\
0 & 0 & 100 & 0\\
0 & 0 & 0 & 1
\end{array}\right)\right)$ for $\theta=\left(\psi_{\delta_{S}},\psi_{\phi_{S}},\psi_{\phi_{L}},\psi_{\delta_{Ab}}\right)$
as the one defined in the NIMROD software. We choose our mesh-size
such that we get $200$ discretization points for each subject on
the observation interval and we use $U=10$ i.e. a value lower than
in the simulated data case because of the model error presence. We
also proceed to the log-transformation of the data to stabilize the
measurement noise variance. This drives us to use the nonlinear model:
\begin{equation}
\dot{\widetilde{Ab_{i}}}(t)=\frac{1}{\ln(10)}\left(\phi_{S,i}e^{-\delta_{S}t}+\phi_{L,i}e^{-\delta_{L}t}\right)10^{-\widetilde{Ab_{i}}(t)}-\frac{\delta_{Ab,i}}{\ln(10)}\label{eq:log10_transformed_Ab_model}
\end{equation}
describing the dynamic of $\widetilde{Ab_{i}}(t):=\log_{10}Ab_{i}(t)$
for parameter estimation purpose. We use $A_{\theta,b_{i}}(t,x,z_{i}(t))=\frac{1}{\ln(10)}\left(\phi_{S,i}e^{-\delta_{S}t}+\phi_{L,i}e^{-\delta_{L}t}\right)\frac{10^{-x}}{x}$
and $r_{\theta,b_{i}}(t,z_{i}(t))=-\frac{\delta_{Ab,i}}{\ln(10)}$
for the pseudo-linear formulation of the model. Our estimations and
the ones from the original paper \cite{Pasin2019} are presented in
Table \ref{tab:Compare_Pasin_OCA} for the sake of comparison. In
the following, we denote $\left(\widehat{\theta}^{P},\widehat{b_{i}}^{P}\right)$
(respectively $\left(\widehat{\theta},\widehat{b_{i}}\right)$) the
estimation obtained by \cite{Pasin2019} (respectively our approach)
for the mean population parameter and subject specific ones. 
\begin{table}
\caption{\label{tab:Compare_Pasin_OCA}Estimation presented in \cite{Pasin2019}
(left) and via our approach (right)}
\begin{tabular}{|c|c|c||c|c|}
\hline 
 & Pasin et al. & CI (95\%) & OCA & CI (95\%)\tabularnewline
\hline 
\hline 
$\psi_{\delta_{S}}$ & -0.57 & {[}-1.02, -0.02{]} & -0.18 & {[}-0.58, 0.22{]}\tabularnewline
\hline 
$\psi_{\phi_{S}}$ & 7.92 & {[}7.52, 8.30{]} & 7.45 & {[}6.85, 7.96{]}\tabularnewline
\hline 
$\psi_{\phi_{L}}$ & 2.78 & {[}2.62, 3.01{]} & 2.58 & {[}2.15, 3.01{]}\tabularnewline
\hline 
$\psi_{\delta_{Ab}}$  & -3.54 & {[}-3.62, -3.45{]} & -3.48 & {[}-3.95, -3.01{]}\tabularnewline
\hline 
{\footnotesize{}$\Psi_{\phi_{S}}$} & 0.92 & {[}0.83, 1.01{]} & 0.64 & {[}0.60, 0.70{]}\tabularnewline
\hline 
{\footnotesize{}$\Psi_{\phi_{L}}$} & 0.85 & {[}0.78, 0.92{]} & 0.70 & {[}0.55, 0.90{]}\tabularnewline
\hline 
{\footnotesize{}$\Psi_{\delta_{Ab}}$} & 0.3 & {[}0.24, 0.36{]} & 0.25 & {[}0.19, 0.31{]}\tabularnewline
\hline 
\end{tabular}
\end{table}
Both methods produce estimations with overlapping confidence intervals
for $\theta$. Still, significant differences appear for $\left(\Psi_{\phi_{S}},\Psi_{\phi_{L}},\Psi_{\delta_{Ab}}\right)$
estimation which quantifies the dispersion of random effects. This
is explained by the fact that we only consider a subset of the subjects
used in \cite{Pasin2019} for estimation. This has an effect on the
observed diversity within the cohort of patients and thus on $\left(\Psi_{\phi_{S}},\Psi_{\phi_{L}},\Psi_{\delta_{Ab}}\right)$
estimation. Regarding the predictions, we present in figure \ref{fig:Compare_trajectories_Nimrod_OCA}
examples of estimated trajectories. 
The confidence intervals are computed via Monte-Carlo sampling from
the approximated normal laws $\mathcal{N}(\widehat{\theta},V(\widehat{\theta}))$
and $\mathcal{N}(\widehat{\theta}^{P},V(\widehat{\theta}^{P}))$ to
quantify the effect of estimation uncertainy on $\theta$ on the predicted
trajectories. For NIMROD estimation, for a given sampled value $\widetilde{\theta}^{P}\sim\mathcal{N}(\widehat{\theta}^{P},V(\widehat{\theta}^{P}))$
and subject $i$, the sampled regression function $X_{\widetilde{\theta}^{P},\widehat{b_{i}}^{P},y_{0,i}}$
is obtained by solving ODE (\ref{eq:log10_transformed_Ab_model})
for parameter values $\left(\theta,b_{i},x_{0,i}\right)=\left(\widetilde{\theta}^{P},\widehat{b_{i}}^{P},y_{0,i}\right).$
Regarding our approach we recall the regression functions are now
defined as optimal trajectories. So, for $\widetilde{\theta}\sim\mathcal{N}(\widehat{\theta},V(\widehat{\theta}))$
the sampled regression function for subject $i$ is the optimal trajectory
$\overline{X}_{\widetilde{\theta},\widehat{b_{i}}}$ obtained via
the minimization of the cost function $\mathcal{C}_{i}(\widehat{b_{i}},x_{i,0},u_{i}\mid\widetilde{\theta},U)$.
This explain the differences between the two confidence intervals
in terms of shape and width. Our method gives narrower intervals because
for each sampled value an optimal control problem is solved to obtain
the related optimal trajectory. This imposes a common goal of data
fidelity to each sampled $\overline{X}_{\widetilde{\theta},\widehat{b_{i}}}$
which limits their inter-variability. Still, despite these differences
in shapes, both prediction intervals cover the same points. Morever,
on the long-term our intervals are nearly always contained in the
ones given by NIMROD.

\begin{figure}
\includegraphics[scale=0.5]{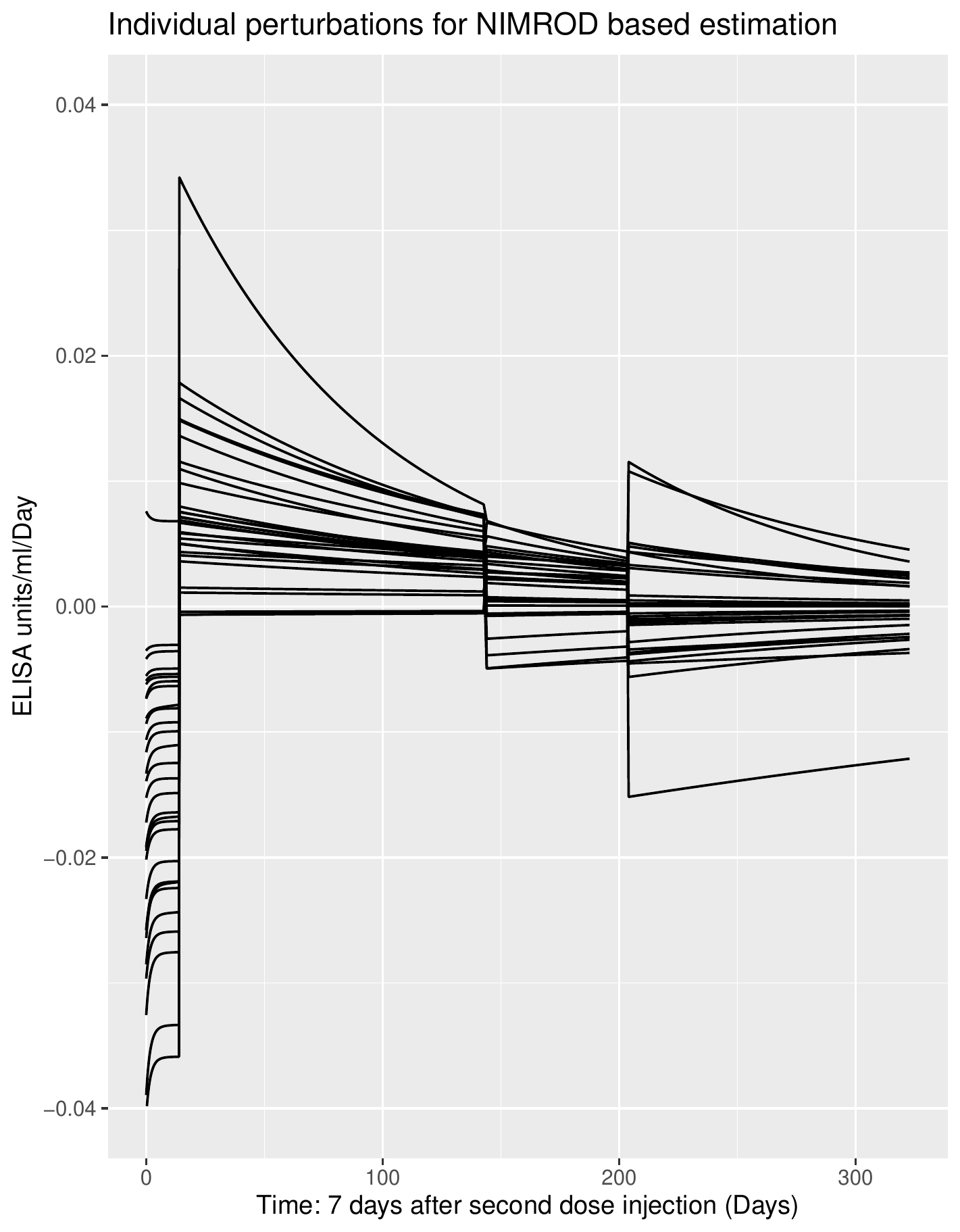}\includegraphics[scale=0.5]{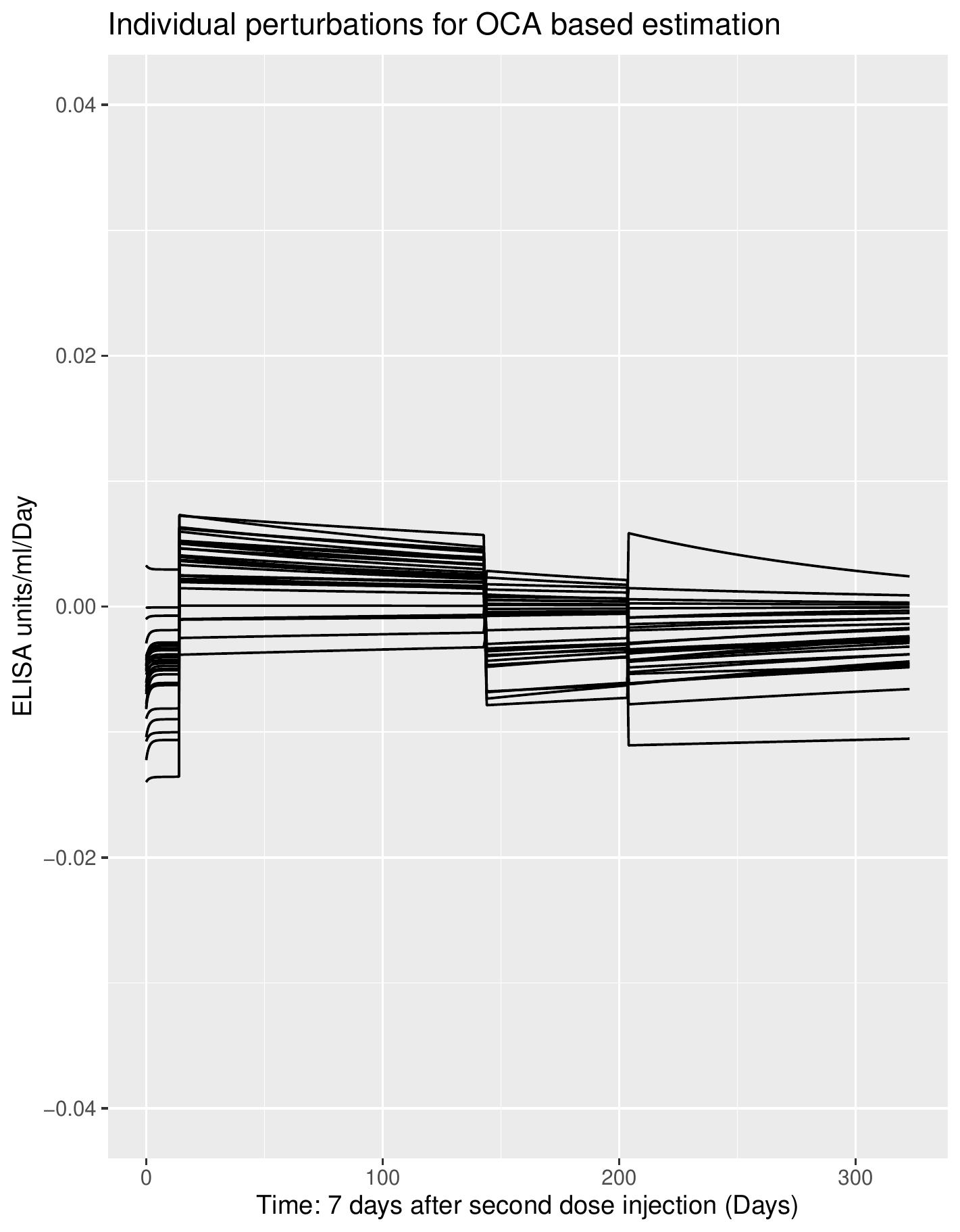}

\includegraphics[scale=0.5]{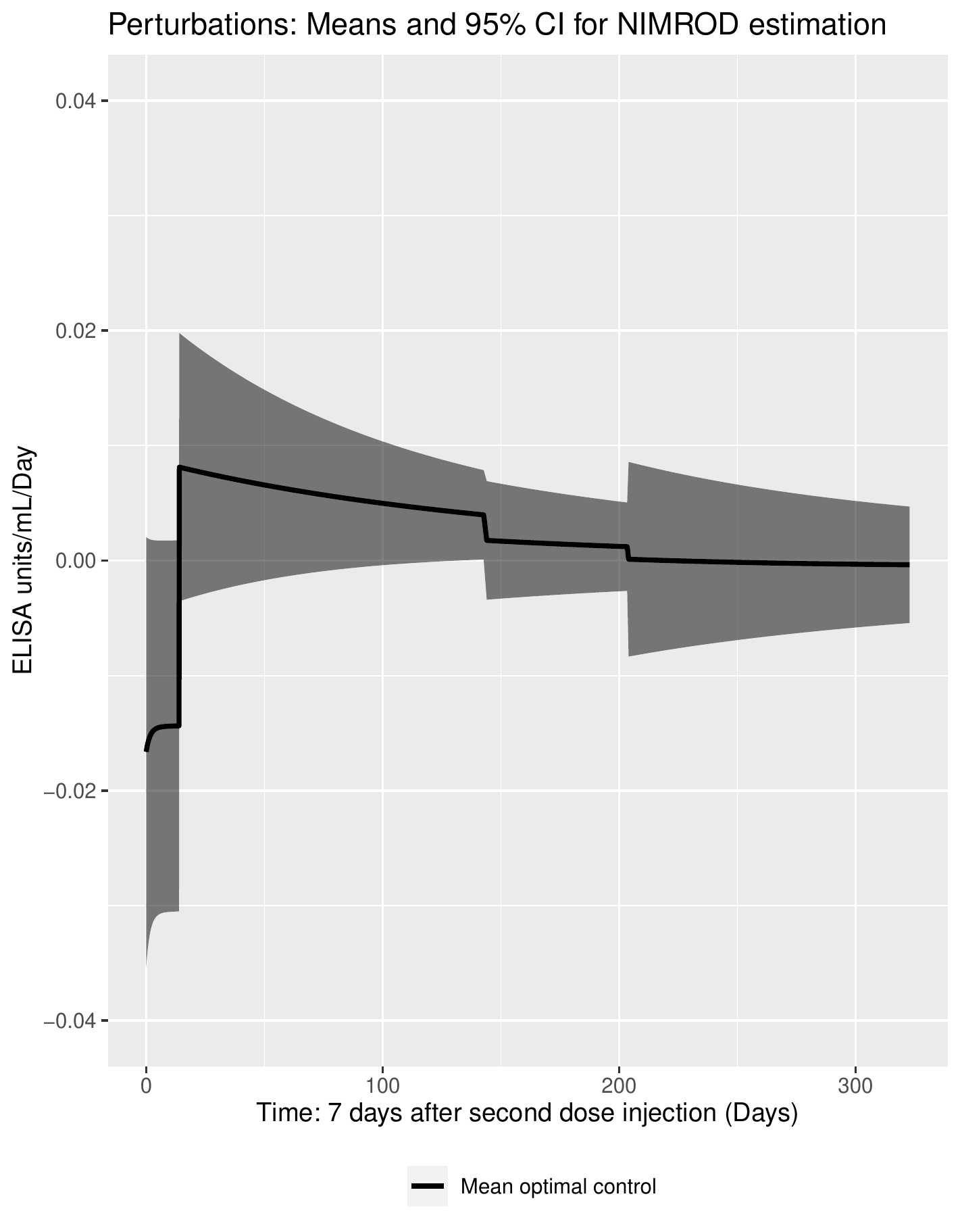}\includegraphics[scale=0.5]{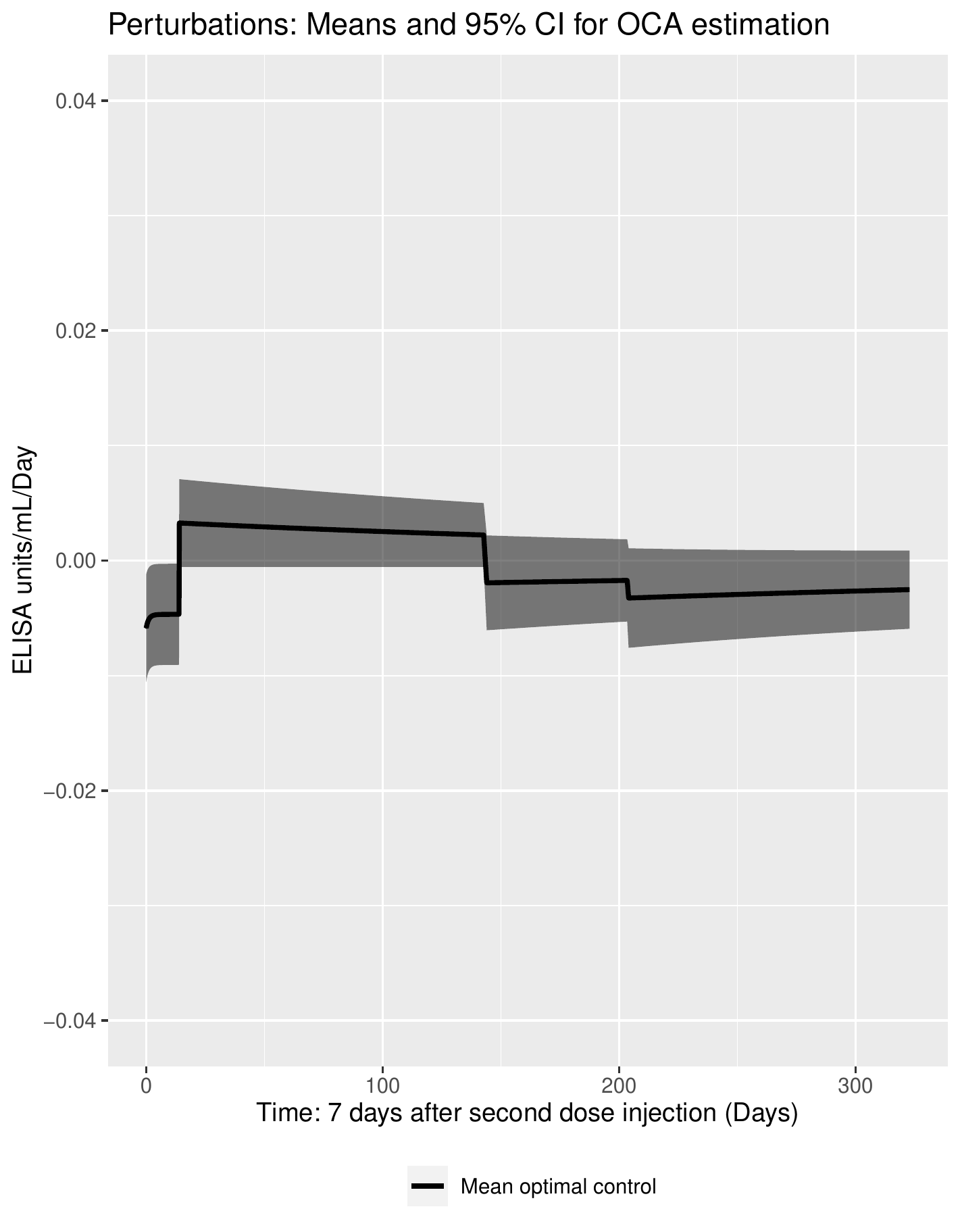}\caption{{\label{fig:Residual_controls_oca_method}1) Up: Estimated residual
controls for each subject, 2) bottom: mean optimal control and 95\%
confidence interval for the optimal controls a) left: $\overline{u}_{i}^{P}$
obtained from parameter estimation in \cite{Pasin2019}, b) right:
$\overline{u}_{i,\widehat{\theta},b_{i}(\widehat{\theta})}$ obtained
from our estimation.}}
\end{figure}

Our estimation of $\theta$ supports the parameter inference obtained
in \cite{Pasin2019} via another method and the subsequent analysis
made on the antibody concentration dynamics. In addition to this parametric
comparison, we want to assess the model adequacy via the temporal
evolution analysis of the optimal controls $\overline{u}_{i,\widehat{\theta},b_{i}(\widehat{\theta})}$
estimated as byproducts of our method. Indeed, they quantify the exogenous
perturbations $u_{i}$ we need to add to model (\ref{eq:log10_transformed_Ab_model})
so that the solution of its perturbed counterpart,
\begin{equation}
\dot{\widetilde{Ab_{i,u}}}(t)=\frac{1}{\ln(10)}\left(\phi_{S,i}e^{-\delta_{S}t}+\phi_{L,i}e^{-\delta_{L}t}\right)10^{-\widetilde{Ab_{i,u}}(t)}-\frac{\delta_{Ab,i}}{\ln(10)}+u_{i}\label{eq:pert_logAb}
\end{equation}
reproduce the observations. This approach is similar to the one developed
in \cite{HookerEllner2013} where control theory replaces non-parametric
procedures to estimate $u_{i}.$ Still, their approach relies on a
finite basis approximation of $\widetilde{Ab_{i,u}}$ which requires
to specify a basis function family, its dimension as well as a penalization
parameter similar to $U$. At the contrary, our method avoids this
complex step of hyper-parameter selection and only needs $U$. For
comparison, we also quantify the committed model error for $\left(\widehat{\theta}^{P},\widehat{b_{i}}^{P}\right)$.
To do so we compute $\overline{u}_{i}^{P}$, the solution of the optimal
control problem: $\overline{u}_{i}^{P}=\arg\min_{u_{i}}\left\{ \sum_{j}\left\Vert \widetilde{Ab}_{i,\widehat{\theta}^{P},\widehat{b_{i}}^{P},y_{i0},u_{i}}(t_{ij})-y_{ij}\right\Vert _{2}^{2}+\left\Vert u_{i}\right\Vert _{U,L^{2}}^{2}\right\} $
by using the procedure described in section \ref{sec:OCA_method_presentation}
for non-linear models. In the last expression $\widetilde{Ab}_{i,\widehat{\theta}^{P},\widehat{b_{i}}^{P},y_{i0},u_{i}}$
is the solution of the perturbed ODE (\ref{eq:pert_logAb}) for $\left(\theta,b_{i}\right)=\left(\widehat{\theta}^{P},\widehat{b_{i}}^{P}\right)$
and $y_{i0}$ is the measured concentration at $t=0$ used a surrogate
value for the initial condition (as they did in \cite{Pasin2019}).
We still use $U=10$ for this optimal control problem to allow for
the same level of perturbation magnitude for both methods. In figure
(\ref{fig:Residual_controls_oca_method}), we plot $\overline{u}_{i,\widehat{\theta},b_{i}(\widehat{\theta})}$
and $\overline{u}_{i}^{P}$ as well as their mean values and confidence
intervals.
 Our method leads to residual perturbations of smaller magnitudes
and narrower confidence intervals. This means our approach produces
an estimation which minimizes the committed model error for each subject
comparing to a method based only on a data fitting criteria. This
is particularly clear at the beginning of the observation interval
when the influence of the initial conditions is the highest. In this
case our narrower confidence interval clearly excludes a null perturbation
and advocates for an over-estimation of the predicted antibody concentration
by the model. This makes sense because model (\ref{eq:Chloe_original_antibody_ODE})
assumes that both populations of antibody secreting cells decrease
with time, and that is probably not completely true at the beginning
of the dynamic. Thus, despite similar results regarding parameter
values between our estimation and \cite{Pasin2019}, the insight given
by our method at the dynamic scale leads us to the additional conclusion
of model misspecification presence at the beginning of the observation
interval.

\section{\label{sec:Conclusion}Conclusion}

In this work, we propose an estimation method addressing issues encountered
by classic approaches for the problem of parameter estimation in NLME-ODEs.
We identify three potential sources of problems for exact methods
such as likelihood based inference: their difficulties in presence
of model error, their need to estimate initial conditions and their
dramatic performance degradation when facing poorly identifiable parameters.
We propose here a method based on control theory accounting for the
presence of potential model uncertainty at the subject level and which
can be easily profiled on the initial conditions. Simulations with
both presence and absence of model errors illustrate the benefits
of regularization techniques for estimating poorly identifiable parameters,
subject specific parameters as well as their variances in NLME-ODEs.
In addition, bypassing estimation of initial conditions represents
a clear advantage for partially observed systems comparing to likelihood
based approaches, as emphasized in simulations.

Still, this benefit in term of estimation accuracy comes with a computational
price. On a server with the parallelization package Snow in R language,
it takes approximately 10-15 minutes to obtain an estimation for the
two-dimensional linear model, 30 minutes for the insulin model and
3-4 hour for the antibody concentration evolution one, whereas it
was a matter of minutes for the other approaches. Nevertheless, the
use of compiled languages and proper parallelization could reduce
the computation time. Moreover, we have willingly separated the formal
definition of the optimal control problem required by our method and
the numerical procedure used to solve it, in case it may exist better
suited approaches for this specific control problem. Right now, our
current strategy allows us to profile on initial conditions, so looking
for another numerical procedure is beyond the scope of this paper. 

An under-exploited feature of the method so far is the obtained optimal
controls. The qualitative based analysis exposed in section \ref{sec:Real-data-analysis}
can be made more rigorous. For example, to stay in a Bayesian setting,
we can specify a prior distribution for the controls and then compare
it with the obtained posterior once the inference is made. This would
lead to a semi-parametric inference problem for which an optimal control
based approach has already been proven useful (see \cite{BrunelClairon_Pontryagin2017,Clairon2019}).
This is a subject for further work. 

\section*{Software}

Our estimation method is implemented in R and a code reproducing the examples of Section \ref{sec:Simulation} is available on a GitHub repository located \href{https://github.com/QuentinClairon/NLME_ODE_estimation_via_optimal_control.git}{here}.

\section*{Acknowledgement}

Experiments presented in this paper were carried out using the PlaFRIM
experimental testbed, supported by Inria, CNRS (LABRI and IMB), Universit\'e
de Bordeaux, Bordeaux INP and Conseil R\'egional d\textquoteright Aquitaine
(see https://www.plafrim.fr/).

This manuscript was developed under WP4 of EBOVAC3. This work has
received funding from the Innovative Medicines Initiative 2 Joint
Undertaking under projects EBOVAC1 and EBOVAC3 (respectively grant
agreement No 115854 and No 800176). The IMI2 Joint Undertaking receives
support from the European Union\textquoteright s Horizon 2020 research
and innovation programme and the European Federation of Pharmaceutical
Industries and Association.

\bibliographystyle{rss}
\bibliography{biblio_OCA_iterative_approach}

\begin{thebibliography}{67}
\expandafter\ifx\csname natexlab\endcsname\relax\def\natexlab#1{#1}\fi
\expandafter\ifx\csname url\endcsname\relax
  \def\url#1{\texttt{#1}}\fi
\expandafter\ifx\csname urlprefix\endcsname\relax\def\urlprefix{URL: }\fi

\bibitem[{Agusto and Adekunle(2014)}]{Agusto2014}
Agusto, F. and Adekunle, A. (2014) Optimal control of a two-strain
  tuberculosis-hiv/aidsco-infection model.
\newblock \textit{BioSystems}, \textbf{119}, 20--44.

\bibitem[{Aliyu(2011)}]{Aliyu2011}
Aliyu, M. (2011) \textit{Nonlinear H-Infinity Control, Hamiltonian Systems and
  Hamilton-Jacobi Equations}.
\newblock CRC Press.

\bibitem[{Andersen(1970)}]{Andersen1970}
Andersen, E. (1970) Asymptotic properties of conditional maximum-likelihood
  estimators.
\newblock \textit{Journal of the Royal Statistical Society}, \textbf{32},
  283--301.

\bibitem[{Andraud et~al.(2012)Andraud, Lejeune, Musoro, Ogunjimi, Beutels and
  Hens}]{Andraud2012}
Andraud, M., Lejeune, O., Musoro, J., Ogunjimi, B., Beutels, P. and Hens, N.
  (2012) Living on three time scales: the dynamics of plasma cell and antibody
  populations illustrated for hepatitis a virus.
\newblock \textit{Plos Computational Biology}, \textbf{63}.

\bibitem[{Balelli et~al.(2020)Balelli, Pasin, Prague, Crauste, Van~Effelterre,
  Bockstal, Solforosi and Thi{\'e}baut}]{Balelli2019}
Balelli, I., Pasin, C., Prague, M., Crauste, F., Van~Effelterre, T., Bockstal,
  V., Solforosi, L. and Thi{\'e}baut, R. (2020) A model for establishment,
  maintenance and reactivation of the immune response after vaccination against
  ebola virus.
\newblock \textit{Journal of Theoretical Biology}, 110254.

\bibitem[{Bowsher and Swain(2012)}]{Bowsher2012}
Bowsher, C.~G. and Swain, P. (2012) Identifying source of variation and the
  flow of information in biochemical networks.
\newblock \textit{PNAS}, \textbf{109}, 1320--1328.

\bibitem[{Brynjarsdottir and O'Hagan(2014)}]{Brynjarsdottir2014}
Brynjarsdottir, J. and O'Hagan, A. (2014) Learning about physical parameters:
  The importance of model discrepancy.
\newblock \textit{Inverse Problems}, \textbf{30}, 24.

\bibitem[{Campbell(2007)}]{campbell2007}
Campbell, D. (2007) \textit{Bayesian Collocation Tempering and Generalized
  Profiling for Estimation of Parameters from Differential Equation Models}.
\newblock Ph.D. thesis, McGill University Montreal,Quebec.

\bibitem[{Cimen(2008)}]{Cimen2008}
Cimen, T. (2008) State-dependent riccati equation (sdre) control: A survey.
\newblock \textit{IFAC Proceedings}, \textbf{41}, 3761--3775.

\bibitem[{Cimen and Banks(2004{\natexlab{a}})}]{CimenBanks2004}
Cimen, T. and Banks, S. (2004{\natexlab{a}}) Global optimal feedback control
  for general nonlinear systems with nonquadratic performance criteria.
\newblock \textit{Systems and Control Letters}, \textbf{53}, 327--346.

\bibitem[{Cimen and Banks(2004{\natexlab{b}})}]{Cimen2004}
--- (2004{\natexlab{b}}) Nonlinear optimal tracking control with application to
  super-tankers for autopilot design.
\newblock \textit{Automatica}, \textbf{40}, 1845--1863.

\bibitem[{Clairon(2020)}]{Clairon2019}
Clairon, Q. (2020) A regularization method for the parameter estimation problem
  in ordinary differential equations via discrete optimal control theory.
\newblock \textit{Journal of Statistical Planning and Inference}.

\bibitem[{Clairon and Brunel(2018)}]{BrunelClairon_Pontryagin2017}
Clairon, Q. and Brunel, N. J.-B. (2018) Optimal control and additive
  perturbations help in estimating ill-posed and uncertain dynamical systems.
\newblock \textit{Journal of the American Statistical Association},
  \textbf{113}, 1195--1209.

\bibitem[{Clarke(2013)}]{clarke2013variationalcalculus}
Clarke, F. (2013) \textit{Functional Analysis, Calculus of Variations and
  Optimal Control}.
\newblock Graduate Texts in Mathematics. Springer-Verlag London.

\bibitem[{Comets et~al.(2017)Comets, Lavenu and Lavielle}]{Comets2017}
Comets, E., Lavenu, A. and Lavielle, M. (2017) Parameter estimation in
  nonlinear mixed effect models using saemix, an r implementation of the saem
  algorithm.
\newblock \textit{Journal of Statistical Software}, \textbf{80}, 1--42.

\bibitem[{Dashti et~al.(2013)Dashti, Law, Stuart and Voss}]{Dashti2013}
Dashti, M., Law, K. J.~H., Stuart, A. and Voss, J. (2013) Map estimators and
  their consistency in bayesian nonparametric inverse problems.
\newblock \textit{Inverse Problems}, \textbf{29}.

\bibitem[{Donnet and Samson(2006)}]{Donnet2006}
Donnet, S. and Samson, A. (2006) Estimation of parameters in incomplete data
  models defined by dynamical systems.
\newblock \textit{Journal of Statistical Planning and Inference}, \textbf{137},
  2815--2831.

\bibitem[{Engl et~al.(2009)Engl, Flamm, K{\"u}gler, Lu, M{\"u}ller and
  Schuster}]{Engl2009}
Engl, H., Flamm, C., K{\"u}gler, P., Lu, J., M{\"u}ller, S. and Schuster, P.
  (2009) Inverse problems in systems biology.
\newblock \textit{Inverse Problems}, \textbf{25}.

\bibitem[{G.~Hooker and Earn(2011)}]{Hooker2011}
G.~Hooker, S. P.~Ellner, L. D. V.~R. and Earn, D. J.~D. (2011) Parameterizing
  state-space models for infectious disease dynamics by generalized profiling:
  measles in ontario.
\newblock \textit{Journal of the Royal Society}, \textbf{8}, 961--974.

\bibitem[{Gillespie(2000)}]{Gillespie2000}
Gillespie, D. (2000) The chemical langevin equation.
\newblock \textit{Journal of Chemical Physics}, \textbf{113}, 297--306.

\bibitem[{Guedj et~al.(2007)Guedj, Thiebaut and Commenges}]{Guedj2007}
Guedj, J., Thiebaut, R. and Commenges, D. (2007) Maximum likelihood estimation
  in dynamical models of hiv.
\newblock \textit{Biometrics}, \textbf{63}, 1198--206.

\bibitem[{Guo and Sun(2012)}]{Guo2012}
Guo, B. and Sun, B. (2012) Dynamic programming approach to the numerical
  solution of optimal control with paradigm by a mathematical model for drug
  therapies.
\newblock \textit{Optimization and Engineering}, 1--18.

\bibitem[{Gutenkunst et~al.(2007)Gutenkunst, Waterfall, Casey, Brown, Myers and
  Sethna}]{Gutenkunst2007}
Gutenkunst, R.~N., Waterfall, J., Casey, F., Brown, K., Myers, C. and Sethna,
  J. (2007) Universally sloppy parameter sensitivities in systems biology
  models.
\newblock \textit{Public Library of Science Computational Biology}, \textbf{3},
  e189.

\bibitem[{Hooker et~al.(2015)Hooker, Ellner et~al.}]{HookerEllner2013}
Hooker, G., Ellner, S.~P. et~al. (2015) Goodness of fit in nonlinear dynamics:
  misspecified rates or misspecified states?
\newblock \textit{The Annals of Applied Statistics}, \textbf{9}, 754--776.

\bibitem[{Huang and Dagne(2011)}]{Huang2011}
Huang, Y. and Dagne, G. (2011) A bayesian approach to joint mixed-effects
  models with a skew normal distribution and measurement errors in covariates.
\newblock \textit{Biometrics}, \textbf{67}, 260--269.

\bibitem[{Huang et~al.(2006)Huang, Liu and Wu}]{Huang2006b}
Huang, Y., Liu, D. and Wu, H. (2006) Hierachical bayesian methods for
  estimation of parameters in a longitudinal hiv dynamic system.
\newblock \textit{Biometrics}, \textbf{62}, 413--423.

\bibitem[{Huang and Lu(2008)}]{Huang2008}
Huang, Y. and Lu, T. (2008) Modeling long-term longitudinal hiv dynamics with
  application to an aids clinical study.
\newblock \textit{Annal of Applied Statistics}, \textbf{2}, 1348--1408.

\bibitem[{Huang et~al.(2010)Huang, Wu and Acosta}]{Huang2010}
Huang, Y., Wu, H. and Acosta, E.~P. (2010) Hierarchical bayesian inference for
  hiv dynamic differential equation models incorporating multiple treatment
  factors.
\newblock \textit{Biom J}, \textbf{52}, 470--486.

\bibitem[{Kampen(1992)}]{Kampen1992}
Kampen, N.~V. (1992) \textit{Stochastic Process in Physics and Chemistry}.
\newblock Elsevier.

\bibitem[{Kennedy and O'Hagan(2001)}]{Kennedy2001}
Kennedy, M.~C. and O'Hagan, A. (2001) Bayesian calibration of computer models.
\newblock \textit{Journal of the Royal Statistical Society: Series B
  (Statistical Methodology)}, \textbf{63}, 425--464.

\bibitem[{Kirk(1998)}]{Kirk1998optimalcontrol}
Kirk, D.~E. (1998) \textit{Optimal Control Theory: An Introduction}.
\newblock Dover Publication.

\bibitem[{Kirk et~al.(2016)Kirk, Silk and Michael}]{kirk2016reverse}
Kirk, P., Silk, D. and Michael, M. (2016) Reverse engineering under
  uncertainty.
\newblock In \textit{Uncertainty in Biology}, 15--32. Springer.

\bibitem[{Komorowski et~al.(2013)Komorowski, Miekisz and
  Stumpf}]{Komorowski2013}
Komorowski, M., Miekisz, J. and Stumpf, M. (2013) Decomposing noise in
  biochemical signaling systems highlights the role of protein degradation.
\newblock \textit{Biophysical journal}, \textbf{10}, 1783--1793.

\bibitem[{Kuhn and Lavielle(2005)}]{Lavielle2005}
Kuhn, E. and Lavielle, M. (2005) Maximum likelihood estimation in nonlinear
  mixed effects models.
\newblock \textit{Computational Statistics and Data Analysis}, \textbf{49},
  1020--1038.

\bibitem[{Kurtz(1978)}]{Kurtz1978}
Kurtz, T. (1978) Strong approximation theorems for density dependent markov
  chains.
\newblock \textit{Stochastic Processes and their Applications}, \textbf{6},
  223--240.

\bibitem[{Lavielle and Aarons(2015)}]{Lavielle2015}
Lavielle, M. and Aarons, L. (2015) What do we mean by identifiability in mixed
  effects models?
\newblock \textit{Journal of pharmacokinetics and pharmacodynamics}.

\bibitem[{Lavielle and Mentr\'e(2007)}]{Lavielle2007}
Lavielle, M. and Mentr\'e, F. (2007) Estimation of population pharmacokinetic
  parameters of saquinavir in hiv patients with the monolix software.
\newblock \textit{Journal of Pharmacokinetics and Pharmacodynamics},
  \textbf{34}.

\bibitem[{Le et~al.(2015)Le, Miller and Ganusov}]{Le2015}
Le, D., Miller, J. and Ganusov, V. (2015) Mathematical modeling provides
  kinetic details of the human immune response to vaccination.
\newblock \textit{Frontiers in Cellular and Infection Microbiology},
  \textbf{4}, 177.

\bibitem[{Leary et~al.(2015)Leary, Sutton and Marder}]{OLeary2015}
Leary, T.~O., Sutton, A. and Marder, E. (2015) Computational models in the age
  of large datasets.
\newblock \textit{Current Opinion in Neurobiology}, \textbf{32}, 87--94.

\bibitem[{Lindstrom and Bates(1990)}]{Lindstrom1990}
Lindstrom, M.~J. and Bates, D.~M. (1990) Nonlinear mixed effects models for
  repeated measures data.
\newblock \textit{Biometrics}, \textbf{46}, 673--687.

\bibitem[{Lunn et~al.(2000)Lunn, A.Thomas, Best and Spiegelhalter}]{Lunn2000}
Lunn, D., A.Thomas, Best, N. and Spiegelhalter, D. (2000) Winbugs - a bayesian
  modelling framework: Concepts, structure and extensibility.
\newblock \textit{Statistics and Computing}, \textbf{10}, 325--337.

\bibitem[{M.~Lavielle and Mentre(2011)}]{Lavielle2011}
M.~Lavielle, A.~Samson, A.~F. and Mentre, F. (2011) Maximum likelihood
  estimation of long terms hiv dynamic models and antiviral response.
\newblock \textit{Biometrics}, \textbf{67}, 250--259.

\bibitem[{Murphy and der Vaart(2000)}]{Murphy2000}
Murphy, S. and der Vaart, A.~V. (2000) On profile likelihood.
\newblock \textit{Journal of American Statistical Association}, \textbf{95},
  449--465.

\bibitem[{Nash(2016)}]{nash2016using}
Nash, J.~C. (2016) Using and extending the optimr package.

\bibitem[{Pasin et~al.(2019)Pasin, Balelli, Van~Effelterre, Bockstal,
  Solforosi, Prague, Douoguih and Thi{\'e}baut}]{Pasin2019}
Pasin, C., Balelli, I., Van~Effelterre, T., Bockstal, V., Solforosi, L.,
  Prague, M., Douoguih, M. and Thi{\'e}baut, R. (2019) Dynamics of the humoral
  immune response to a prime-boost ebola vaccine: quantification and sources of
  variation.
\newblock \textit{Journal of virology}, \textbf{93}, e00579--19.

\bibitem[{Pasin et~al.(2018)Pasin, Dufour, Villain, Zhang and
  Thiebaut}]{Pasin2018}
Pasin, C., Dufour, F., Villain, L., Zhang, H. and Thiebaut, R. (2018)
  Controlling il-7 injections in hiv-infected patients.
\newblock \textit{Bulletin of Mathematical Biology}, \textbf{80}, 2349--2377.

\bibitem[{Perelson et~al.(1996)Perelson, Neumann, Markowitz, Leonard and
  Ho}]{Perelson1996}
Perelson, A., Neumann, A., Markowitz, M., Leonard, J. and Ho, D. (1996) Hiv-1
  dynamics in vivo: virion clearance rate, infected cell life-span, and viral
  generation time.
\newblock \textit{Science}, \textbf{271}, 1582--1586.

\bibitem[{Pinheiro and Bates(1994)}]{Pinheiro1994}
Pinheiro, J. and Bates, D.~M. (1994) Approximations to the loglikelihood
  function in the nonlinear mixed effects model.
\newblock \textit{Journal of the Computational and Graphical Statistics},
  \textbf{4}, 12--35.

\bibitem[{Prague et~al.(2013)Prague, Commengues, Guedj, Drylewicz and
  Thi\'ebaut}]{Pragues2013}
Prague, M., Commengues, D., Guedj, J., Drylewicz, J. and Thi\'ebaut, R. (2013)
  Nimrod: A program for inference via a normal approximation of the posterior
  in models with random effects based on ordinary differential equations.
\newblock \textit{Computer Methods and Programs in Biomedicine}, \textbf{111},
  447--458.

\bibitem[{Raftery and Bao(2010)}]{Raftery2010}
Raftery, A. and Bao, L. (2010) Estimating and projecting trends in hiv/aids
  generalized epidemics using incremental mixture importance sampling.
\newblock \textit{Biometrics}, \textbf{66}, 1162--1173.

\bibitem[{Ramsay et~al.(2007)Ramsay, Hooker, Cao and Campbell}]{Ramsay2007}
Ramsay, J., Hooker, G., Cao, J. and Campbell, D. (2007) Parameter estimation
  for differential equations: A generalized smoothing approach.
\newblock \textit{Journal of the Royal Statistical Society (B)}, \textbf{69},
  741--796.

\bibitem[{Sartori(2003)}]{Sartori2003}
Sartori, N. (2003) Modified profile likelihood in models with stratum nuisance
  parameters.
\newblock \textit{Biometrika}, \textbf{90}, 553--549.

\bibitem[{Sontag(1998)}]{Sontag1998}
Sontag, E. (1998) \textit{Mathematical Control Theory: Deterministic
  finite-dimensional systems}.
\newblock Springer-Verlag (New-York).

\bibitem[{Stein et~al.(2013)Stein, Bucci, Toussaint, Buffie, Ratsch, Pamer,
  Sander and Xavier}]{Stein2013}
Stein, R., Bucci, V., Toussaint, N., Buffie, C., Ratsch, G., Pamer, E., Sander,
  C. and Xavier, J. (2013) Ecological modeling from time-series inference:
  Insight into dynamics and stability of intestinal microbiota.
\newblock \textit{Public Library of Science Computational Biology}, \textbf{9},
  12.

\bibitem[{Stuart(2010)}]{Stuart2010}
Stuart, A. (2010) Inverse problems: A bayesian perspective.
\newblock \textit{Acta Numerica}, 451--559.

\bibitem[{Thiebaut et~al.(2014)Thiebaut, Drylewicz, Prague, Lacabaratz and
  et~al.}]{Thiebaut2014}
Thiebaut, R., Drylewicz, J., Prague, M., Lacabaratz, C. and et~al., S.~B.
  (2014) Quantifying and predicting the effect of exogenous interleukin on
  cd4+t cells in hiv-1 infection.
\newblock \textit{Plos Computational Biology}, \textbf{10 (5)}.

\bibitem[{Tornoe et~al.(2004)Tornoe, Agerso, Jonsson, Madsen and
  Nielsen}]{Tornoe2004}
Tornoe, C., Agerso, H., Jonsson, E.~N., Madsen, H. and Nielsen, H.~A. (2004)
  Non-linear mixed-effects pharmacokinetic/pharmacodynamic modelling in nlme
  using differential equations.
\newblock \textit{Computer Methods and Programs in Biomedicine}, \textbf{76},
  31--41.

\bibitem[{Transtrum et~al.(2011)Transtrum, Machta and Sethna}]{Transtrum2011}
Transtrum, M., Machta, B. and Sethna, J. (2011) Geometry of nonlinear least
  squares with applications to sloppy models and optimization.
\newblock \textit{Physical Review}, \textbf{83}, 35.

\bibitem[{Transtrum et~al.(2015)Transtrum, Machta, Brown, Daniels, Myers and
  Sethna}]{Transtrum2015}
Transtrum, M.~K., Machta, B.~B., Brown, K.~S., Daniels, B.~C., Myers, C.~R. and
  Sethna, J.~P. (2015) Perspective: Sloppiness and emergent theories in
  physics, biology, and beyond.
\newblock \textit{The Journal of chemical physics}, \textbf{143}, 07B201\_1.

\bibitem[{Tuo and Wu(2015)}]{TuoWu2015}
Tuo, R. and Wu, C. (2015) Efficient calibration for imperfect computer models.
\newblock \textit{Annals of Statistics}.

\bibitem[{van~der Vaart(1998)}]{Vaart1998}
van~der Vaart, A. (1998) \textit{Asymptotic Statistics}.
\newblock Cambridge Series in Statistical and Probabilities Mathematics.
  Cambridge University Press.

\bibitem[{Varah(1982)}]{Varah1982}
Varah, J.~M. (1982) A spline least squares method for numerical parameter
  estimation in differential equations.
\newblock \textit{SIAM J.sci. Stat. Comput.}, \textbf{3}, 28--46.

\bibitem[{Villain et~al.(2019)Villain, Commenges, Pasin, Prague and
  Thi{\'e}baut}]{villain2019adaptive}
Villain, L., Commenges, D., Pasin, C., Prague, M. and Thi{\'e}baut, R. (2019)
  Adaptive protocols based on predictions from a mechanistic model of the
  effect of il7 on cd4 counts.
\newblock \textit{Statistics in medicine}, \textbf{38}, 221--235.

\bibitem[{Wakefield and Racine-Poon(1995)}]{Wakefield1995}
Wakefield, J. and Racine-Poon, A. (1995) An application of bayesian population
  pharmacokinetic/pharmacodynamic models to dose recommendation.
\newblock \textit{Statistics in Medicine}, \textbf{14}, 971--986.

\bibitem[{Wang et~al.(2014)Wang, Cao, Ramsay, Burger, Laporte and
  Rockstroh}]{Wang2014}
Wang, L., Cao, J., Ramsay, J., Burger, D., Laporte, C. and Rockstroh, J. (2014)
  Estimating mixed-effects differential equation models.
\newblock \textit{Statistics and Computing}, \textbf{24}, 111--121.

\bibitem[{Wu et~al.(2014)Wu, Lu, Xue and Liang}]{Wu2014}
Wu, H., Lu, T., Xue, H. and Liang, H. (2014) Sparse additive odes for dynamic
  gene regulatory network modeling.
\newblock \textit{Journal of the American Statistical Association},
  \textbf{109}, 700--716.

\bibitem[{Zhang and Xu(2016)}]{Zhang2016}
Zhang, S. and Xu, X. (2016) Dynamic analysis and optimal control for a model of
  hepatitis c with treatment.
\newblock \textit{Communications in Nonlinear Science and Numerical
  Simulation}, \textbf{46}, 14--25.

\end{thebibliography}
\end{document}

\section*{Appendix A: \label{sec:outer_criteria_derivation} Outer criteria
derivation}

We recall the definition of $\overline{G}^{(2)}$:
\[
\overline{G}^{(2)}\left[\theta,\Delta,\sigma\mid\mathbf{y}\right]=\sum_{i}\left(\ln\widetilde{\mathbb{P}}\left[\mathbf{y_{i}}\mid\theta,\Delta,\sigma,\widehat{b_{i}}\left(\theta,\Delta\right)\right]+\ln\mathbb{P}\left[\widehat{b_{i}}\left(\theta,\Delta\right)\mid\theta,\Delta,\sigma\right]\right)+\ln\mathbb{P}\left[\theta,\Delta\right].
\]
By using 
\[
\widetilde{\mathbb{P}}\left[\mathbf{y_{i}}\mid\theta,\Delta,\sigma,\widehat{b_{i}}\left(\theta,\Delta\right)\right]=\prod_{j}p(y_{ij}\mid\widehat{b_{i}}\left(\theta,\Delta\right),\theta,\Delta,\sigma)=\prod_{j}\left(2\pi\right)^{-d^{o}/2}\sigma^{-d^{o}}e^{-0.5\left\Vert C\overline{X}_{\theta,\widehat{b_{i}}\left(\theta,\Delta\right)}(t_{ij})-y_{ij}\right\Vert _{2}^{2}/\sigma^{2}},
\]
$\mathbb{P}\left[\widehat{b_{i}}\left(\theta,\Delta\right)\mid\theta,\Delta,\sigma\right]=\left(2\pi\right)^{-q/2}\left|\Psi\right|^{-1/2}e^{-0.5\widehat{b_{i}}\left(\theta,\Delta\right)\Psi^{-1}\widehat{b_{i}}\left(\theta,\Delta\right)}$,
we get
\[
\begin{array}{lll}
\ln\widetilde{\mathbb{P}}\left[\mathbf{y_{i}}\mid\theta,\Delta,\sigma,\widehat{b_{i}}\left(\theta,\Delta\right)\right] & = & \sum_{j}\left(-d^{o}\ln\sigma-d^{o}\ln\left(2\pi\right)/2-0.5/\sigma^{2}\left\Vert C\overline{X}_{\theta,\widehat{b_{i}}\left(\theta,\Delta\right)}(t_{ij})-y_{ij}\right\Vert _{2}^{2}\right)\\
 & = & -d^{o}n_{i}\left(\ln\sigma+\ln\left(2\pi\right)/2\right)-0.5/\sigma^{2}\sum_{j}\left\Vert C\overline{X}_{\theta,\widehat{b_{i}}\left(\theta,\Delta\right)}(t_{ij})-y_{ij}\right\Vert _{2}^{2}\\
\ln\mathbb{P}\left[\widehat{b_{i}}\left(\theta,\Delta\right)\mid\theta,\Delta,\sigma\right] & = & -q/2\ln\left(2\pi\right)-0.5\ln\left|\Psi\right|-0.5\widehat{b_{i}}\left(\theta,\Delta\right)^{T}\Psi^{-1}\widehat{b_{i}}\left(\theta,\Delta\right)
\end{array}
\]
so by reinjecting in $\overline{G}^{(2)}$ we obtain: {\small{}
\[
\begin{array}{l}
\overline{G}^{(2)}\left[\theta,\Delta,\sigma\mid\mathbf{y}\right]\\
=-\frac{0.5}{\sigma^{2}}\sum_{i}\sum_{j}\left\Vert C\overline{X}_{\theta,\widehat{b_{i}}\left(\theta,\Delta\right)}(t_{ij})-y_{ij}\right\Vert _{2}^{2}\\
-\sum_{i}\left(0.5\widehat{b_{i}}\left(\theta,\Delta\right)^{T}\Psi^{-1}\widehat{b_{i}}\left(\theta,\Delta\right)+d^{o}n_{i}\left(\ln\sigma+\ln\left(2\pi\right)/2\right)+q/2\ln\left(2\pi\right)+0.5\ln\left|\Psi\right|\right)+\ln\mathbb{P}\left[\theta,\Delta\right].
\end{array}
\]
}From this, we have $\arg\max_{\left(\theta,\Delta,\sigma\right)}\left\{ \overline{G}^{(2)}\left[\theta,\Delta,\sigma\mid\mathbf{y}\right]\right\} =\arg\max_{\left(\theta,\Delta,\sigma\right)}\left\{ \overline{G}^{(3)}\left[\theta,\Delta,\sigma\mid\mathbf{y}\right]\right\} $
with
\[
\begin{array}{lll}
\overline{G}^{(3)}\left[\theta,\Delta,\sigma\mid\mathbf{y}\right] & = & -0.5\sum_{i}\left(\sum_{j}\left\Vert C\overline{X}_{\theta,\widehat{b_{i}}\left(\theta,\Delta\right)}(t_{ij})-y_{ij}\right\Vert _{2}^{2}/\sigma^{2}+\widehat{b_{i}}\left(\theta,\Delta\right)^{T}\Psi^{-1}\widehat{b_{i}}\left(\theta,\Delta\right)\right)\\
 & - & \left(d^{o}\sum_{i}n_{i}\right)\ln\sigma-0.5n\ln\left|\Psi\right|+\ln\mathbb{P}\left[\theta,\Delta\right].
\end{array}
\]
Now let us use the relationship $\sigma^{2}\Psi^{-1}=\triangle^{T}\triangle$
(or equivalently $\Psi=\sigma^{2}\left(\triangle^{T}\triangle\right)^{-1}$),
we get: {\small{}
\[
\begin{array}{l}
\overline{G}^{(3)}\left[\theta,\Delta,\sigma\mid\mathbf{y}\right]\\
=-0.5\sum_{i}\left(\sum_{j}\left\Vert C\overline{X}_{\theta,\widehat{b_{i}}\left(\theta,\Delta\right)}(t_{ij})-y_{ij}\right\Vert _{2}^{2}/\sigma^{2}+\left\Vert \triangle\widehat{b_{i}}\left(\theta,\Delta\right)\right\Vert _{2}^{2}/\sigma^{2}\right)\\
-\left(d^{o}\sum_{i}n_{i}\right)\ln\sigma-0.5n\ln\left(\left|\sigma^{2}\left(\triangle^{T}\triangle\right)^{-1}\right|\right)+\ln\mathbb{P}\left[\theta,\Delta\right]\\
=-\frac{0.5}{\sigma^{2}}\sum_{i}\left(\sum_{j}\left\Vert C\overline{X}_{\theta,\widehat{b_{i}}\left(\theta,\Delta\right)}(t_{ij})-y_{ij}\right\Vert _{2}^{2}+\left\Vert \triangle\widehat{b_{i}}\left(\theta,\Delta\right)\right\Vert _{2}^{2}\right)-\left(d^{o}\sum_{i}n_{i}\right)\ln\sigma\\
-0.5n\left(\ln\left(\left|\left(\triangle^{T}\triangle\right)^{-1}\right|\right)+\ln\left(\sigma^{2q}\right)\right)+\ln\mathbb{P}\left[\theta,\Delta\right]\\
=-\frac{0.5}{\sigma^{2}}\sum_{i}\left(\sum_{j}\left\Vert C\overline{X}_{\theta,\widehat{b_{i}}\left(\theta,\Delta\right)}(t_{ij})-y_{ij}\right\Vert _{2}^{2}+\left\Vert \triangle\widehat{b_{i}}\left(\theta,\Delta\right)\right\Vert _{2}^{2}\right)-0.5\left(d^{o}\sum_{i}n_{i}+qn\right)\ln\left(\sigma^{2}\right)\\
-0.5n\ln\left(\left|\left(\triangle^{T}\triangle\right)^{-1}\right|\right)+\ln\mathbb{P}\left[\theta,\Delta\right]
\end{array}
\]
}From this, we derive:
\[
\nabla_{\sigma^{2}}\overline{G}^{(3)}\left[\theta,\Delta,\sigma\mid\mathbf{y}\right]=\frac{0.5}{\sigma^{4}}\sum_{i}\left(\sum_{j}\left\Vert C\overline{X}_{\theta,\widehat{b_{i}}\left(\theta,\Delta\right)}(t_{ij})-y_{ij}\right\Vert _{2}^{2}+\left\Vert \triangle\widehat{b_{i}}\left(\theta,\Delta\right)\right\Vert _{2}^{2}\right)-\frac{0.5}{\sigma^{2}}\left(d^{o}\sum_{i}n_{i}+qn\right)
\]
and we have $\nabla_{\sigma^{2}}\overline{G}^{(3)}\left[\theta,\Delta,\sigma\mid\mathbf{y}\right]=0$
when: 
\[
\sigma^{2}:=\sigma^{2}\left(\theta,\Delta\right)=\sum_{i}\left(\sum_{j}\left\Vert C\overline{X}_{\theta,\widehat{b_{i}}\left(\theta,\Delta\right)}(t_{ij})-y_{ij}\right\Vert _{2}^{2}+\left\Vert \triangle\widehat{b_{i}}\left(\theta,\Delta\right)\right\Vert _{2}^{2}\right)/\left(d^{o}\sum_{i}n_{i}+qn\right).
\]
By re-injecting, we get:
\[
\begin{array}{l}
\overline{G}^{(3)}\left[\theta,\Delta,\sigma^{2}\left(\theta,\Delta\right)\mid\mathbf{y}\right]\\
=-0.5/\left(\sum_{i}\left(\sum_{j}\left\Vert C\overline{X}_{\theta,\widehat{b_{i}}\left(\theta,\Delta\right)}(t_{ij})-y_{ij}\right\Vert _{2}^{2}+\left\Vert \triangle\widehat{b_{i}}\left(\theta,\Delta\right)\right\Vert _{2}^{2}\right)/\left(d^{o}\sum_{i}n_{i}+qn\right)\right)\\
\times\sum_{i}\left(\sum_{j}\left\Vert C\overline{X}_{\theta,\widehat{b_{i}}\left(\theta,\Delta\right)}(t_{ij})-y_{ij}\right\Vert _{2}^{2}+\left\Vert \triangle\widehat{b_{i}}\left(\theta,\Delta\right)\right\Vert _{2}^{2}\right)\\
-0.5\left(d^{o}\sum_{i}n_{i}+qn\right)\ln\left(\sigma^{2}\left(\theta,\Delta\right)\right)-0.5n\ln\left(\left|\left(\triangle^{T}\triangle\right)^{-1}\right|\right)+\ln\mathbb{P}\left[\theta,\Delta\right]\\
=-0.5\left(d^{o}\sum_{i}n_{i}+qn\right)-0.5\left(d^{o}\sum_{i}n_{i}+qn\right)\ln\left(\sigma^{2}\left(\theta,\Delta\right)\right)+n\ln\left|\triangle\right|+\ln\mathbb{P}\left[\theta,\Delta\right]
\end{array}
\]
Thus $\begin{array}{lll}
\arg\max_{\left(\theta,\Delta\right)}\left\{ \overline{G}^{(3)}\left[\theta,\Delta,\sigma^{2}\left(\theta,\Delta\right)\mid\mathbf{y}\right]\right\}  & = & \arg\max_{\left(\theta,\Delta\right)}\left\{ G\left[\theta,\Delta\mid\mathbf{y}\right]\right\} \end{array}$with: 
\[
G\left[\theta,\Delta\mid\mathbf{y}\right]=-0.5\left(d^{o}\sum_{i}n_{i}+qn\right)\ln\left(\sigma^{2}\left(\theta,\Delta\right)\right)+n\ln\left|\triangle\right|+\ln\mathbb{P}\left[\theta,\Delta\right].
\]

\section*{Appendix B: Estimation results for $n=20$}

In this section, we resume examples of section 4 for $n=20.$ We explore
the effect of small sample size on the relative accuracy of our method
with respect to ML. In particular, we want to estimate the coverage
rate of our 95\%-confidence interval in a situation where the asymptotic
justification for our variance estimator (and ML variance estimator
as well) is questionable. Interestingly, the regularization brought
by our method limits the degradation of estimation accuracy when we
move from $n=50$ to $n=20$ comparing to ML in all examples below
both in well and misspecified cases. Also, our confidence intervals
generally have a better coverage rate and closer to 95\% than the
ML ones, the exception for $\Psi^{*}$ being explained by the high
variance of SAEM estimator for these parameters.

\subsection*{Partially observed linear model}

We present the estimation results in table \ref{tab:lin2dim_Bias_Var_pop_param_n20}.
{\footnotesize{}}

\begin{table}
\caption{\label{tab:lin2dim_Bias_Var_pop_param_n20} Results of estimation
for model (4.1) for $n=20$}
{\footnotesize{}}%
\begin{tabular}{l|lllllll}
\hline 
\multicolumn{8}{l}{Well-specified model}\tabularnewline
\hline 
\multicolumn{1}{l}{} &  & {\footnotesize{}MSE} & {\footnotesize{}Bias} & {\footnotesize{}Emp. Var} & {\footnotesize{}Est. Var} & {\footnotesize{}Cov. Rate} & {\footnotesize{}MSE $b_{i}$}\tabularnewline
\hline 
\hline 
\multirow{4}{*}{{\footnotesize{}$\theta_{1}$}} & {\footnotesize{}$\widehat{\theta}_{ML,x_{0,1},x_{0,2}}$} & 0.02 & -0.01 & 0.02 & 0.02 & 0.89 & \tabularnewline
 & {\footnotesize{}$\widehat{\theta}_{ML,x_{0,2}}$} & 0.03 & -0.01 & 0.03 & 0.03 & 0.91 & \tabularnewline
 & {\footnotesize{}$\widehat{\theta}_{ML}$} & 0.20 & 0.01 & 0.20 & 0.04 & 0.80 & \tabularnewline
 & {\footnotesize{}$\widehat{\theta}$} & \textbf{0.02} & \textbf{-0.02} & \textbf{0.02} & \textbf{0.03} & \textbf{0.96} & \tabularnewline
\hline 
\multirow{4}{*}{{\footnotesize{}$\theta_{2}$}} & {\footnotesize{}$\widehat{\theta}_{ML,x_{0,1},x_{0,2}}$} & 0.02 & 0.02 & 0.02 & 0.001 & 0.94 & \tabularnewline
 & {\footnotesize{}$\widehat{\theta}_{ML,x_{0,2}}$} & 0.02 & 0.02 & 0.02 & 3e-4 & 0.89 & \tabularnewline
 & {\footnotesize{}$\widehat{\theta}_{ML}$} & 0.03 & 0.05 & 0.02 & 2e-3 & 0.69 & \tabularnewline
 & {\footnotesize{}$\widehat{\theta}$} & \textbf{1e-4} & \textbf{2e-3} & \textbf{1e-4} & \textbf{1e-4} & \textbf{0.92} & \tabularnewline
\hline 
\multirow{4}{*}{{\footnotesize{}$\Psi$}} & {\footnotesize{}$\widehat{\theta}_{ML,x_{0,1},x_{0,2}}$} & 0.02 & 0.01 & 0.02 & 0.02 & 1 & 0.01\tabularnewline
 & {\footnotesize{}$\widehat{\theta}_{ML,x_{0,2}}$} & 0.03 & -0.01 & 0.03 & 0.02 & 1 & 0.01\tabularnewline
 & {\footnotesize{}$\widehat{\theta}_{ML}$} & 0.24 & 0.26 & 0.14 & 0.11 & 1 & 0.15\tabularnewline
 & {\footnotesize{}$\widehat{\theta}$} & \textbf{0.03} & \textbf{-0.03} & \textbf{0.03} & \textbf{0.03} & \textbf{0.88} & \textbf{0.01}\tabularnewline
\hline 
\multicolumn{8}{l}{Misspecified model}\tabularnewline
\hline 
\multicolumn{1}{l}{} &  & {\footnotesize{}MSE} & {\footnotesize{}Bias} & {\footnotesize{}Emp. Var} & {\footnotesize{}Est. Var} & {\footnotesize{}Cov. Rate} & {\footnotesize{}MSE $b_{i}$}\tabularnewline
\hline 
\hline 
\multirow{4}{*}{{\footnotesize{}$\theta_{1}$}} & {\footnotesize{}$\widehat{\theta}_{ML,x_{0,1},x_{0,2}}$} & 0.01 & -0.01 & 0.01 & 0.02 & 0.94 & \tabularnewline
 & {\footnotesize{}$\widehat{\theta}_{ML,x_{0,2}}$} & 0.01 & -0.01 & 0.01 & 0.02 & 0.88 & \tabularnewline
 & {\footnotesize{}$\widehat{\theta}_{ML}$} & 0.30 & 0.09 & 0.29 & 0.24 & 0.70 & \tabularnewline
 & {\footnotesize{}$\widehat{\theta}$} & \textbf{0.03} & \textbf{1e-3} & \textbf{0.03} & \textbf{0.24} & \textbf{1} & \tabularnewline
\hline 
\multirow{4}{*}{{\footnotesize{}$\theta_{2}$}} & {\footnotesize{}$\widehat{\theta}_{ML,x_{0,1},x_{0,2}}$} & 5e-4 & -5e-4 & 5e-4 & 4e-3 & 0.70 & \tabularnewline
 & {\footnotesize{}$\widehat{\theta}_{ML,x_{0,2}}$} & 6e-4 & -1e-3 & 6e-4 & 5e-4 & 0.65 & \tabularnewline
 & {\footnotesize{}$\widehat{\theta}_{ML}$} & 0.03 & 0.07 & 0.03 & 0.002 & 0.62 & \tabularnewline
 & {\footnotesize{}$\widehat{\theta}$} & \textbf{4e-4} & \textbf{6e-4} & \textbf{4e-4} & \textbf{4e-4} & \textbf{0.92} & \tabularnewline
\hline 
\multirow{4}{*}{{\footnotesize{}$\Psi$}} & {\footnotesize{}$\widehat{\theta}_{ML,x_{0,1},x_{0,2}}$} & 0.02 & -0.02 & 0.02 & 0.02 & 1 & 0.01\tabularnewline
 & {\footnotesize{}$\widehat{\theta}_{ML,x_{0,2}}$} & 0.02 & -0.03 & 0.02 & 0.02 & 1 & 0.02\tabularnewline
 & {\footnotesize{}$\widehat{\theta}_{ML}$} & 0.46 & 0.37 & 0.33 & 0.05 & 1 & 0.18\tabularnewline
 & {\footnotesize{}$\widehat{\theta}$} & \textbf{0.04} & \textbf{-0.05} & \textbf{0.03} & \textbf{0.02} & \textbf{0.92} & \textbf{0.01}\tabularnewline
\end{tabular}{\footnotesize\par}
{\footnotesize\par}
\end{table}

\subsection*{Partially observed nonlinear model}

We present the estimation results in table \ref{tab:Insulin_Misspe_model_Bias_Var_pop_param_n20}.{\footnotesize{}
}
\begin{table}
\caption{\label{tab:Insulin_Misspe_model_Bias_Var_pop_param_n20} Results of
estimation for model (4.3) for $n=20$}
{\footnotesize{}}%
\begin{tabular}{l|lllllll}
\hline 
\multicolumn{8}{l}{Well-specified model}\tabularnewline
\hline 
\multicolumn{1}{l}{} &  & {\footnotesize{}MSE} & {\footnotesize{}Bias} & {\footnotesize{}Emp. Var} & {\footnotesize{}Est. Var} & {\footnotesize{}Cov. Rate} & {\footnotesize{}MSE $b_{i}$}\tabularnewline
\hline 
\hline 
\multirow{4}{*}{{\footnotesize{}$\theta_{S_{G}}$}} & {\footnotesize{}$\widehat{\theta}_{ML,S_{i}}$} & 1.7e-4 & 5e-3 & 1.4e-4 & 4.5e-5 & 0.74 & \tabularnewline
 & {\footnotesize{}$\widehat{\theta}_{ML}$} & 0.01 & 0.06 & 6.5e-3 & 9.7e-4 & 0.62 & \tabularnewline
 & {\footnotesize{}$\widehat{\theta}_{S_{i}}$} & \textbf{6.7e-5} & \textbf{9.8e-4} & \textbf{6.6e-5} & \textbf{6.6e-5} & \textbf{0.92} & \tabularnewline
 & {\footnotesize{}$\widehat{\theta}$} & \textbf{4e-3} & \textbf{2.4e-3} & \textbf{3.8e-3} & \textbf{3.5e-3} & \textbf{0.92} & \tabularnewline
\hline 
\multirow{4}{*}{{\footnotesize{}$\theta_{S_{I}}$}} & {\footnotesize{}$\widehat{\theta}_{ML,S_{i}}$} & \multicolumn{5}{l}{known} & \tabularnewline
 & {\footnotesize{}$\widehat{\theta}_{ML}$} & 0.02 & 0.07 & 0.02 & 1.4e-3 & 0.67 & \tabularnewline
 & {\footnotesize{}$\widehat{\theta}_{S_{i}}$} & \multicolumn{5}{l}{\textbf{known}} & \tabularnewline
 & {\footnotesize{}$\widehat{\theta}$} & \textbf{3.6e-3} & \textbf{4.7e-3} & \textbf{3.6e-3} & \textbf{6.1e-3} & \textbf{0.92} & \tabularnewline
\hline 
\multirow{4}{*}{{\footnotesize{}$\theta_{n}$}} & {\footnotesize{}$\widehat{\theta}_{ML,S_{i}}$} & 1.8e-3 & -4.9e-3 & 1.8e-3 & 1.3e-4 & 0.86 & \tabularnewline
 & {\footnotesize{}$\widehat{\theta}_{ML}$} & 0.01 & -3e-3 & 0.01 & 1.4e-3 & 0.82 & \tabularnewline
 & {\footnotesize{}$\widehat{\theta}_{S_{i}}$} & \textbf{1.0e-3} & \textbf{5.4e-3} & \textbf{1.0e-3} & \textbf{1.2e-3} & \textbf{0.92} & \tabularnewline
 & {\footnotesize{}$\widehat{\theta}$} & \textbf{1.1e-3} & \textbf{4.7e-3} & \textbf{1.1e-3} & \textbf{1.4e-3} & \textbf{0.90} & \tabularnewline
\hline 
\multirow{4}{*}{{\footnotesize{}$\Psi$}} & {\footnotesize{}$\widehat{\theta}_{ML,S_{i}}$} & 0.05 & -0.04 & 0.04 & 0.05 & 0.92 & 0.03\tabularnewline
 & {\footnotesize{}$\widehat{\theta}_{ML}$} & 0.06 & -0.04 & 0.05 & 0.05 & 0.90 & 0.03\tabularnewline
 & {\footnotesize{}$\widehat{\theta}_{S_{i}}$} & \textbf{0.03} & \textbf{-0.04} & \textbf{0.03} & \textbf{0.02} & \textbf{0.90} & \textbf{0.01}\tabularnewline
 & {\footnotesize{}$\widehat{\theta}$} & \textbf{0.04} & \textbf{-0.04} & \textbf{0.04} & \textbf{0.04} & \textbf{0.90} & \textbf{0.02}\tabularnewline
\hline 
\multicolumn{8}{l}{Misspecified model}\tabularnewline
\hline 
\multicolumn{1}{l}{} &  & {\footnotesize{}MSE} & {\footnotesize{}Bias} & {\footnotesize{}Emp. Var} & {\footnotesize{}Est. Var} & {\footnotesize{}Cov. Rate} & {\footnotesize{}MSE $b_{i}$}\tabularnewline
\hline 
\hline 
\multirow{4}{*}{{\footnotesize{}$\theta_{S_{G}}$}} & {\footnotesize{}$\widehat{\theta}_{ML,S_{i}}$} & 1.1e-4 & 6.8e-3 & 7.1e-5 & 1.7e-5 & 0.72 & \tabularnewline
 & {\footnotesize{}$\widehat{\theta}_{ML}$} & 0.02 & 0.04 & 0.01 & 1.7e-4 & 0.50 & \tabularnewline
 & {\footnotesize{}$\widehat{\theta}_{S_{i}}$} & \textbf{3.2e-5} & \textbf{2.3e-3} & \textbf{2.7e-5} & \textbf{2.7e-5} & \textbf{0.94} & \tabularnewline
 & {\footnotesize{}$\widehat{\theta}$} & \textbf{4.7e-4} & \textbf{7.7e-4} & \textbf{4.7e-4} & \textbf{4.5e-4} & \textbf{0.93} & \tabularnewline
\hline 
\multirow{4}{*}{{\footnotesize{}$\theta_{S_{I}}$}} & {\footnotesize{}$\widehat{\theta}_{ML,S_{i}}$} & known &  &  &  &  & \tabularnewline
 & {\footnotesize{}$\widehat{\theta}_{ML}$} & 0.02 & 0.02 & 0.01 & 1.0e-4 & 0.50 & \tabularnewline
 & {\footnotesize{}$\widehat{\theta}_{S_{i}}$} & \textbf{known} &  &  &  &  & \tabularnewline
 & {\footnotesize{}$\widehat{\theta}$} & \textbf{2.3e-4} & \textbf{-1.0e-4} & \textbf{2.3e-4} & \textbf{2.8e-4} & \textbf{0.92} & \tabularnewline
\hline 
\multirow{4}{*}{{\footnotesize{}$\theta_{n}$}} & {\footnotesize{}$\widehat{\theta}_{ML,S_{i}}$} & 1.4e-3 & -3.2e-4 & 1.3e-3 & 1.1e-3 & 0.89 & \tabularnewline
 & {\footnotesize{}$\widehat{\theta}_{ML}$} & 5.3e-3 & -5.0e-3 & 5.0e-3 & 1.1e-3 & 0.50 & \tabularnewline
 & {\footnotesize{}$\widehat{\theta}_{S_{i}}$} & \textbf{1.3e-3} & \textbf{3.4e-3} & \textbf{1.2e-3} & \textbf{1.1e-3} & \textbf{0.90} & \tabularnewline
 & {\footnotesize{}$\widehat{\theta}$} & \textbf{1.3e-3} & \textbf{4.5e-3} & \textbf{1.3e-3} & \textbf{1.2e-3} & \textbf{0.90} & \tabularnewline
\hline 
\multirow{4}{*}{{\footnotesize{}$\Psi$}} & {\footnotesize{}$\widehat{\theta}_{ML,S_{i}}$} & 0.03 & -0.05 & 0.02 & 0.04 & 0.91 & 0.02\tabularnewline
 & {\footnotesize{}$\widehat{\theta}_{ML}$} & 0.05 & -0.07 & 0.05 & 0.05 & 0.87 & 0.02\tabularnewline
 & {\footnotesize{}$\widehat{\theta}_{S_{i}}$} & \textbf{0.02} & \textbf{-0.05} & \textbf{0.02} & \textbf{0.02} & \textbf{0.93} & \textbf{0.01}\tabularnewline
 & {\footnotesize{}$\widehat{\theta}$} & \textbf{0.03} & \textbf{-0.06} & \textbf{0.03} & \textbf{0.02} & \textbf{0.92} & \textbf{0.01}\tabularnewline
\end{tabular}{\footnotesize\par}
{\footnotesize\par}
\end{table}
{\footnotesize\par}

\subsection*{Antibody concentration evolution model}

Estimation results are presented in tables \ref{tab:Chloe_model_Bias_Var_pop_param_n20}
and \ref{tab:Chloe_model_misspe_Bias_Var_pop_param_n20}. {\footnotesize{}}
\begin{table}
\caption{\label{tab:Chloe_model_Bias_Var_pop_param_n20} Results of estimation
for model (4.6) for $n=20$.}
{\footnotesize{}}%
\begin{tabular}{l|lllllll}
\hline 
\multicolumn{8}{l}{Well-specified model}\tabularnewline
\hline 
\multicolumn{1}{l}{} &  & {\footnotesize{}MSE} & {\footnotesize{}Bias} & {\footnotesize{}Emp. Var} & {\footnotesize{}Est. Var} & {\footnotesize{}Cov. Rate} & {\footnotesize{}MSE $b_{i}$}\tabularnewline
\hline 
\hline 
\multirow{4}{*}{{\footnotesize{}$\psi_{\delta_{s}}$}} & {\footnotesize{}$\widehat{\theta}_{ML,\psi_{\delta_{s}}}$} & \multicolumn{5}{l}{known} & \tabularnewline
 & {\footnotesize{}$\widehat{\theta}_{ML}$} & 3.58 & 1.35 & 1.75 & 76.83 & 0.87 & \tabularnewline
 & {\footnotesize{}$\widehat{\theta}_{\psi_{\delta_{s}}}$} & \multicolumn{5}{l}{\textbf{known}} & \tabularnewline
 & {\footnotesize{}$\widehat{\theta}$} & \textbf{1.10} & \textbf{-0.40} & \textbf{0.97} & \textbf{1.14} & \textbf{0.87} & \tabularnewline
\hline 
\multirow{4}{*}{{\footnotesize{}$\psi_{\phi_{S}}$}} & {\footnotesize{}$\widehat{\theta}_{ML,\psi_{\delta_{s}}}$} & 3e-3 & 0.02 & 3e-3 & 1e-3 & 0.89 & \tabularnewline
 & {\footnotesize{}$\widehat{\theta}_{ML}$} & 0.02 & -0.08 & 7e-3 & 0.43 & 0.90 & \tabularnewline
 & {\footnotesize{}$\widehat{\theta}_{\psi_{\delta_{s}}}$} & \textbf{3e-3} & \textbf{-0.04} & \textbf{1e-3} & \textbf{2e-3} & \textbf{0.90} & \tabularnewline
 & {\footnotesize{}$\widehat{\theta}$} & \textbf{7e-3} & \textbf{-0.03} & \textbf{6e-3} & \textbf{4e-3} & \textbf{0.90} & \tabularnewline
\hline 
\multirow{4}{*}{{\footnotesize{}$\psi_{\phi_{L}}$}} & {\footnotesize{}$\widehat{\theta}_{ML,\psi_{\delta_{s}}}$} & 0.01 & 0.02 & 0.01 & 6e-3 & 0.91 & \tabularnewline
 & {\footnotesize{}$\widehat{\theta}_{ML}$} & 0.02 & 0.05 & 0.02 & 7e-3 & 0.84 & \tabularnewline
 & {\footnotesize{}$\widehat{\theta}_{\psi_{\delta_{s}}}$} & \textbf{4e-3} & \textbf{-0.01} & \textbf{4e-3} & \textbf{0.01} & \textbf{0.95} & \tabularnewline
 & {\footnotesize{}$\widehat{\theta}$} & \textbf{0.01} & \textbf{4e-3} & \textbf{0.01} & \textbf{0.01} & \textbf{0.85} & \tabularnewline
\hline 
\multirow{4}{*}{{\footnotesize{}$\psi_{\delta_{Ab}}$}} & {\footnotesize{}$\widehat{\theta}_{ML,\psi_{\delta_{s}}}$} & 3e-3 & -0.02 & 3e-3 & 1e-3 & 0.82 & \tabularnewline
 & {\footnotesize{}$\widehat{\theta}_{ML}$} & 0.01 & -0.03 & 0.02 & 0.01 & 0.76 & \tabularnewline
 & {\footnotesize{}$\widehat{\theta}_{\psi_{\delta_{s}}}$} & \textbf{1e-3} & \textbf{0.01} & \textbf{4e-4} & \textbf{1e-3} & \textbf{0.93} & \tabularnewline
 & {\footnotesize{}$\widehat{\theta}$} & \textbf{6e-4} & \textbf{0.01} & \textbf{5e-4} & \textbf{6e-4} & \textbf{0.86} & \tabularnewline
\hline 
\multirow{4}{*}{{\footnotesize{}$\Psi_{\phi_{S}}$}} & {\footnotesize{}$\widehat{\theta}_{ML,\psi_{\delta_{s}}}$} & 0.04 & 0.05 & 0.03 & 0.13 & 1 & 0.15\tabularnewline
 & {\footnotesize{}$\widehat{\theta}_{ML}$} & 0.16 & 0.05 & 0.15 & 0.28 & 1 & 0.18\tabularnewline
 & {\footnotesize{}$\widehat{\theta}_{\psi_{\delta_{s}}}$} & \textbf{0.02} & \textbf{0.02} & \textbf{0.03} & \textbf{0.05} & \textbf{0.93} & \textbf{0.10}\tabularnewline
 & {\footnotesize{}$\widehat{\theta}$} & \textbf{0.04} & \textbf{-0.01} & \textbf{0.04} & \textbf{0.03} & \textbf{0.90} & \textbf{0.11}\tabularnewline
\hline 
\multirow{4}{*}{{\footnotesize{}$\Psi_{\phi_{L}}$}} & {\footnotesize{}$\widehat{\theta}_{ML,\psi_{\delta_{s}}}$} & 0.05 & 0.03 & 0.05 & 0.08 & 1 & 0.30\tabularnewline
 & {\footnotesize{}$\widehat{\theta}_{ML}$} & 0.09 & 0.05 & 0.08 & 0.18 & 1 & 0.64\tabularnewline
 & {\footnotesize{}$\widehat{\theta}_{\psi_{\delta_{s}}}$} & \textbf{0.03} & \textbf{-0.03} & \textbf{0.03} & \textbf{0.03} & \textbf{0.86} & \textbf{0.10}\tabularnewline
 & {\footnotesize{}$\widehat{\theta}$} & \textbf{0.04} & \textbf{-0.10} & \textbf{0.03} & \textbf{0.03} & \textbf{0.90} & \textbf{0.12}\tabularnewline
\hline 
\multirow{4}{*}{{\footnotesize{}$\Psi_{\delta_{Ab}}$}} & {\footnotesize{}$\widehat{\theta}_{ML,\psi_{\delta_{s}}}$} & 0.23 & 0.10 & 0.20 & 0.12 & 1 & 0.38\tabularnewline
 & {\footnotesize{}$\widehat{\theta}_{ML}$} & 0.46 & 0.30 & 0.37 & 0.17 & 1 & 0.55\tabularnewline
 & {\footnotesize{}$\widehat{\theta}_{\psi_{\delta_{s}}}$} & \textbf{0.09} & \textbf{-0.21} & \textbf{0.04} & \textbf{0.11} & \textbf{0.85} & \textbf{0.03}\tabularnewline
 & {\footnotesize{}$\widehat{\theta}$} & \textbf{0.18} & \textbf{-0.34} & \textbf{0.07} & \textbf{0.10} & \textbf{0.83} & \textbf{0.04}\tabularnewline
\end{tabular}{\footnotesize\par}

{\footnotesize\par}
\end{table}

\begin{table}
\caption{\label{tab:Chloe_model_misspe_Bias_Var_pop_param_n20} Results of
estimation for misspecified model (4.3) for $n=20$.}
{\footnotesize{}}%
\begin{tabular}{l|lllllll}
\hline 
\multicolumn{8}{l}{Misspecified model}\tabularnewline
\hline 
\multicolumn{1}{l}{} &  & {\footnotesize{}MSE} & {\footnotesize{}Bias} & {\footnotesize{}Emp. Var} & {\footnotesize{}Est. Var} & {\footnotesize{}Cov. Rate} & {\footnotesize{}MSE $b_{i}$}\tabularnewline
\hline 
\hline 
\multirow{4}{*}{{\footnotesize{}$\psi_{\delta_{s}}$}} & {\footnotesize{}$\widehat{\theta}_{ML,\psi_{\delta_{s}}}$} & \multicolumn{5}{l}{known} & \tabularnewline
 & {\footnotesize{}$\widehat{\theta}_{ML}$} & 4.64 & 1.69 & 1.76 & 63.9 & 0.70 & \tabularnewline
 & {\footnotesize{}$\widehat{\theta}_{\psi_{\delta_{s}}}$} & \multicolumn{5}{l}{\textbf{known}} & \tabularnewline
 & {\footnotesize{}$\widehat{\theta}$} & \textbf{1.32} & \textbf{-0.70} & \textbf{0.92} & \textbf{0.60} & \textbf{0.77} & \tabularnewline
\hline 
\multirow{4}{*}{{\footnotesize{}$\psi_{\phi_{S}}$}} & {\footnotesize{}$\widehat{\theta}_{ML,\psi_{\delta_{s}}}$} & 1e-3 & 0.02 & 1e-3 & 1e-3 & 0.84 & \tabularnewline
 & {\footnotesize{}$\widehat{\theta}_{ML}$} & 0.02 & -0.09 & 8e-3 & 0.37 & 0.85 & \tabularnewline
 & {\footnotesize{}$\widehat{\theta}_{\psi_{\delta_{s}}}$} & \textbf{1e-3} & \textbf{-0.02} & \textbf{7e-4} & \textbf{3e-3} & \textbf{0.91} & \tabularnewline
 & {\footnotesize{}$\widehat{\theta}$} & \textbf{0.01} & \textbf{-0.01} & \textbf{0.01} & \textbf{2e-3} & \textbf{0.89} & \tabularnewline
\hline 
\multirow{4}{*}{{\footnotesize{}$\psi_{\phi_{L}}$}} & {\footnotesize{}$\widehat{\theta}_{ML,\psi_{\delta_{s}}}$} & 7e-3 & 0.02 & 7e-3 & 5e-3 & 0.87 & \tabularnewline
 & {\footnotesize{}$\widehat{\theta}_{ML}$} & 0.01 & 0.04 & 0.01 & 6e-3 & 0.86 & \tabularnewline
 & {\footnotesize{}$\widehat{\theta}_{\psi_{\delta_{s}}}$} & \textbf{7e-3} & \textbf{7e-3} & \textbf{7e-3} & \textbf{9e-3} & \textbf{0.95} & \tabularnewline
 & {\footnotesize{}$\widehat{\theta}$} & \textbf{0.06} & \textbf{-0.03} & \textbf{0.06} & \textbf{4e-3} & \textbf{0.91} & \tabularnewline
\hline 
\multirow{4}{*}{{\footnotesize{}$\psi_{\delta_{Ab}}$}} & {\footnotesize{}$\widehat{\theta}_{ML,\psi_{\delta_{s}}}$} & 2e-3 & -0.01 & 1e-3 & 1e-3 & 0.77 & \tabularnewline
 & {\footnotesize{}$\widehat{\theta}_{ML}$} & 4e-3 & -0.03 & 4e-3 & 1e-3 & 0.85 & \tabularnewline
 & {\footnotesize{}$\widehat{\theta}_{\psi_{\delta_{s}}}$} & \textbf{6e-4} & \textbf{7e-3} & \textbf{5e-4} & \textbf{1e-3} & \textbf{0.94} & \tabularnewline
 & {\footnotesize{}$\widehat{\theta}$} & \textbf{8e-4} & \textbf{0.01} & \textbf{7e-4} & \textbf{6e-4} & \textbf{0.90} & \tabularnewline
\hline 
\multirow{4}{*}{{\footnotesize{}$\Psi_{\phi_{S}}$}} & {\footnotesize{}$\widehat{\theta}_{ML,\psi_{\delta_{s}}}$} & 0.09 & 0.05 & 0.09 & 2.04 & 1 & 0.21\tabularnewline
 & {\footnotesize{}$\widehat{\theta}_{ML}$} & 0.15 & 0.11 & 0.15 & 1.22 & 1 & 0.25\tabularnewline
 & {\footnotesize{}$\widehat{\theta}_{\psi_{\delta_{s}}}$} & \textbf{0.03} & \textbf{-0.04} & \textbf{0.03} & \textbf{0.03} & \textbf{0.88} & \textbf{0.11}\tabularnewline
 & {\footnotesize{}$\widehat{\theta}$} & \textbf{0.05} & \textbf{-0.06} & \textbf{0.04} & \textbf{0.03} & \textbf{0.85} & \textbf{0.15}\tabularnewline
\hline 
\multirow{4}{*}{{\footnotesize{}$\Psi_{\phi_{L}}$}} & {\footnotesize{}$\widehat{\theta}_{ML,\psi_{\delta_{s}}}$} & 0.08 & -0.20 & 0.04 & 0.10 & 1 & 0.95\tabularnewline
 & {\footnotesize{}$\widehat{\theta}_{ML}$} & 0.10 & 0.08 & 0.08 & 0.14 & 1 & 0.96\tabularnewline
 & {\footnotesize{}$\widehat{\theta}_{\psi_{\delta_{s}}}$} & \textbf{0.04} & \textbf{-0.11} & \textbf{0.03} & \textbf{0.03} & \textbf{0.88} & \textbf{0.12}\tabularnewline
 & {\footnotesize{}$\widehat{\theta}$} & \textbf{0.06} & \textbf{-0.16} & \textbf{0.03} & \textbf{0.03} & \textbf{0.90} & \textbf{0.15}\tabularnewline
\hline 
\multirow{4}{*}{{\footnotesize{}$\Psi_{\delta_{Ab}}$}} & {\footnotesize{}$\widehat{\theta}_{ML,\psi_{\delta_{s}}}$} & 0.27 & 0.27 & 0.19 & 0.09 & 1 & 0.46\tabularnewline
 & {\footnotesize{}$\widehat{\theta}_{ML}$} & 0.51 & 0.64 & 0.28 & 0.14 & 1 & 0.74\tabularnewline
 & {\footnotesize{}$\widehat{\theta}_{\psi_{\delta_{s}}}$} & \textbf{0.11} & \textbf{-0.21} & \textbf{0.07} & \textbf{0.08} & \textbf{0.88} & \textbf{0.05}\tabularnewline
 & {\footnotesize{}$\widehat{\theta}$} & \textbf{0.16} & \textbf{-0.31} & \textbf{0.07} & \textbf{0.08} & \textbf{0.88} & \textbf{0.05}\tabularnewline
\end{tabular}{\footnotesize\par}

{\footnotesize\par}
\end{table}

\section*{Appendix C:\label{sec:Appendix-C:Variance_Covariance_estimator}
Formal details for Variance-Covariance estimator}

Before presenting the proof, let us introduce notations used in the
following lemma:
\begin{enumerate}
\item The operator $\otimes$ defined by the relationship $\left(\frac{d^{2}}{d\gamma_{1}d\gamma_{2}}Q\right)^{T}\otimes R=\sum_{i}R_{i}\frac{d^{2}}{d\gamma_{1}d\gamma_{2}}Q_{i}$
for two same dimensional vectors $Q$ and $R$,
\item $\overline{\partial_{\alpha_{1},\ldots\alpha_{l}}f}=\sup_{\left(\theta,b_{i},t,x\right)}\left\Vert \frac{\partial^{l}f_{\theta,b_{i}}}{\partial\alpha_{1},\ldots\partial\alpha_{l}}(t,x)\right\Vert _{2}$
for $\left(\theta,b_{i},t,x\right)\in\Theta\times\Theta_{b}\times\left[0,\,T\right]\times\chi.$
\end{enumerate}
\begin{theorem}
Under conditions 1-7, there is a model dependent lower bound $\lambda$
such that if $\left\Vert U\right\Vert _{2}>\lambda$ then the estimator
$\left(\widehat{\theta},\widehat{\delta}\right)$ is asymptotically
normal and:
\[
\sqrt{n}(\widehat{\theta}-\overline{\theta},\widehat{\delta}-\overline{\delta})\rightsquigarrow N\left(0,A(\overline{\theta},\overline{\delta})^{-1}B(\overline{\theta},\overline{\delta})\left(A(\overline{\theta},\overline{\delta})^{-1}\right)^{T}\right)
\]
where $A(\overline{\theta},\overline{\delta})=\lim_{n}\frac{1}{n}\sum_{i=1}^{n}\left[\frac{\partial\widetilde{J}(\overline{\theta},\overline{\delta},\mathbf{y_{i}})}{\partial(\theta,\delta)}\right]$,
$B(\overline{\theta},\overline{\delta})=\lim_{n}\frac{1}{n}\left[\sum_{i}\widetilde{J}(\overline{\theta},\overline{\delta},\mathbf{y_{i}})\widetilde{J}(\overline{\theta},\overline{\delta},\mathbf{y_{i}})^{T}\right]$
and the vector valued function $\widetilde{J}(\theta,\delta,\mathbf{y_{i}})=\left(\begin{array}{l}
\widetilde{J}_{\theta}(\theta,\delta,\mathbf{y_{i}})\\
\widetilde{J}_{\delta}(\theta,\delta,\mathbf{y_{i}})
\end{array}\right)$ is given by:
\[
\begin{array}{l}
\widetilde{J}_{\theta}(\theta,\delta,\mathbf{y_{i}})=\frac{d}{d\theta}h(\widehat{b}(\theta,\Delta(\delta)),\theta,\Delta(\delta),y_{i})\\
\widetilde{J}_{\delta}(\theta,\delta,\mathbf{y_{i}})=\frac{d}{d\delta}h(\widehat{b}_{i}(\theta,\Delta(\delta)),\theta,\Delta(\delta),y_{i})-\frac{2}{d^{o}\mathbb{E}\left[n_{1}\right]+q}Tr\left(\triangle(\delta)^{-1}\frac{\partial\triangle(\delta)}{\partial\delta_{k}}\right)h(\widehat{b}_{i}(\theta,\Delta(\delta)),\theta,\Delta(\delta),y_{i}).
\end{array}
\]
\end{theorem}
\begin{proof}
From lemma \ref{lem:h_uniform_boundedness}, we can use the law of
large number presented theorem 5.1.2 in Wang (1968) for the independent,
possibly non identically distributed, random variables $h(\widehat{b}(\theta,\Delta(\delta)),\theta,\Delta(\delta),y_{i})$
to derive the almost surely uniform convergence of 
\[
\frac{1}{n}\sum_{i}^{n}h(\widehat{b}(\theta,\Delta(\delta)),\theta,\Delta(\delta),y_{i})\longrightarrow\lim_{n}\frac{1}{n}\sum_{i}^{n}\mathbb{E}\left[h(\widehat{b}(\theta,\Delta(\delta)),\theta,\Delta(\delta),y_{i})\right]
\]
 on $\Theta\times\Omega$. This is in turn sufficient to derive the
same result for $\frac{1}{n}G\left[\theta,\Delta(\delta)\mid\mathbf{y}\right]$
toward $\widetilde{G}\left[\theta,\Delta(\delta)\right].$ Now, by
relying on condition 1 regarding the well-separateness property of
$\left(\overline{\theta},\overline{\delta}\right)$, we can apply
theorem 5.3 in \cite{Vaart1998} to conclude that $\left(\widehat{\theta},\widehat{\delta}\right)\longrightarrow\left(\overline{\theta},\overline{\delta}\right)$
in probability when $n\longrightarrow\infty$.

Since $\left(\widehat{\theta},\widehat{\delta}\right)$ is defined
as an M-estimator, we can obtain an asymptotic Variance-Covariance
estimator by using a Taylor expansion of equation $\nabla_{\left(\theta,\delta\right)}G\left[\widehat{\theta},\Delta(\widehat{\delta})\mid\mathbf{y}\right]=0.$
By developing the previous gradient expression, we have $\sum_{i=1}^{n}J(\widehat{\theta},\widehat{\delta},\mathbf{y_{i}})=M_{n}(\widehat{\theta},\widehat{\delta})$
where 
\[
M_{n}(\theta,\delta)=\frac{2}{\mathbb{P}(\theta,\Delta(\delta))}\frac{\partial\mathbb{P}(\theta,\Delta(\delta))}{\partial\left(\theta,\delta\right)}\frac{\sum_{i}h(\widehat{b}_{i}(\theta,\Delta(\delta)),\theta,\Delta(\delta),\mathbf{y_{i}})}{d^{o}\sum_{i}n_{i}+qn}
\]
 and the $(p+q')$ components of the vector valued function $J$ for
$1\leq k\leq p$ are given by
\[
J_{k}(\theta,\delta,y_{i})=\frac{d}{d\theta_{k}}h(\widehat{b}_{i}(\theta,\Delta(\delta)),\theta,\Delta(\delta),y_{i})
\]
and for $p+1\leq k\leq p+q'$ by
\[
J_{k}(\theta,\delta,y_{i})=\frac{d}{d\delta_{k}}h(\widehat{b}_{i}(\theta,\Delta(\delta)),\theta,\Delta(\delta),y_{i})-\frac{2}{d^{o}\frac{\sum_{i}n_{i}}{n}+q}Tr\left(\triangle(\delta)^{-1}\frac{\partial\triangle(\delta)}{\partial\delta_{k}}\right)h(\widehat{b}_{i}(\theta,\Delta(\delta)),\theta,\Delta(\delta),y_{i}).
\]
In order to use asymptotic results, we want to consider quantities
independents of $n$ and $n_{i}$, the $n_{i}$ being i.i.d by using
the central limit theorem we get $\sqrt{n}\left(\frac{\sum_{i}n_{i}}{n}-\mathbb{E}\left[n_{1}\right]\right)\rightsquigarrow0$,
the last convergence in distribution is also a convergence in probability
which can be re-expressed as $\frac{\sum_{i}n_{i}}{n}=\mathbb{E}\left[n_{1}\right]+\sqrt{n}^{-1}o_{p,n}(1)$.
From this, we can introduce $\widetilde{J}$ given by $\widetilde{J_{k}(}\theta,\delta,y_{i})=J_{k}(\theta,\delta,y_{i})$
for $1\leq k\leq q'$ and for $p+1\leq k\leq p+q'$ by
\[
\widetilde{J_{k}}(\theta,\delta,y_{i})=\frac{d}{d\delta_{k}}h(\widehat{b}_{i}(\theta,\Delta(\delta)),\theta,\Delta(\delta),y_{i})-\frac{2}{d^{o}\mathbb{E}\left[n_{1}\right]+q}Tr\left(\triangle(\delta)^{-1}\frac{\partial\triangle(\delta)}{\partial\delta_{k}}\right)h(\widehat{b}_{i}(\theta,\Delta(\delta)),\theta,\Delta(\delta),y_{i})
\]
which respects the approximation $J_{k}(\theta,\delta,y_{i})=\widetilde{J_{k}}(\theta,\delta,y_{i})+\sqrt{n}^{-1}o_{p,n}(1)$
and 
\[
\sum_{i=1}^{n}\widetilde{J}(\widehat{\theta},\widehat{\delta},\mathbf{y_{i}})=\widetilde{M_{n}}(\widehat{\theta},\widehat{\delta})+\sqrt{n}o_{p,n}(1)
\]
with $\widetilde{M_{n}}(\theta,\delta)=\frac{2}{\mathbb{P}(\theta,\Delta(\delta))}\frac{\partial\mathbb{P}(\theta,\Delta(\delta))}{\partial\left(\theta,\delta\right)}\frac{1}{n}\frac{\sum_{i}h(\widehat{b}_{i}(\theta,\Delta(\delta)),\theta,\Delta(\delta),\mathbf{y_{i}})}{d^{o}\mathbb{E}\left[n_{1}\right]+q}.$
Now we want to proceed to the Taylor expansion of $K_{n}(\widehat{\theta},\widehat{\delta})=n^{-1}\sum_{i=1}^{n}\widetilde{J}(\widehat{\theta},\widehat{\delta},\mathbf{y_{i}})$
around $\left(\overline{\theta},\overline{\delta}\right)$, but first
we need to ensure $K_{n}\in C^{2}(\Theta_{\overline{\theta}}^{(1)}\times\Omega_{\overline{\delta}}^{(1)},\mathbb{R}^{(p+q')})$
for a neighborhood $\Theta_{\overline{\theta}}^{(1)}\times\Omega_{\overline{\delta}}^{(1)}$
of $\left(\overline{\theta},\overline{\delta}\right)$ almost surely
for every sequence $\mathbf{y_{i}}$. From the definition of $\widetilde{J}$,
it is obvious this holds if $\left(\theta,\delta\right)\longmapsto h(\widehat{b}_{i}(\theta,\Delta(\delta)),\theta,\Delta(\delta),\mathbf{y_{i}})\in C^{3}(\Theta_{\overline{\theta}}^{(1)}\times\Omega_{\overline{\delta}}^{(1)},\mathbb{R}).$
From the conditions 6 and 7 on $f_{\theta,b_{i}}$and $\mathcal{C}_{i}$
invertibility and part 2) of lemma \ref{lem:X_u_bar_continuity_boundedness},
we derive that it exists a neighborhood of $\widetilde{\Theta}_{\overline{\theta}}\times\widetilde{\Theta}_{\widehat{b}_{i}(\overline{\theta},\Delta(\overline{\delta}))}\subset\Theta_{\overline{\theta}}\times\mathbb{R}^{q}$
such that $\left(\theta,b_{i},t\right)\longmapsto\overline{X}_{\theta,b_{i}}(t)\in C^{3}(\widetilde{\Theta}_{\overline{\theta}}\times\widetilde{\Theta}_{\widehat{b}_{i}(\overline{\theta},\Delta(\overline{\delta}))}\times\left[0,\,T\right],\mathbb{R}^{d})$
and $\left(\theta,b_{i},t\right)\longmapsto\overline{u}_{\theta,b_{i}}(t)\in C^{3}(\widetilde{\Theta}_{\overline{\theta}}\times\widetilde{\Theta}_{\widehat{b}_{i}(\overline{\theta},\Delta(\overline{\delta}))}\times\left[0,\,T\right],\mathbb{R}^{d_{u}})$.
From this, it is easy to see that $\left(\theta,b_{i},\delta\right)\longmapsto h(b_{i},\theta,\Delta(\delta),\mathbf{y_{i}})$
and $\left(\theta,b_{i},\delta\right)\longmapsto g_{i}(b_{i}\mid\theta,\Delta(\delta),U)$
both belong to $C^{3}(\widetilde{\Theta}_{\overline{\theta}}\times\widetilde{\Theta}_{\widehat{b}_{i}(\overline{\theta},\Delta(\overline{\delta}))}\times\Omega,\mathbb{R})$
almost surely for every sequence $\mathbf{y_{i}}$. Since $\nabla_{b_{i}}g_{i}(\widehat{b_{i}}\left(\theta,\Delta(\delta)\right)\mid\theta,\Delta(\delta),U)=0$
and we assume$\frac{\partial^{2}}{\partial^{2}b_{i}}g_{i}(\widehat{b_{i}}\left(\overline{\theta},\Delta(\overline{\delta})\right)\mid\overline{\theta},\Delta(\overline{\delta}),U)$
is of full rank (condition 6), we use the implicit function theorem
to conclude about the existence of an open subset $\Theta_{\overline{\theta}}^{(1)}\times\Omega_{\overline{\delta}}^{(1)}\subset\widetilde{\Theta}_{\overline{\theta}}\times\Omega$
such that $\left(\theta,\delta\right)\longmapsto\widehat{b_{i}}\left(\theta,\delta\right)\in C^{3}(\Theta_{\overline{\theta}}^{(1)}\times\Omega_{\overline{\delta}}^{(1)},\widetilde{\Theta}_{\widehat{b}_{i}(\overline{\theta},\Delta(\overline{\delta}))}).$
From this, classic results on the regularity of composed functions
gives us $\left(\theta,\delta\right)\longrightarrow h(\widehat{b}_{i}(\theta,\Delta(\delta)),\theta,\Delta(\delta),\mathbf{y_{i}})\in C^{3}(\Theta_{\overline{\theta}}^{(1)}\times\Omega_{\overline{\delta}}^{(1)},\mathbb{R}^{p+q})$.
Since $\left(\widehat{\theta},\widehat{\delta}\right)\longrightarrow\left(\overline{\theta},\overline{\delta}\right)$,
the event $\left(\widehat{\theta},\widehat{\delta}\right)\in\Theta_{\overline{\theta}}^{(1)}\times\Omega_{\overline{\delta}}^{(1)}$
has a probability tending to one and we can use the Taylor expansion
of $K_{n}(\widehat{\theta},\widehat{\delta})$ around $\left(\overline{\theta},\overline{\delta}\right)$
on $\Theta_{\overline{\theta}}^{(1)}\times\Omega_{\overline{\delta}}^{(1)}$
which gives us:
\[
\frac{1}{n}\widetilde{M_{n}}(\widehat{\theta},\widehat{\delta})+\sqrt{n}^{-1}o_{p,n}(1)=K_{n}(\overline{\theta},\overline{\delta})+((\widehat{\theta},\widehat{\delta})-(\overline{\theta},\overline{\delta}))^{T}\left(\frac{\partial K_{n}(\overline{\theta},\overline{\delta})}{\partial(\theta,\delta)}+\frac{\partial^{2}K_{n}(\widetilde{\theta}_{n},\widetilde{\delta_{n}})}{\partial^{2}(\theta,\delta)}((\widehat{\theta},\widehat{\delta})-(\overline{\theta},\overline{\delta}))\right)
\]
with $\left(\widetilde{\theta}_{n},\widetilde{\delta_{n}}\right)$
defined such that $\left\Vert (\widetilde{\theta}_{n},\widetilde{\delta_{n}})-(\overline{\theta},\overline{\delta})\right\Vert \leq\left\Vert (\widehat{\theta},\widehat{\delta})-(\overline{\theta},\overline{\delta})\right\Vert $.
By rearranging, we get: 
\begin{equation}
\sqrt{n}\left(\frac{\partial K_{n}(\overline{\theta},\overline{\delta})}{\partial(\theta,\delta)}^{T}+((\widehat{\theta},\widehat{\delta})-(\overline{\theta},\overline{\delta}))^{T}\frac{\partial^{2}K_{n}(\widetilde{\theta}_{n},\widetilde{\delta_{n}})}{\partial^{2}(\theta,\delta)}\right)((\widehat{\theta},\widehat{\delta})-(\overline{\theta},\overline{\delta}))=-\sqrt{n}K_{n}(\overline{\theta},\overline{\delta})+\frac{\widetilde{M_{n}}(\widehat{\theta},\widehat{\delta})}{\sqrt{n}}+o_{p,n}(1).\label{eq:Taylor_exp_rewritting}
\end{equation}
As before, by using the weak law of large numbers, we derive: 
\[
\frac{\sum_{i}h(\widehat{b}_{i}(\theta,\Delta(\delta)),\theta,\Delta(\delta),\mathbf{y_{i}})}{d^{o}\sum_{i}n_{i}+qn}\longrightarrow\left(\frac{1}{q+d^{o}\mathbb{E}\left[n_{1}\right]}\right)\lim_{n}\frac{1}{n}\sum_{i}^{n}\mathbb{E}\left[h(\widehat{b}(\theta,\Delta(\delta)),\theta,\Delta(\delta),y_{i})\right]
\]
 in probability. From this and the continuity mapping theorem it is
easy to see that 
\[
\widetilde{M_{n}}(\widehat{\theta},\widehat{\delta})\longrightarrow\frac{2}{\mathbb{P}(\overline{\theta},\Delta(\overline{\delta}))}\frac{\partial\mathbb{P}(\overline{\theta},\Delta(\overline{\delta}))}{\partial\left(\theta,\delta\right)}\left(\frac{1}{q+d^{o}\mathbb{E}\left[n_{1}\right]}\right)\mathbb{E}_{Y}\left[h(\widehat{b}(\overline{\theta},\Delta(\overline{\delta})),\overline{\theta},\Delta(\overline{\delta}),Y)\right]
\]
in probability thus $\widetilde{M_{n}}(\widehat{\theta},\widehat{\delta})$
is an uniformly tight sequence and so $\frac{1}{\sqrt{n}}\widetilde{M_{n}}(\widehat{\theta},\widehat{\delta})=o_{p,n}(1)$,
the continuity of $\mathbb{E}_{Y}\left[h(\widehat{b}(\overline{\theta},\Delta(\overline{\delta})),\overline{\theta},\Delta(\overline{\delta}),Y)\right]$
being ensured by the uniform convergence. From these results, we can
use the approximation $-\sqrt{n}K_{n}(\overline{\theta},\overline{\delta})+\frac{1}{\sqrt{n}}M_{n}(\widehat{\theta},\widehat{\delta})+o_{p,n}(1)=-\sqrt{n}K_{n}(\overline{\theta},\overline{\delta})+o_{p,n}(1)$
in the right-hand side of equation (\ref{eq:Taylor_exp_rewritting}).
Based on condition 2, if the initial conditions $\left\{ x_{0,i}^{*}\right\} _{i\in\left\llbracket 1,n\right\rrbracket }\in\left\llbracket 1,\,n\right\rrbracket $
are i.i.d so are the $\widetilde{J}(\overline{\theta},\overline{\delta},\mathbf{y_{i}})$
and we derive $-\sqrt{n}K_{n}(\overline{\theta},\overline{\delta})\rightsquigarrow N(0,B(\overline{\theta},\overline{\delta}))$
with $B(\overline{\theta},\overline{\delta})=\lim_{n}\frac{1}{n}\left[\sum_{i}\widetilde{J}(\overline{\theta},\overline{\delta},\mathbf{y_{i}})\widetilde{J}(\overline{\theta},\overline{\delta},\mathbf{y_{i}})^{T}\right]$
according to the central limit theorem, otherwise with the Lindeberg-Feller
theorem, bound conditions for $\frac{d}{d(\theta,\delta)}h(\widehat{b}(\overline{\theta},\Delta(\overline{\delta})),\overline{\theta},\Delta(\overline{\delta}),\mathbf{y_{i}})$
are proven in lemma \ref{lem:dh_d2h_derivative_boundedness}. Now,
regarding the left-hand side of (\ref{eq:Taylor_exp_rewritting}),
by using the convergence of $\left(\widetilde{\theta}_{n},\widetilde{\delta_{n}}\right)$
toward $(\overline{\theta},\overline{\delta})$ we know the event
$\left(\widetilde{\theta}_{n},\widetilde{\delta_{n}}\right)\in\Theta_{\overline{\theta}}^{(1)}\times\Omega_{\overline{\delta}}^{(1)}$
has a probability tending to one. From this, we derive the matrix
$\frac{\partial^{2}K_{n}(\widetilde{\theta}_{n},\widetilde{\delta_{n}})}{\partial^{2}(\theta,\delta)}$,
involving third derivatives of $h$ which are continuous on $\Theta_{\overline{\theta}}^{(1)}\times\Omega_{\overline{\delta}}^{(1)}$,
is bounded in probability, thus $((\widehat{\theta},\widehat{\delta})-(\overline{\theta},\overline{\delta}))^{T}\frac{\partial^{2}K_{n}(\widetilde{\theta}_{n},\widetilde{\delta_{n}})}{\partial^{2}(\theta,\delta)}=o_{p,n}(1)$.
From this, equation (\ref{eq:Taylor_exp_rewritting}) can be re-expressed
as: 
\[
\sqrt{n}\left(\frac{\partial K_{n}(\overline{\theta},\overline{\delta})}{\partial(\theta,\delta)}^{T}+o_{p,n}(1)\right)((\widehat{\theta},\widehat{\delta})-(\overline{\theta},\overline{\delta}))=-\sqrt{n}K_{n}(\overline{\theta},\overline{\delta})+o_{p,n}(1)
\]
and we can use the weak law of large numbers to get the convergence
in probability of $\frac{\partial K_{n}(\overline{\theta},\overline{\delta})}{\partial(\theta,\delta)}=n^{-1}\sum_{i=1}^{n}\left[\frac{\partial\widetilde{J}(\overline{\theta},\overline{\delta},\mathbf{y_{i}})}{\partial(\theta,\delta)}\right]\rightarrow\lim_{n}\frac{1}{n}\sum_{i=1}^{n}\left[\frac{\partial\widetilde{J}(\overline{\theta},\overline{\delta},\mathbf{y_{i}})}{\partial(\theta,\delta)}\right]=A(\overline{\theta},\overline{\delta})$
with moment boundedness conditions on $\frac{d}{d(\theta,\delta)}h(\widehat{b}(\overline{\theta},\Delta(\overline{\delta})),\overline{\theta},\Delta(\overline{\delta}),\mathbf{y_{i}})$
proven in lemma \ref{lem:dh_d2h_derivative_boundedness}. If the matrix
$A(\overline{\theta},\overline{\delta})$ is non-singular, we can
then use the multivariate version of Slutsky's lemma to conclude that:
\[
\sqrt{n}(\widehat{\theta},\widehat{\delta})-(\overline{\theta},\overline{\delta})\rightsquigarrow N\left(0,A(\overline{\theta},\overline{\delta})^{-1}B(\overline{\theta},\overline{\delta})\left(A(\overline{\theta},\overline{\delta})^{-1}\right)^{T}\right).
\]
\end{proof}
\begin{lem}
\label{lem:h_uniform_boundedness}Under conditions 3,4,5, $(\theta,\delta)\longmapsto h(\widehat{b}_{i}(\theta,\Delta(\delta)),\theta,\Delta(\delta),\mathbf{y_{i}})$
is uniformly bounded on $\Theta\times\Omega$ and has a finite moment
of order two uniformly bounded on $i\in\left\llbracket 1,\,n\right\rrbracket $
\end{lem}

\begin{proof}
By definition of our optimal control problem, we derive the following
bound by sequentially using $\left\Vert \overline{u}_{\theta,b_{i}}\right\Vert _{U,L^{2}}^{2}>0$,
upper-bounding the minimum in $b_{i}$ by the true subject specific
parameter values $b_{i}^{*}$, upper-bounding the minimum in $x_{0,i}^{u}$
by the true initial condition value $x_{0,i}^{*}$ and upper-bounding
in $u_{i}$ by $u_{i}^{0}=0$:
\[
\begin{array}{l}
h(\widehat{b}_{i}(\theta,\Delta(\delta)),\theta,\Delta(\delta),\mathbf{y_{i}})=\left\Vert \Delta(\delta)\widehat{b}_{i}(\theta,\Delta(\delta))\right\Vert _{2}^{2}+\sum_{j}\left\Vert C\overline{X}_{\theta,\widehat{b}_{i}(\theta,\Delta(\delta))}(t_{ij})-y_{ij}\right\Vert _{2}^{2}\\
=\left\Vert \Delta(\delta)\widehat{b}_{i}(\theta,\Delta(\delta))\right\Vert _{2}^{2}+\sum_{j}\left\Vert C\overline{X}_{\theta,\widehat{b}_{i}(\theta,\Delta(\delta))}(t_{ij})-y_{ij}\right\Vert _{2}^{2}\\
\leq\left\Vert \Delta(\delta)\widehat{b}_{i}(\theta,\Delta(\delta))\right\Vert _{2}^{2}+\sum_{j}\left\Vert C\overline{X}_{\theta,\widehat{b}_{i}(\theta,\Delta(\delta))}(t_{ij})-y_{ij}\right\Vert _{2}^{2}+\left\Vert \overline{u}_{\theta,\widehat{b}_{i}(\theta,\Delta(\delta))}\right\Vert _{U,L^{2}}^{2}\\
=g_{i}(\widehat{b}_{i}(\theta,\Delta(\delta))\mid\theta,\Delta(\delta),U)\\
=\min_{b_{i}}\left\{ \left\Vert \Delta(\delta)b_{i}\right\Vert _{2}^{2}+\min_{x_{0,i}^{u}}\min_{u_{i}}\left\{ \sum_{j}\left\Vert CX_{\theta,b_{i},x_{0,i},u_{i}}(t_{ij})-y_{ij}\right\Vert _{2}^{2}+\left\Vert u_{i}\right\Vert _{U,L^{2}}^{2}\right\} \right\} \\
\leq\left\Vert \Delta(\delta)b_{i}^{*}\right\Vert _{2}^{2}+\min_{x_{0,i}^{u}}\min_{u_{i}}\left\{ \sum_{j}\left\Vert CX_{\theta,b_{i}^{*},x_{0,i},u_{i}}(t_{ij})-y_{ij}\right\Vert _{2}^{2}+\left\Vert u_{i}\right\Vert _{U,L^{2}}^{2}\right\} \\
\leq\left\Vert \Delta(\delta)b_{i}^{*}\right\Vert _{2}^{2}+\min_{u_{i}}\left\{ \sum_{j}\left\Vert CX_{\theta,b_{i}^{*},x_{0,i}^{*},u_{i}}(t_{ij})-y_{ij}\right\Vert _{2}^{2}+\left\Vert u_{i}\right\Vert _{U,L^{2}}^{2}\right\} \\
\leq\left\Vert \Delta(\delta)b_{i}^{*}\right\Vert _{2}^{2}+\sum_{j}\left\Vert CX_{\theta,b_{i}^{*},x_{0,i}^{*}}(t_{ij})-y_{ij}\right\Vert _{2}^{2}
\end{array}
\]
where $X_{\theta,b_{i},x_{0,i}^{*}}$ is the solution of the original ODE.
Since $y_{ij}=CX_{\theta,b_{i}^{*},x_{0,i}^{*}}(t_{ij})+\epsilon_{i,j}$,
we get: 
\[
h(\widehat{b}_{i}(\theta,\Delta(\delta)),\theta,\Delta(\delta),\mathbf{y_{i}})\leq\left\Vert \Delta(\delta)b_{i}^{*}\right\Vert _{2}^{2}+\left\Vert C\right\Vert _{2}^{2}\sum_{j}\left\Vert X_{\theta,b_{i}^{*},x_{0,i}^{*}}(t_{ij})-X_{\theta^{*},b_{i}^{*},x_{0,i}^{*}}(t_{ij})\right\Vert _{2}^{2}+\sum_{j}\left\Vert \epsilon_{i,j}\right\Vert _{2}^{2}.
\]
Now, let us control the behavior of: 
\[
\begin{array}{lll}
X_{\theta,b_{i}^{*},x_{0,i}^{*}}(t)-X_{\theta^{*},b_{i}^{*},x_{0,i}^{*}}(t) & = & \int_{0}^{t}(f_{\theta,b_{i}^{*}}(t,X_{\theta,b_{i}^{*},x_{0,i}^{*}}(t))-f_{\theta^{*},b_{i}^{*}}(t,X_{\theta,b_{i}^{*},x_{0,i}^{*}}(t))dt\\
 & + & \int_{0}^{t}(f_{\theta^{*},b_{i}^{*}}(t,X_{\theta,b_{i}^{*},x_{0,i}^{*}}(t))-f_{\theta^{*},b_{i}^{*}}(t,X_{\theta^{*},b_{i}^{*},x_{0,i}^{*}}(t))dt
\end{array}
\]
by triangular inequality, we obtain 
\[
\begin{array}{lll}
\left\Vert X_{\theta,b_{i}^{*},x_{0,i}^{*}}(t)-X_{\theta^{*},b_{i}^{*},x_{0,i}^{*}}(t)\right\Vert _{2}^{2} & \leq & 2\overline{\partial_{\theta}f}T\left\Vert \theta-\theta^{*}\right\Vert _{2}^{2}+2\overline{\partial_{x}f}\int_{0}^{t}\left\Vert X_{\theta,b_{i}^{*},x_{0,i}^{*}}(t)-X_{\theta^{*},b_{i}^{*},x_{0,i}^{*}}(t)\right\Vert _{2}^{2}dt\end{array}
\]
 and from this we can use the Gronwall lemma to derive the following
bound uniform in time: $\left\Vert X_{\theta,b_{i}^{*},x_{0,i}^{*}}(t)-X_{\theta^{*},b_{i}^{*},x_{0,i}^{*}}(t)\right\Vert _{2}^{2}\leq2\overline{\partial_{\theta}f}T\left\Vert \theta-\theta^{*}\right\Vert _{2}^{2}e^{2\overline{\partial_{x}f}T}$
and finally obtain $h(\widehat{b}_{i}(\theta,\Delta(\delta)),\theta,\Delta(\delta),\mathbf{y_{i}})\leq\left\Vert \Delta(\delta)b_{i}^{*}\right\Vert _{2}^{2}+2\left\Vert C\right\Vert _{2}^{2}T\left\Vert \theta-\theta^{*}\right\Vert _{2}^{2}n_{i}\overline{\partial_{\theta}f}e^{2\overline{\partial_{x}f}T}+\sum_{j}\left\Vert \epsilon_{i,j}\right\Vert _{2}^{2}$
since $\left(\delta,\theta\right)$ belongs to a compact, we derive
from this last inequality the uniform boundedness of $h(\widehat{b}_{i}(\theta,\Delta(\delta)),\theta,\Delta(\delta),\mathbf{y_{i}})$
on $\Theta\times\Omega$. Moreover, $\left\Vert \Delta(\delta)b_{i}^{*}\right\Vert _{2}^{2}$
and $\sum_{j}\left\Vert \epsilon_{i,j}\right\Vert _{2}^{2}$ follow
$\chi^{2}$ law and the $\left\{ n_{i}\right\} _{i\in\left\llbracket 1,\,n\right\rrbracket }$
are bounded so they both have moment of order two uniformly bounded.
By hypothesis on the compact support of $f_{\theta,b_{i}}$ with respect
to $b$, both $\overline{\partial_{\theta}f_{b_{i}^{*}}}$ and $e^{2\overline{\partial_{x}f_{b_{i}^{*}}}T}$
are bounded in probability, the same holds for $n_{i}\overline{\partial_{\theta}f_{b_{i}^{*}}}e^{2\overline{\partial_{x}f_{b_{i}^{*}}}T}$
and so the previous upper bound we obtain for $h(\widehat{b}_{i}(\theta,\Delta(\delta)),\theta,\Delta(\delta),\mathbf{y_{i}})$
has a uniformly bounded moment of order two.
\end{proof}
\begin{lem}
\label{lem:dh_d2h_derivative_boundedness}Under conditions 4,5,6,7
the second moment of $\frac{dh}{d\left(\theta,\delta\right)}(\widehat{b_{i}}\left(\overline{\theta},\Delta(\overline{\delta})\right),\overline{\theta},\Delta(\overline{\delta}),\mathbf{y_{i}})$
and $\frac{d^{2}h}{d^{2}\left(\theta,\delta\right)}(\widehat{b_{i}}\left(\overline{\theta},\Delta(\overline{\delta})\right),\overline{\theta},\Delta(\overline{\delta}),\mathbf{y_{i}})$
are uniformly bounded
\end{lem}

\begin{proof}
In the following, we denote $\eta=(\theta,\delta)$ and $\overline{\eta}=(\overline{\theta},\overline{\delta})$
and replace dependence for $\widehat{b_{i}}$ with respect to parameter
by $\widehat{b_{i}}\left(\overline{\eta}\right)$. Also, we have to
emphasize at some point of the lemma the dependence of the estimated
initial conditions with respect to the parameters $\left(\theta,b\right)$,
i.e $\widehat{x_{0}}:=\widehat{x_{0}}(\overline{\theta},\widehat{b_{i}}\left(\overline{\eta}\right))$
and that for the sake of correctness the optimal trajectory and control
should be written $\overline{X}_{\overline{\theta},\widehat{b}_{i}(\overline{\eta})}:=\overline{X}_{\overline{\theta},\widehat{b}_{i}(\overline{\eta}),\widehat{x_{0}}(\overline{\eta},\widehat{b_{i}}\left(\overline{\eta}\right))}$
and $\overline{u}_{\overline{\theta},\widehat{b}_{i}(\overline{\eta})}:=\overline{u}_{\overline{\theta},\widehat{b}_{i}(\overline{\eta}),\widehat{x_{0}}(\overline{\eta},\widehat{b_{i}}\left(\overline{\eta}\right))}.$

First of all, let us derive that $\mathbb{E}_{b_{i}}\left[\left\Vert \overline{u}_{\overline{\theta},\widehat{b}_{i}(\overline{\eta})}\right\Vert _{U,L^{2}}^{2}\right]$
and $\mathbb{E}_{b_{i}}\left[\left\Vert \overline{X}_{\overline{\theta},\widehat{b}_{i}(\overline{\eta})}(t)\right\Vert _{2}^{2}\right]$
are finite. Similarly as the bound derived for $h$, we get:
\[
\begin{array}{l}
\left\Vert \overline{u}_{\overline{\theta},\widehat{b}_{i}(\overline{\eta})}\right\Vert _{U,L^{2}}^{2}+\left\Vert \Delta(\overline{\delta})\widehat{b}_{i}(\overline{\eta})\right\Vert _{2}^{2}\\
\leq\sum_{j}\left\Vert C\overline{X}_{\overline{\theta},\widehat{b}_{i}(\overline{\eta})}(t_{ij})-y_{ij}\right\Vert _{2}^{2}+\left\Vert \overline{u}_{\overline{\theta},\widehat{b}_{i}(\overline{\eta})}\right\Vert _{U,L^{2}}^{2}+\left\Vert \Delta(\overline{\delta})\widehat{b}_{i}(\overline{\eta})\right\Vert _{2}^{2}\\
=\min_{b_{i}}\left\{ \left\Vert \Delta(\overline{\delta})b_{i}\right\Vert _{2}^{2}+\min_{x_{0,i}^{u}}\min_{u_{i}}\left\{ \sum_{j}\left\Vert CX_{\overline{\theta},b_{i},x_{0,i},u_{i}}(t_{ij})-y_{ij}\right\Vert _{2}^{2}+\left\Vert u_{i}\right\Vert _{U,L^{2}}^{2}\right\} \right\} \\
\leq\left\Vert \Delta(\overline{\delta})b_{i}^{*}\right\Vert _{2}^{2}+\sum_{j}\left\Vert CX_{\overline{\theta},b_{i}^{*},x_{0,i}^{*}}(t_{ij})-y_{ij}\right\Vert _{2}^{2}
\end{array}
\]
and as in the previous lemma, we derive from this inequality that
$\mathbb{E}\left[\left\Vert \overline{u}_{\overline{\theta},\widehat{b}_{i}(\overline{\eta})}\right\Vert _{L^{2}}^{2}\right]$
and $\mathbb{E}\left[\left\Vert \widehat{b}_{i}(\overline{\eta})\right\Vert _{2}^{2}\right]$
are finite. We bound now $\overline{X}_{\overline{\theta},\widehat{b}_{i}(\overline{\eta})}$:
\[
\begin{array}{l}
\left\Vert \overline{X}_{\overline{\theta},\widehat{b}_{i}(\overline{\eta})}(t)-X_{\overline{\theta},\widehat{b}_{i}(\overline{\eta}),\widehat{x_{0,i}}}(t)\right\Vert _{2}^{2}\\
=\left\Vert \int_{0}^{t}f_{\overline{\theta},\widehat{b}_{i}(\overline{\eta})}(s,\overline{X}_{\overline{\theta},\widehat{b}_{i}(\overline{\eta})}(s))-f_{\overline{\theta},\widehat{b}_{i}(\overline{\eta})}(s,X_{\overline{\theta},\widehat{b}_{i}(\overline{\eta}),\widehat{x_{0,i}}}(s))ds+\int_{0}^{t}\overline{u}_{\overline{\theta},\widehat{b}_{i}(\overline{\eta})}(s)ds\right\Vert \\
\leq2\int_{0}^{t}\left\Vert f_{\overline{\theta},\widehat{b}_{i}(\overline{\eta})}(s,\overline{X}_{\overline{\theta},\widehat{b}_{i}(\overline{\eta})}(s))-f_{\overline{\theta},\widehat{b}_{i}(\overline{\eta})}(s,X_{\overline{\theta},\widehat{b}_{i}(\overline{\eta}),\widehat{x_{0,i}}}(s))\right\Vert _{2}^{2}ds+2\left\Vert \overline{u}_{\overline{\theta},\widehat{b}_{i}(\overline{\eta})}\right\Vert _{L^{2}}^{2}\\
\leq2\overline{\partial_{x}f}\int_{0}^{t}\left\Vert \overline{X}_{\overline{\theta},\widehat{b}_{i}(\overline{\eta})}(s)-X_{\overline{\theta},\widehat{b}_{i}(\overline{\eta}),\widehat{x_{0,i}}}(s)\right\Vert _{2}^{2}ds+2\left\Vert \overline{u}_{\overline{\theta},\widehat{b}_{i}(\overline{\eta})}\right\Vert _{L^{2}}^{2}
\end{array}
\]
and by using the Gronwall lemma, we derive the bound $\left\Vert \overline{X}_{\overline{\theta},\widehat{b}_{i}(\overline{\eta})}(t)-X_{\overline{\theta},\widehat{b}_{i}(\overline{\eta}),\widehat{x_{0,i}}}(t)\right\Vert _{2}^{2}\leq2\left\Vert \overline{u}_{\overline{\theta},\widehat{b}_{i}(\overline{\eta})}\right\Vert _{L^{2}}^{2}e^{2\overline{\partial_{x}f}T}.$
Now by using triangular inequality $\left\Vert \overline{X}_{\overline{\theta},\widehat{b}_{i}(\overline{\eta})}(t)\right\Vert _{2}^{2}\leq\left\Vert \overline{X}_{\overline{\theta},\widehat{b}_{i}(\overline{\eta})}(t)-X_{\overline{\theta},\widehat{b}_{i}(\overline{\eta}),\widehat{x_{0,i}}}(t)\right\Vert _{2}^{2}+\left\Vert X_{\overline{\theta},\widehat{b}_{i}(\overline{\eta}),\widehat{x_{0,i}}}(t)\right\Vert _{2}^{2}$
and condition 4 we derive $\mathbb{E}\left[\left\Vert \overline{X}_{\overline{\theta},\widehat{b}_{i}(\overline{\eta})}(t)\right\Vert _{2}^{2}\right]$
is finite. Now, we derive the expression of $\frac{dh}{d\left(\theta,\delta\right)}$,
by differentiation we got:
\[
\begin{array}{lll}
\frac{dh}{d\theta}(\widehat{b}_{i}(\overline{\eta}),\overline{\eta},\mathbf{y_{i}}) & = & \frac{d}{d\theta}\left(\left\Vert \Delta(\overline{\delta})\widehat{b}_{i}(\overline{\eta})\right\Vert _{2}^{2}+\sum_{j}\left\Vert C\overline{X}_{\overline{\theta},\widehat{b}_{i}(\overline{\eta})}(t_{ij})-y_{ij}\right\Vert _{2}^{2}\right)\\
 & = & 2\frac{\partial\widehat{b}_{i}(\overline{\eta})}{\partial\theta}^{T}\Delta(\overline{\delta})^{T}\Delta(\overline{\delta})\widehat{b}_{i}(\overline{\eta})+\sum_{j}\left(C\frac{d}{d\theta}\overline{X}_{\overline{\theta},\widehat{b}_{i}(\overline{\eta})}(t_{ij})\right)^{T}\left(C\overline{X}_{\overline{\theta},\widehat{b}_{i}(\overline{\eta})}(t_{ij})-y_{ij}\right)
\end{array}
\]
and similarly 
\[
\begin{array}{lll}
\frac{dh}{d\delta}(\widehat{b}_{i}(\overline{\eta}),\overline{\eta},\mathbf{y_{i}}) & = & \frac{d}{d\delta}\left(\left\Vert \Delta(\overline{\delta})\widehat{b}_{i}(\overline{\eta})\right\Vert _{2}^{2}+\sum_{j}\left\Vert C\overline{X}_{\overline{\theta},\widehat{b}_{i}(\overline{\eta})}(t_{ij})-y_{ij}\right\Vert _{2}^{2}\right)\\
 & = & 2\frac{\partial\Delta(\overline{\delta})\widehat{b}_{i}(\overline{\eta})}{\partial\delta}^{T}\Delta(\overline{\delta})\widehat{b}_{i}(\overline{\eta})+2\sum_{j}\left(C\frac{d}{d\delta}\overline{X}_{\overline{\theta},\widehat{b}_{i}(\overline{\eta})}(t_{ij})\right)^{T}\left(C\overline{X}_{\overline{\theta},\widehat{b}_{i}(\overline{\eta})}(t_{ij})-y_{ij}\right)
\end{array}
\]
with
\[
\begin{array}{lll}
\frac{d}{d\theta}\overline{X}_{\overline{\theta},\widehat{b}_{i}(\overline{\eta})}(t) & = & \frac{\partial}{\partial\theta}\overline{X}_{\overline{\theta},\widehat{b}_{i}(\overline{\eta})}(t)+\frac{\partial}{\partial b_{i}}\overline{X}_{\overline{\theta},\widehat{b}_{i}(\overline{\eta})}(t)\frac{\partial}{\partial\theta}\widehat{b}_{i}(\overline{\eta})\\
 & + & \frac{\partial}{\partial x_{0}}\overline{X}_{\overline{\theta},\widehat{b}_{i}(\overline{\eta})}(t)\left(\frac{\partial}{\partial\theta}\widehat{x_{0}}(\overline{\theta},\widehat{b_{i}}\left(\overline{\eta}\right))+\frac{\partial}{\partial b_{i}}\widehat{x_{0}}(\overline{\theta},\widehat{b_{i}}\left(\overline{\eta}\right))\frac{\partial\widehat{b_{i}}\left(\overline{\eta}\right)}{\partial\theta}\right)
\end{array}
\]
and $\frac{d}{d\delta}\overline{X}_{\overline{\theta},\widehat{b}_{i}(\overline{\eta})}(t)=\frac{\partial}{\partial b_{i}}\overline{X}_{\overline{\theta},\widehat{b}_{i}(\overline{\eta})}(t)\frac{\partial}{\partial\delta}\widehat{b}_{i}(\overline{\eta})+\frac{\partial}{\partial x_{0}}\overline{X}_{\overline{\theta},\widehat{b}_{i}(\overline{\eta})}(t)\frac{\partial}{\partial b_{i}}\widehat{x_{0}}(\overline{\theta},\widehat{b_{i}}\left(\overline{\eta}\right))\frac{\partial\widehat{b_{i}}\left(\overline{\eta}\right)}{\partial\delta}$.
From this, it is clear we need prove that $\mathbb{E}\left[\left\Vert \frac{\partial}{\partial\theta}\overline{X}_{\overline{\theta},\widehat{b}_{i}(\overline{\eta})}(t)\right\Vert _{2}^{2}\right]$,
$\mathbb{E}\left[\left\Vert \frac{\partial}{\partial b_{i}}\overline{X}_{\overline{\theta},\widehat{b}_{i}(\overline{\eta})}(t)\right\Vert _{2}^{2}\right]$,
$\mathbb{E}\left[\left\Vert \frac{\partial}{\partial x_{0}}\overline{X}_{\overline{\theta},\widehat{b}_{i}(\overline{\eta})}(t)\right\Vert _{2}^{2}\right]$,
$\mathbb{E}\left[\left\Vert \frac{\partial}{\partial\theta}\widehat{x_{0}}(\overline{\theta},\widehat{b_{i}}\left(\overline{\eta}\right))\right\Vert _{2}^{2}\right]$,
$\mathbb{E}\left[\left\Vert \frac{\partial}{\partial b_{i}}\widehat{x_{0}}(\overline{\theta},\widehat{b_{i}}\left(\overline{\eta}\right))\right\Vert _{2}^{2}\right]$
and $\mathbb{E}\left[\left\Vert \frac{\partial}{\partial\eta}\widehat{b}_{i}(\overline{\eta})\right\Vert _{2}^{2}\right]$
are finite to derive the same result for $\frac{dh}{d\left(\theta,\delta\right)}(\widehat{b}_{i}(\overline{\eta}),\overline{\eta},\mathbf{y_{i}})$.
For this, we firstly derive a suitable expression for $\frac{\partial}{\partial\eta}\widehat{b}_{i}(\overline{\eta})$.
Thanks to condition 6 and 7, we can use the implicit function theorem
which gives us $\frac{\partial}{\partial\eta}\widehat{b}_{i}(\overline{\eta})=-\frac{\partial^{2}}{\partial^{2}b_{i}}g_{i}(\widehat{b}_{i}(\overline{\eta})\mid\overline{\eta},U)^{-1}\frac{\partial}{\partial\eta}\nabla_{b_{i}}g_{i}(\widehat{b}_{i}(\overline{\eta})\mid\overline{\eta},U)$
with 
\[
\begin{array}{l}
\frac{1}{2}\nabla_{b_{i}}g_{i}(b_{i}\mid\eta,U)=\Delta(\overline{\delta})^{T}\Delta(\overline{\delta})b_{i}\\
+\sum_{j}\left(\frac{d}{db_{i}}C\overline{X}_{\theta,b_{i}}(t_{ij})\right)^{T}\left(C\overline{X}_{\theta,b_{i}}(t_{ij})-y_{ij}\right)+\int_{0}^{T}\left(\frac{d}{db_{i}}\overline{u}_{\theta,b_{i}}(t_{ij})\right)^{T}\overline{u}_{\theta,b_{i}}(t)dt\\
\frac{1}{2}\frac{\partial^{2}}{\partial^{2}b_{i}}g_{i}(b_{i}\mid\eta,U)=\sum_{j}\left(\frac{d^{2}}{d^{2}b_{i}}C\overline{X}_{\theta,b_{i}}(t_{ij})\right)^{T}\otimes\left(C\overline{X}_{\theta,b_{i}}(t_{ij})-y_{ij}\right)\\
+\sum_{j}\left(C\frac{d}{db_{i}}\overline{X}_{\theta,b_{i}}(t_{ij})\right)^{T}C\frac{d}{db_{i}}\overline{X}_{\theta,b_{i}}(t_{ij})\\
+\int_{0}^{T}\left(\frac{d^{2}}{d^{2}b_{i}}\overline{u}_{\theta,b_{i}}(t)\right)^{T}\otimes\overline{u}_{\theta,b_{i}}(t)dt+\int_{0}^{T}\left(\frac{d}{db_{i}}\overline{u}_{\theta,b_{i}}(t)\right)^{T}\frac{d}{db_{i}}\overline{u}_{\theta,b_{i}}(t)dt+\Delta(\overline{\delta})^{T}\Delta(\overline{\delta})\\
\frac{1}{2}\frac{\partial}{\partial\delta}\nabla_{b_{i}}g_{i}(b_{i}\mid\eta,U)=\frac{\partial}{\partial\delta}\left(\Delta(\overline{\delta})^{T}\Delta(\overline{\delta})b_{i}\right)\\
\frac{1}{2}\frac{\partial}{\partial\theta}\nabla_{b_{i}}g_{i}(b_{i}\mid\eta,U)=\sum_{j}\left(\frac{d^{2}}{d\theta db_{i}}C\overline{X}_{\theta,b_{i}}(t_{ij})\right)^{T}\otimes\left(C\overline{X}_{\theta,b_{i}}(t_{ij})-y_{ij}\right)\\
+\sum_{j}\left(C\frac{d}{db_{i}}\overline{X}_{\theta,b_{i}}(t_{ij})\right)^{T}\left(C\frac{d}{d\theta}\overline{X}_{\theta,b_{i}}(t_{ij})\right)\\
+\int_{0}^{T}\left(\left(\frac{d^{2}}{d\theta db_{i}}\overline{u}_{\overline{\theta},\widehat{b}_{i}(\overline{\eta})}(t)\right)^{T}\otimes\overline{u}_{\overline{\theta},\widehat{b}_{i}(\overline{\eta})}(t)+\left(\frac{d}{db_{i}}\overline{u}_{\theta,b_{i}}\right)^{T}\frac{d}{d\theta}\overline{u}_{\theta,b_{i}}(t)\right)d
\end{array}
\]
and $\frac{d}{db_{i}}\overline{X}_{\theta,b_{i}}(t)=\frac{\partial}{\partial b_{i}}\overline{X}_{\theta,b_{i},\widehat{x_{0}}\left(\theta,b_{i}\right)}(t)+\frac{\partial}{\partial x_{0}}\overline{X}_{\theta,b_{i},\widehat{x_{0}}\left(\theta,b_{i}\right)}(t)\frac{\partial\widehat{x_{0}}\left(\theta,b_{i}\right)}{\partial b_{i}}$,
$\frac{d}{db_{i}}\overline{u}_{\theta,b_{i}}(t)=\frac{\partial}{\partial b_{i}}\overline{u}_{\theta,b_{i},\widehat{x_{0}}\left(\theta,b_{i}\right)}(t)+\frac{\partial}{\partial x_{0}}\overline{u}_{\theta,b_{i},\widehat{x_{0}}\left(\theta,b_{i}\right)}(t)\frac{\partial\widehat{x_{0}}\left(\theta,b_{i}\right)}{\partial b_{i}}$,
$\frac{d}{d\theta}\overline{X}_{\theta,b_{i}}(t)=\frac{\partial}{\partial\theta}\overline{X}_{\theta,b_{i},\widehat{x_{0}}\left(\theta,b_{i}\right)}(t)+\frac{\partial}{\partial x_{0}}\overline{X}_{\theta,b_{i},\widehat{x_{0}}\left(\theta,b_{i}\right)}(t)\frac{\partial\widehat{x_{0}}\left(\theta,b_{i}\right)}{\partial\theta}$
and $\frac{d}{d\theta}\overline{u}_{\theta,b_{i}}(t)=\frac{\partial}{\partial\theta}\overline{u}_{\theta,b_{i},\widehat{x_{0}}\left(\theta,b_{i}\right)}(t)+\frac{\partial}{\partial x_{0}}\overline{u}_{\theta,b_{i},\widehat{x_{0}}\left(\theta,b_{i}\right)}(t)\frac{\partial\widehat{x_{0}}\left(\theta,b_{i}\right)}{\partial\theta}.$
This lead us in turn to control the behavior of the partial derivatives
$\frac{\partial}{\partial b_{i}}\overline{X}_{\overline{\theta},\widehat{b}_{i}(\overline{\eta})},\frac{\partial}{\partial\theta}\overline{X}_{\overline{\theta},\widehat{b}_{i}(\overline{\eta})}$,
$\frac{\partial^{2}}{\partial^{2}b_{i}}\overline{X}_{\overline{\theta},\widehat{b}_{i}(\overline{\eta})},\frac{\partial^{2}}{\partial\theta\partial b_{i}}\overline{X}_{\overline{\theta},\widehat{b}_{i}(\overline{\eta})}$,$\frac{\partial}{\partial b_{i}}\overline{u}_{\overline{\theta},\widehat{b}_{i}(\overline{\eta})}$,
$\frac{\partial}{\partial\theta}\overline{u}_{\overline{\theta},\widehat{b}_{i}(\overline{\eta})}$,
$\frac{\partial^{2}}{\partial^{2}b_{i}}\overline{u}_{\overline{\theta},\widehat{b}_{i}(\overline{\eta})}$,
$\frac{\partial^{2}}{\partial\theta\partial b_{i}}\overline{u}_{\overline{\theta},\widehat{b}_{i}(\overline{\eta})}$,$\frac{\partial}{\partial b_{i}}\widehat{x_{0}}\left(\overline{\theta},\widehat{b}_{i}(\overline{\eta})\right)$,
$\frac{\partial}{\partial\theta}\widehat{x_{0}}\left(\overline{\theta},\widehat{b}_{i}(\overline{\eta})\right)$,
$\frac{\partial^{2}}{\partial^{2}b_{i}}\widehat{x_{0}}\left(\overline{\theta},\widehat{b}_{i}(\overline{\eta})\right)$,
$\frac{\partial^{2}}{\partial\theta\partial b_{i}}\widehat{x_{0}}\left(\overline{\theta},\widehat{b}_{i}(\overline{\eta})\right)$,
this is done in lemma \ref{lem:X_u_bar_continuity_boundedness} part
3). From this, the conclusion holds as well for $\frac{\partial}{\partial\eta}\widehat{b}_{i}(\overline{\eta})$
composed of products and sums of random variables of finite bounded
moment of order two. The same results holds then for $\frac{d}{d\theta}\overline{X}_{\overline{\theta},\widehat{b}_{i}(\overline{\eta})}$
and $\frac{d}{d\delta}\overline{X}_{\overline{\theta},\widehat{b}_{i}(\overline{\eta})}$
and we can conclude for $\frac{dh}{d\left(\theta,\delta\right)}(\widehat{b}_{i}(\overline{\eta}),\overline{\eta},\mathbf{y_{i}}).$
The proof for $\frac{d^{2}h}{d^{2}\left(\theta,\delta\right)}(\widehat{b}_{i}(\overline{\eta}),\overline{\eta},\mathbf{y_{i}})$
follows the same steps.
\end{proof}

\begin{lem}
\label{lem:X_u_bar_continuity_boundedness}Let us assume $\left(\theta,b,t,x\right)\longmapsto f_{\theta,b}(t,x)$
is $(k+1)$ continuously differentiable and bounded on a subset $\widetilde{\Theta}\times\mathbb{R}^{q}\times\left[0,\,T\right]\times\widetilde{\chi}$.
1)If $K_{1}(U)=2T\left\Vert C^{T}C\right\Vert _{2}^{2}\left\Vert BU^{-1}B^{T}\right\Vert e^{6T\overline{\partial_{x}f}}<1$
and $K_{2}(U)=\frac{\left\Vert BU^{-1}B^{T}\right\Vert }{1-K_{1}(U)}4T\overline{\partial_{xx}f}\sup_{\left(\theta,b,x_{0}\right)}\left\Vert C^{T}\left(C\overline{X}_{\theta,b,x_{0}}^{(j)}(1)-y_{j}\right)\right\Vert _{2}^{2}e^{7T\overline{\partial_{x}f}}<1$,
then $\overline{u}_{\theta,b,x_{0}}=\arg\min_{u}\left\{ \sum_{j}\left\Vert CX_{\theta,b,x_{0}}(t_{j})-y_{j}\right\Vert _{2}^{2}+\left\Vert u\right\Vert _{U,L^{2}}^{2}\right\} $
and the related optimal trajectory $\overline{X}_{\theta,b,x_{0}}$
are $k$ continuously differentiable on $\widetilde{\Theta}\times\mathbb{R}^{q}\times\left[0,\,T\right]\times\widetilde{\chi}.$ 

2) For every $\left(\widetilde{\theta},\widetilde{b}_{i}\right)\in\widetilde{\Theta}\times\mathbb{R}^{q}$
such that $\frac{\partial^{2}\mathcal{C}_{i}}{\partial^{2}b_{i}}(\widetilde{b}_{i},\widehat{x_{i,0}}(\widetilde{\theta},\widetilde{b_{i}}),\overline{u}_{\widetilde{\theta},\widetilde{b_{i}},\widehat{x_{i,0}}(\widetilde{\theta},\widetilde{b_{i}})}\mid\widetilde{\theta},U)$
is of full rank and $\widehat{x_{i,0}}(\widetilde{\theta},\widetilde{b_{i}})\in\widetilde{\chi}$,
there is a neighborhood of $\widetilde{\Theta}_{\widetilde{\theta}}\times\widetilde{\Theta}_{\widetilde{b_{i}}}$
such that $\left(\theta,b_{i}\right)\longmapsto\widehat{x_{i,0}}\left(\theta,b_{i}\right),$
$\overline{X}_{\theta,b}$ and $\overline{u}_{\theta,b}$ are $(k-1)$
differentiable on $\widetilde{\Theta}_{\widetilde{\theta}}\times\widetilde{\Theta}_{\widetilde{b_{i}}}.$ 

3)If $K_{3}(U)=2\left\Vert C^{T}C\right\Vert _{2}^{2}\left\Vert BU^{-1}B\right\Vert _{2}^{2}e^{2T\overline{\partial_{x}f}}<1$,
$K_{4}(U)=\frac{4T\left\Vert BU^{-1}B\right\Vert _{2}^{2}e^{6T\overline{\partial_{x}f}}}{1-K_{3}(U)}\left(2\left\Vert C^{T}C\right\Vert _{2}^{2}\overline{\partial_{x}f}+\frac{\overline{\partial_{xx}f}}{\left|p^{0}\right|}\right)<1$
and $K_{5}(U)=\frac{4T}{1-K_{3}(U)}\left\Vert BU^{-1}B\right\Vert _{2}^{2}\left\Vert C^{T}C\right\Vert _{2}^{2}\overline{\partial_{x}f}e^{6T\overline{\partial_{x}f}}<1$,
then $\frac{\partial^{k}}{\partial^{k}\left(\theta,b\right)}\overline{X}_{\overline{\theta},b(\overline{\theta},\triangle(\overline{\delta}))}(t)$
, $\frac{\partial^{k}}{\partial^{k}\left(\theta,b\right)}\overline{u}_{i,\overline{\theta},b(\overline{\theta},\triangle(\overline{\delta}))}(t)$
and $\frac{\partial^{k}}{\partial^{k}\left(\theta,b\right)}\widehat{x_{i,0}}\left(\overline{\theta},b(\overline{\theta},\triangle(\overline{\delta}))\right)$
have finite moment of order two.
\end{lem}

\begin{proof}
We focus here on one subject, so we omit the dependence in $i$. Let
us introduce the $n_{i}$ state variables $X_{\theta,b,x_{0}}^{(j)}(t)=X_{\theta,b,x_{0}}(t_{j}+t(t_{j+1}-t_{j}))$
such that $X_{\theta,b,x_{0}}^{(j)}(0)=X_{\theta,b,x_{0}}(t_{j})$
and $X_{\theta,b,x_{0}}^{(j)}(1)=X_{\theta,b,x_{0}}(t_{j})$ to reformulate
our control problem as a final time one:
\[
\begin{array}{l}
\min_{u}\left\{ \sum_{j}\left\Vert CX_{\theta,b,x_{0}}^{(j)}(1)-y_{j}\right\Vert _{2}^{2}+\sum_{j}\int_{0}^{1}u^{(j)}(t)Uu^{(j)}(t)dt\right\} \\
\left\{ \begin{array}{l}
\frac{d}{dt}X_{\theta,b,x_{0}}^{(j)}=(t_{j+1}-t_{j})f_{\theta,b}(t_{j}+t(t_{j+1}-t_{j}),X_{\theta,b,x_{0}}^{(j)})+Bu^{(j)}(t)\\
X_{\theta,b,x_{0}}^{(0)}(0)=x_{0},\,X_{\theta,b,x_{0}}^{(j)}(0)=X_{\theta,b,x_{0}}^{(j-1)}(1)\\
u^{(j)}(0)=u^{(j-1)}(1).
\end{array}\right.
\end{array}
\]
Now we use the Pontryagin maximum principle \cite{pontryagin1962minimum}
to characterize the dependence of the optimal control and trajectory
with respect to $\left(\theta,b,x_{0}\right)$ and measurement noise.
We formulate the Hamiltonian of the controlled system: $H_{\theta,b}(t,x,p,p^{0},u)=\sum_{j}\left(p^{(j)}\right)^{T}\left((t_{j+1}-t_{j})f_{\theta,b}(t_{j}+t(t_{j+1}-t_{j}),x^{(j)})+Bu^{(j)}(t)\right)+p^{0}\left(u^{(j)}\right)^{T}Uu^{(j)}$
and derive the conjugate equation:
\[
\begin{array}{lll}
\dot{x}^{(j)} & = & \frac{\partial H_{\theta,b}}{\partial p^{(j)}}(t,x,p,p^{0},u)=(t_{j+1}-t_{j})f_{\theta,b}(t_{j}+t(t_{j+1}-t_{j}),x^{(j)})+Bu^{(j)}(t)\\
\dot{p}^{(j)} & = & -\frac{\partial H_{\theta,b}}{\partial x^{(j)}}(t,x,p,p^{0},u)=-(t_{j+1}-t_{j})\frac{\partial f_{\theta,b}}{\partial x}(t_{j}+t(t_{j+1}-t_{j}),x^{(j)})^{T}p^{(j)}.
\end{array}
\]
The maximum condition on the Hamiltonian $H_{\theta,b}(t,x(t),p(t),p^{0},\overline{u_{\theta,b}}(t))=\max_{u}H_{\theta,b}(t,x(t),p(t),p^{0},u)$
allows us to derive the following expression for the optimal control:
\[
\nabla_{u^{(j)}}H_{\theta,b}(t,x(t),p(t),p^{0},\overline{u}_{\theta,b,x_{0}}^{(j)}(t))=B^{T}p^{(j)}(t)+2p^{0}U\overline{u}_{\theta,b,x_{0}}^{(j)}(t)=0\iff\overline{u}_{\theta,b,x_{0}}^{(j)}(t)=-\frac{U^{-1}B^{T}}{2p^{0}}p^{(j)}(t).
\]
Now regarding the final time value for the adjoint system imposed
by the transversality conditions, we get $p^{j}(1)=2p^{0}C^{T}\left(C\overline{X}_{\theta,b,x_{0}}^{(j)}(1)-y_{j}\right).$
We end up with following boundary value problem: 
\[
\left\{ \begin{array}{l}
\frac{d}{dt}\overline{X}_{\theta,b,x_{0}}^{(j)}=(t_{j+1}-t_{j})f_{\theta,b}(t_{j}+t(t_{j+1}-t_{j}),\overline{X}_{\theta,b,x_{0}}^{(j)})-\frac{1}{2p^{0}}BU^{-1}B^{T}p_{\theta,b,x_{0}}^{(j)}(t)\\
\frac{d}{dt}p_{\theta,b,x_{0}}^{(j)}=-(t_{j+1}-t_{j})\frac{\partial f_{\theta,b}}{\partial x}(t_{j}+t(t_{j+1}-t_{j}),\overline{X}_{\theta,b,x_{0}}^{(j)})^{T}p_{\theta,b,x_{0}}^{(j)}\\
X_{\theta,b,x_{0}}^{(0)}(0)=x_{0},\,X_{\theta,b,x_{0}}^{(j)}(0)=X_{\theta,b,x_{0}}^{(j-1)}(1)\\
p_{\theta,b,x_{0}}^{(j)}(1)=2p^{0}C^{T}\left(C\overline{X}_{\theta,b,x_{0}}^{(j)}(1)-y_{j}\right)\\
\overline{u}_{\theta,b,x_{0}}^{(j)}(t)=-\frac{1}{2p^{0}}U^{-1}B^{T}p_{\theta,b,x_{0}}^{(j)}(t),\,u_{\theta,b,x_{0}}^{(j)}(0)=u_{\theta,b,x_{0}}^{(j-1)}(1).
\end{array}\right.
\]
So, for two set of parameters $\left(\theta,b,x_{0}\right)$ and $\left(\theta',b',x_{0}^{'}\right)$
we get the inequality: 

\[
\begin{array}{l}
\left\Vert \overline{X}_{\theta,b,x_{0}}^{(j)}(t)-\overline{X}_{\theta',b',x_{0}'}^{(j)}(t)\right\Vert _{2}^{2}\\
\leq8T\left(\overline{\partial_{b}f}\left\Vert b-b'\right\Vert _{2}^{2}+\overline{\partial_{\theta}f}\left\Vert \theta-\theta'\right\Vert _{2}^{2}\right)+4\overline{\partial_{x}f}T\int_{0}^{t}\left\Vert \overline{X}_{\theta,b,x_{0}}^{(j)}(s)-\overline{X}_{\theta',b',x_{0}'}^{(j)}(s)\right\Vert _{2}^{2}ds\\
+\left\Vert \frac{BU^{-1}B^{T}}{p^{0}}\right\Vert \int_{0}^{t}\left\Vert p_{\theta,b,x_{0}}^{(j)}(s)-p_{\theta',b',x_{0}'}^{(j)}(s)\right\Vert _{2}^{2}ds+\left\Vert \overline{X}_{\theta,b,x_{0}}^{(j)}(0)-\overline{X}_{\theta',b',x_{0}'}^{(j)}(0)\right\Vert _{2}^{2}\\
\left\Vert p_{\theta,b,x_{0}}^{(j)}(t)-p_{\theta',b',x_{0}'}^{(j)}(t)\right\Vert _{2}^{2}\\
\leq2\left\Vert p^{0}C^{T}C\right\Vert _{2}^{2}\left\Vert \overline{X}_{\theta,b,x_{0}}^{(j)}(1)-\overline{X}_{\theta',b',x_{0}'}^{(j)}(1)\right\Vert _{2}^{2}+2T\overline{\partial_{x}f}\int_{t}^{1}\left\Vert p_{\theta',b',x_{0}'}^{(j)}(s)-p_{\theta,b,x_{0}}^{(j)}(s)\right\Vert _{2}^{2}ds\\
+8T\left(\overline{\partial_{bx}f}\left\Vert b-b'\right\Vert _{2}^{2}+\overline{\partial_{\theta x}f}\left\Vert \theta-\theta'\right\Vert _{2}^{2}\right)\\
+4T\overline{\partial_{xx}f}\int_{t}^{1}\left\Vert p_{\theta',b',x_{0}'}^{(j)}(s)\right\Vert _{2}^{2}\left\Vert \overline{X}_{\theta,b,x_{0}}^{(j)}(s)-\overline{X}_{\theta',b',x_{0}'}^{(j)}(s)\right\Vert _{2}^{2}ds.
\end{array}
\]
which leads by using Gronwall lemma:
\[
\begin{array}{l}
\left\Vert \overline{X}_{\theta,b,x_{0}}^{(j)}(t)-\overline{X}_{\theta',b',x_{0}'}^{(j)}(t)\right\Vert _{2}^{2}\\
\leq\left\Vert \frac{BU^{-1}B^{T}}{p^{0}}\right\Vert \left\Vert p_{\theta,b,x_{0}}^{(j)}-p_{\theta',b',x_{0}'}^{(j)}\right\Vert _{L^{2}}e^{4T\overline{\partial_{x}f}}\\
+\left(8T\left(\overline{\partial_{b}f}\left\Vert b-b'\right\Vert _{2}^{2}+\overline{\partial_{\theta}f}\left\Vert \theta-\theta'\right\Vert _{2}^{2}\right)+\left\Vert \overline{X}_{\theta,b,x_{0}}^{(j)}(0)-\overline{X}_{\theta',b',x_{0}'}^{(j)}(0)\right\Vert _{2}^{2}\right)e^{4T\overline{\partial_{x}f}}
\end{array}
\]
by re-injecting in $\left\Vert p_{\theta,b,x_{0}}^{(j)}(t)-p_{\theta',b',x_{0}'}^{(j)}(t)\right\Vert _{2}^{2}$
and using Cauchy-Schwarz inequality we get:
\[
\begin{array}{lll}
\left\Vert p_{\theta,b,x_{0}}^{(j)}(t)-p_{\theta',b',x_{0}'}^{(j)}(t)\right\Vert _{2}^{2} & \leq & 2\left\Vert C^{T}C\right\Vert _{2}^{2}\left\Vert BU^{-1}B^{T}\right\Vert \left\Vert p_{\theta,b,x_{0}}^{(j)}-p_{\theta',b',x_{0}'}^{(j)}\right\Vert _{L^{2}}e^{4T\overline{\partial_{x}f}}\\
 & + & 2\left\Vert p^{0}C^{T}C\right\Vert _{2}^{2}8T\left(\overline{\partial_{b}f}\left\Vert b-b'\right\Vert _{2}^{2}+\overline{\partial_{\theta}f}\left\Vert \theta-\theta'\right\Vert _{2}^{2}\right)e^{4T\overline{\partial_{x}f}}\\
 & + & 2\left\Vert p^{0}C^{T}C\right\Vert _{2}^{2}\left\Vert \overline{X}_{\theta,b,x_{0}}^{(j)}(0)-\overline{X}_{\theta',b',x_{0}'}^{(j)}(0)\right\Vert _{2}^{2}e^{4T\overline{\partial_{x}f}}\\
 & + & 4T\overline{\partial_{xx}f}\int_{t}^{1}\left\Vert p_{\theta',b',x_{0}'}^{(j)}(s)\right\Vert _{2}^{2}ds\int_{t}^{1}\left\Vert \overline{X}_{\theta,b,x_{0}}^{(j)}(s)-\overline{X}_{\theta',b',x_{0}'}^{(j)}(s)\right\Vert _{2}^{2}ds\\
 & + & 8T\left(\overline{\partial_{bx}f}\left\Vert b-b'\right\Vert _{2}^{2}+\overline{\partial_{\theta x}f}\left\Vert \theta-\theta'\right\Vert _{2}^{2}\right)\\
 & + & 2T\overline{\partial_{x}f}\int_{t}^{1}\left\Vert p_{\theta',b',x_{0}'}^{(j)}(s)-p_{\theta,b,x_{0}}^{(j)}(s)\right\Vert _{2}^{2}ds
\end{array}
\]
in the same way, we obtain:
\[
\begin{array}{lll}
\left\Vert p_{\theta',b',x_{0}'}^{(j)}(t)\right\Vert _{2}^{2} & = & \left\Vert p_{\theta',b',x_{0}'}^{(j)}(1)+\int_{t}^{1}(t_{j+1}-t_{j})\frac{\partial f_{\theta',b'}}{\partial x}(t_{j}+s(t_{j+1}-t_{j}),\overline{X}_{\theta',b',x_{0}'}^{(j)})^{T}p_{\theta',b',x_{0}'}^{(j)}(s)ds\right\Vert _{2}^{2}\\
 & \leq & \left\Vert p^{0}C^{T}\left(C\overline{X}_{\theta',b',x_{0}'}^{(j)}(1)-y_{j}\right)\right\Vert _{2}^{2}+T\overline{\partial_{x}f}\int_{t}^{1}\left\Vert p_{\theta',b',x_{0}'}^{(j)}(s)\right\Vert _{2}^{2}ds
\end{array}
\]
and $\left\Vert p_{\theta',b',x_{0}'}^{(j)}(t)\right\Vert _{2}^{2}\leq\sup_{\left(\theta,b,x_{0}\right)}\left\Vert p^{0}C^{T}\left(C\overline{X}_{\theta,b,x_{0}}^{(j)}(1)-y_{j}\right)\right\Vert _{2}^{2}e^{T\overline{\partial_{x}f}}$
so the inequality for $\left\Vert p_{\theta,b,x_{0}}^{(j)}(t)-p_{\theta',b',x_{0}'}^{(j)}(t)\right\Vert _{2}^{2}$
can be re-expressed as:
\[
\begin{array}{lll}
\left\Vert p_{\theta,b,x_{0}}^{(j)}(t)-p_{\theta',b',x_{0}'}^{(j)}(t)\right\Vert _{2}^{2} & \leq & 2\left\Vert C^{T}C\right\Vert _{2}^{2}\left\Vert BU^{-1}B^{T}\right\Vert \left\Vert p_{\theta,b,x_{0}}^{(j)}-p_{\theta',b',x_{0}'}^{(j)}\right\Vert _{L^{2}}e^{4T\overline{\partial_{x}f}}\\
 & + & 2\left\Vert p^{0}C^{T}C\right\Vert _{2}^{2}\left(\left\Vert \overline{X}_{\theta,b,x_{0}}^{(j)}(0)-\overline{X}_{\theta',b',x_{0}'}^{(j)}(0)\right\Vert _{2}^{2}\right)e^{4T\overline{\partial_{x}f}}\\
 & + & 16\left\Vert p^{0}C^{T}C\right\Vert _{2}^{2}T\left(\overline{\partial_{b}f}\left\Vert b-b'\right\Vert _{2}^{2}+\overline{\partial_{\theta}f}\left\Vert \theta-\theta'\right\Vert _{2}^{2}\right)e^{4T\overline{\partial_{x}f}}\\
 & + & 4T\overline{\partial_{xx}f}\sup_{\left(\theta,b,x_{0}\right)}\left\Vert p^{0}C^{T}\left(C\overline{X}_{\theta,b,x_{0}}^{(j)}(1)-y_{j}\right)\right\Vert _{2}^{2}\left\Vert \overline{X}_{\theta,b,x_{0}}^{(j)}-\overline{X}_{\theta',b',x_{0}'}^{(j)}\right\Vert _{L^{2}}^{2}e^{T\overline{\partial_{x}f}}\\
 & + & 8T\left(\overline{\partial_{bx}f}\left\Vert b-b'\right\Vert _{2}^{2}+\overline{\partial_{\theta x}f}\left\Vert \theta-\theta'\right\Vert _{2}^{2}\right)+2T\overline{\partial_{x}f}\int_{t}^{T}\left\Vert p_{\theta',b',x_{0}'}^{(j)}(s)-p_{\theta,b,x_{0}}^{(j)}(s)\right\Vert _{2}^{2}ds
\end{array}
\]
and by using Gronwall lemma: 
\[
\begin{array}{lll}
\left\Vert p_{\theta,b,x_{0}}^{(j)}(t)-p_{\theta',b',x_{0}'}^{(j)}\right\Vert _{L^{2}}^{2} & \leq & 2T\left\Vert C^{T}C\right\Vert _{2}^{2}\left\Vert BU^{-1}B^{T}\right\Vert \left\Vert p_{\theta,b,x_{0}}^{(j)}-p_{\theta',b',x_{0}'}^{(j)}\right\Vert _{L^{2}}e^{6T\overline{\partial_{x}f}}\\
 & + & 16T\left\Vert p^{0}C^{T}C\right\Vert _{2}^{2}\left(\overline{\partial_{b}f}\left\Vert b-b'\right\Vert _{2}^{2}+\overline{\partial_{\theta}f}\left\Vert \theta-\theta'\right\Vert _{2}^{2}\right)e^{6T\overline{\partial_{x}f}}\\
 & + & 2\left\Vert p^{0}C^{T}C\right\Vert _{2}^{2}\left\Vert \overline{X}_{\theta,b,x_{0}}^{(j)}(0)-\overline{X}_{\theta',b',x_{0}'}^{(j)}(0)\right\Vert _{2}^{2}e^{6T\overline{\partial_{x}f}}\\
 & + & 8T\left(\overline{\partial_{bx}f}\left\Vert b-b'\right\Vert _{2}^{2}+\overline{\partial_{\theta x}f}\left\Vert \theta-\theta'\right\Vert _{2}^{2}\right)e^{6T\overline{\partial_{x}f}}\\
 & + & 4T\overline{\partial_{xx}f}\sup_{\left(\theta,b,x_{0}\right)}\left\Vert p^{0}C^{T}\left(C\overline{X}_{\theta,b,x_{0}}^{(j)}(1)-y_{j}\right)\right\Vert _{2}^{2}\left\Vert \overline{X}_{\theta,b,x_{0}}^{(j)}-\overline{X}_{\theta',b',x_{0}'}^{(j)}\right\Vert _{L^{2}}^{2}e^{3T\overline{\partial_{x}f}}
\end{array}
\]
so if $U$ is chosen such that $K_{1}(U)=2T\left\Vert C^{T}C\right\Vert _{2}^{2}\left\Vert BU^{-1}B^{T}\right\Vert e^{6T\overline{\partial_{x}f}}<1$,
we derive: 
\[
\begin{array}{lll}
\left\Vert p_{\theta,b,x_{0}}^{(j)}-p_{\theta',b',x_{0}'}^{(j)}\right\Vert _{L^{2}}^{2} & \leq & \frac{2\left\Vert p^{0}C^{T}C\right\Vert _{2}^{2}\left(8T\left(\overline{\partial_{b}f}\left\Vert b-b'\right\Vert _{2}^{2}+\overline{\partial_{\theta}f}\left\Vert \theta-\theta'\right\Vert _{2}^{2}\right)+\left\Vert \overline{X}_{\theta,b,x_{0}}^{(j)}(0)-\overline{X}_{\theta',b',x_{0}'}^{(j)}(0)\right\Vert _{2}^{2}\right)e^{6T\overline{\partial_{x}f}}}{1-K_{1}(U)}\\
 & + & \frac{8T\left(\overline{\partial_{bx}f}\left\Vert b-b'\right\Vert _{2}^{2}+\overline{\partial_{\theta x}f}\left\Vert \theta-\theta'\right\Vert _{2}^{2}\right)e^{6T\overline{\partial_{x}f}}}{1-K_{1}(U)}\\
 & + & \frac{4T\overline{\partial_{xx}f}\sup_{\left(\theta,b,x_{0}\right)}\left\Vert p^{0}C^{T}\left(C\overline{X}_{\theta,b,x_{0}}^{(j)}(1)-y_{j}\right)\right\Vert _{2}^{2}\left\Vert \overline{X}_{\theta,b,x_{0}}^{(j)}-\overline{X}_{\theta',b',x_{0}'}^{(j)}\right\Vert _{L^{2}}^{2}e^{3T\overline{\partial_{x}f}}}{1-K_{1}(U)}
\end{array}
\]
by re-injecting this expression in the inequality ruling $\left\Vert \overline{X}_{\theta,b,x_{0}}^{(j)}(t)-\overline{X}_{\theta',b',x_{0}'}^{(j)}(t)\right\Vert _{2}^{2}$,
we get: 
\[
\begin{array}{l}
\left\Vert \overline{X}_{\theta,b,x_{0}}^{(j)}(t)-\overline{X}_{\theta',b',x_{0}'}^{(j)}(t)\right\Vert _{2}^{2}\\
\leq\left\Vert BU^{-1}B^{T}\right\Vert \frac{2\left\Vert C^{T}C\right\Vert _{2}^{2}\left(8T\left(\overline{\partial_{b}f}\left\Vert b-b'\right\Vert _{2}^{2}+\overline{\partial_{\theta}f}\left\Vert \theta-\theta'\right\Vert _{2}^{2}\right)+\left\Vert \overline{X}_{\theta,b,x_{0}}^{(j)}(0)-\overline{X}_{\theta',b',x_{0}'}^{(j)}(0)\right\Vert _{2}^{2}\right)}{1-K_{1}(U)}e^{10T\overline{\partial_{x}f}}\\
+\left\Vert \frac{BU^{-1}B^{T}}{p^{0}}\right\Vert \frac{8T\left(\overline{\partial_{bx}f}\left\Vert b-b'\right\Vert _{2}^{2}+\overline{\partial_{\theta x}f}\left\Vert \theta-\theta'\right\Vert _{2}^{2}\right)}{1-K_{1}(U)}e^{10T\overline{\partial_{x}f}}\\
+\left\Vert BU^{-1}B^{T}\right\Vert \frac{4T\overline{\partial_{xx}f}\sup_{\left(\theta,b,x_{0}\right)}\left\Vert C^{T}\left(C\overline{X}_{\theta,b,x_{0}}^{(j)}(1)-y_{j}\right)\right\Vert _{2}^{2}\left\Vert \overline{X}_{\theta,b,x_{0}}^{(j)}-\overline{X}_{\theta',b',x_{0}'}^{(j)}\right\Vert _{L^{2}}^{2}}{1-K_{1}(U)}e^{7T\overline{\partial_{x}f}}\\
+\left(8T\left(\overline{\partial_{b}f}\left\Vert b-b'\right\Vert _{2}^{2}+\overline{\partial_{\theta}f}\left\Vert \theta-\theta'\right\Vert _{2}^{2}\right)+\left\Vert \overline{X}_{\theta,b,x_{0}}^{(j)}(0)-\overline{X}_{\theta',b',x_{0}'}^{(j)}(0)\right\Vert _{2}^{2}\right)e^{4T\overline{\partial_{x}f}}
\end{array}
\]
now if $U$ is chosen such that $K_{2}(U)=\frac{\left\Vert BU^{-1}B^{T}\right\Vert }{1-K_{1}(U)}4T\overline{\partial_{xx}f}\sup_{\left(\theta,b,x_{0}\right)}\left\Vert C^{T}\left(C\overline{X}_{\theta,b,x_{0}}^{(j)}(1)-y_{j}\right)\right\Vert _{2}^{2}e^{7T\overline{\partial_{x}f}}<1$,
then the previous inequality becomes
\[
\begin{array}{l}
\left\Vert \overline{X}_{\theta,b,x_{0}}^{(j)}-\overline{X}_{\theta',b',x_{0}'}^{(j)}\right\Vert _{L^{2}}^{2}\\
\leq\left\Vert BU^{-1}B^{T}\right\Vert \frac{2\left\Vert C^{T}C\right\Vert _{2}^{2}\left(8T\left(\overline{\partial_{b}f}\left\Vert b-b'\right\Vert _{2}^{2}+\overline{\partial_{\theta}f}\left\Vert \theta-\theta'\right\Vert _{2}^{2}\right)+\left\Vert \overline{X}_{\theta,b,x_{0}}^{(j)}(0)-\overline{X}_{\theta',b',x_{0}'}^{(j)}(0)\right\Vert _{2}^{2}\right)e^{12T^{2}\overline{\partial_{x}f}}}{\left(1-K_{2}(U)\right)\left(1-K_{1}(U)\right)}\\
+\left\Vert \frac{BU^{-1}B^{T}}{p^{0}}\right\Vert \frac{8T\left(\overline{\partial_{bx}f}\left\Vert b-b'\right\Vert _{2}^{2}+\overline{\partial_{\theta x}f}\left\Vert \theta-\theta'\right\Vert _{2}^{2}\right)}{\left(1-K_{2}(U)\right)\left(1-K_{1}(U)\right)}e^{10T\overline{\partial_{x}f}}\\
+\frac{\left(8T\left(\overline{\partial_{b}f}\left\Vert b-b'\right\Vert _{2}^{2}+\overline{\partial_{\theta}f}\left\Vert \theta-\theta'\right\Vert _{2}^{2}\right)+\left\Vert \overline{X}_{\theta,b,x_{0}}^{(j)}(0)-\overline{X}_{\theta',b',x_{0}'}^{(j)}(0)\right\Vert _{2}^{2}\right)}{1-K_{2}(U)}e^{4T\overline{\partial_{x}f}}.
\end{array}
\]
Since $\overline{X}_{\theta,b,x_{0}}^{(0)}(0)-\overline{X}_{\theta',b',x_{0}'}^{(0)}(0)=x_{0}-x_{0}'$,
we see from the previous inequality, it exists constants $K_{b},K_{\theta}$
and $K_{x_{0}}$ such that $\left\Vert \overline{X}_{\theta,b,x_{0}}^{(0)}-\overline{X}_{\theta',b',x_{0}'}^{(0)}\right\Vert _{L^{2}}^{2}\leq K_{b}\left\Vert b-b'\right\Vert _{2}^{2}+K_{\theta}\left\Vert \theta-\theta'\right\Vert _{2}^{2}+K_{x_{0}}\left\Vert x_{0}-x_{0}'\right\Vert _{2}^{2}$
and so we derive the uniform continuity of $\left(\theta,b,x_{0}\right)\longrightarrow\left\Vert \overline{X}_{\theta,b,x_{0}}^{(0)}\right\Vert _{L^{2}}^{2}$
which in turn leads to the continuity of $\left(\theta,b,x_{0}\right)\longrightarrow\left\Vert p_{\theta,b,x_{0}}^{(0)}\right\Vert _{L^{2}}^{2}.$
The differential constraints imposed by the conjugate equations turn
this continuity in $L^{2}-$norm into the continuity of $\left(\theta,b,x_{0},t\right)\longrightarrow\overline{X}_{\theta,b,x_{0}}^{(0)}(t)$
and $\left(\theta,b,x_{0}\right)\longrightarrow p_{\theta,b,x_{0}}^{(0)}(t).$
Now, for the general case of $j>0$, we notice from the previous inequalities
that $\left\Vert \overline{X}_{\theta,b,x_{0}}^{(j)}-\overline{X}_{\theta',b',x_{0}'}^{(j)}\right\Vert _{L^{2}}^{2}\leq K_{b}\left\Vert b-b'\right\Vert _{2}^{2}+K_{\theta}\left\Vert \theta-\theta'\right\Vert _{2}^{2}+K_{x_{0}}\left\Vert \overline{X}_{\theta,b,x_{0}}^{(j-1)}(1)-\overline{X}_{\theta',b',x_{0}'}^{(j-1)}(1)\right\Vert _{2}^{2}$
and $\left\Vert p_{\theta,b,x_{0}}^{(j)}-p_{\theta,b,x_{0}}^{(j)}\right\Vert _{L^{2}}^{2}$
$\leq K'_{b}\left\Vert b-b'\right\Vert _{2}^{2}+K'_{\theta}\left\Vert \theta-\theta'\right\Vert _{2}^{2}$
$+K'_{x_{1}}\left\Vert \overline{X}_{\theta,b,x_{0}}^{(j)}(1)-\overline{X}_{\theta',b',x_{0}'}^{(j)}(1)\right\Vert _{L^{2}}^{2}$
$+K'_{x_{0}}\left\Vert \overline{X}_{\theta,b,x_{0}}^{(j-1)}(1)-\overline{X}_{\theta',b',x_{0}'}^{(j-1)}(1)\right\Vert _{2}^{2}$
since $\overline{X}_{\theta,b,x_{0}}^{(j)}(0)=\overline{X}_{\theta,b,x_{0}}^{(j-1)}(1).$
So, we can use recursive argument to conclude about the continuity
of $\left(\theta,b,x_{0},t\right)\longrightarrow\overline{X}_{\theta,b,x_{0}}^{(j)}(t)$
and $\left(\theta,b,x_{0}\right)\longrightarrow p_{\theta,b,x_{0}}^{(j)}(t)$
for all $j$, since the property holds for $j=0.$ The dependence
between $\overline{u}_{\theta,b,x_{0}}^{(j)}$ and $p_{\theta,b,x_{0}}^{(j)}$
allows us to conclude for the continuity of $\overline{u}_{\theta,b,x_{0}}^{(j)}$
as well. The proof remains the same for higher order of differentiation
and thus it is omitted.

Now regarding the second part of the lemma, let us denote: $\overline{\mathcal{C}_{i}}(\theta,b_{i},x_{i,0}):=\mathcal{C}_{i}(b_{i},x_{i,0},\overline{u}_{\theta,b_{i},x_{i,0}}\mid\theta,U)=\sum_{j}\left\Vert C\overline{X}_{\theta,b_{i},x_{i,0}}(t_{ij})-y_{ij}\right\Vert _{2}^{2}+\left\Vert \overline{u}_{\theta,b_{i},x_{i,0}}\right\Vert _{U,L^{2}}^{2},$thanks
to the first part of the lemma, we know $\overline{\mathcal{C}_{i}}$
is $k$ continuously differentiable on $\widetilde{\Theta}\times\mathbb{R}^{q}\times\widetilde{\chi}$.
Since $\nabla_{x_{i,0}}\overline{\mathcal{C}_{i}}(\theta,b_{i},\widehat{x_{i,0}}\left(\theta,b_{i}\right))=0$,
we can use the implicit function theorem to derive that for each point
$\left(\widetilde{\theta},\widetilde{b_{i}}\right)\in\widetilde{\Theta}\times\mathbb{R}^{q}$
such that $\frac{\partial^{2}}{\partial^{2}x_{i,0}}\overline{\mathcal{C}_{i}}(\widetilde{\theta},\widetilde{b_{i}},\widehat{x_{i,0}}\left(\widetilde{\theta},\widetilde{b_{i}}\right))$
is of full rank it exist a neighborhood of $\widetilde{\Theta}_{\widetilde{\theta}}\times\widetilde{\Theta}_{\widetilde{b_{i}}}$
such that $\left(\theta,b_{i}\right)\longmapsto\widehat{x_{i,0}}\left(\theta,b_{i}\right)$
is $(k-1)$ differentiable on $\widetilde{\Theta}_{\widetilde{\theta}}\times\widetilde{\Theta}_{\widetilde{b_{i}}}$
and moreover: 
\[
\frac{\partial}{\partial\left(\theta,b_{i}\right)}\widehat{x_{i,0}}\left(\theta,b_{i}\right)=-\frac{\partial^{2}}{\partial^{2}x_{i,0}}\overline{\mathcal{C}_{i}}(\theta,b_{i},\widehat{x_{i,0}}\left(\theta,b_{i}\right))^{-1}\frac{\partial}{\partial\left(\theta,b_{i}\right)}\nabla_{x_{i,0}}\overline{\mathcal{C}_{i}}(\theta,b_{i},\widehat{x_{i,0}}\left(\theta,b_{i}\right)).
\]
 From this, we also derive the $(k-1)$ differentiability of $\overline{X}_{\theta,b}$
and $\overline{u}_{\theta,b_{i}}$ by classic results on the regularity
of composed functions. Since, we get: 
\[
\begin{array}{lll}
\frac{\partial^{2}}{\partial^{2}x_{i,0}}\overline{\mathcal{C}_{i}}(\theta,b_{i},\widehat{x_{i,0}}\left(\theta,b_{i}\right)) & = & \sum_{j}\left(C\frac{\partial^{2}}{\partial^{2}x_{i,0}}\overline{X}_{\theta,b_{i},\widehat{x_{i,0}}\left(\theta,b_{i}\right)}(t_{ij})\right)^{T}\otimes\left(C\overline{X}_{\theta,b_{i},\widehat{x_{i,0}}\left(\theta,b_{i}\right)}-y_{ij}\right)\\
 & + & 2\sum_{j}\left(C\frac{\partial}{\partial x_{i,0}}\overline{X}_{\theta,b_{i},\widehat{x_{i,0}}\left(\theta,b_{i}\right)}(t_{ij})\right)^{T}C\frac{\partial}{\partial x_{i,0}}\overline{X}_{\theta,b_{i},\widehat{x_{i,0}}\left(\theta,b_{i}\right)}(t_{ij})\\
 & + & 2\int_{0}^{T}\left(\frac{\partial^{2}}{\partial^{2}x_{i,0}}\overline{u}_{\theta,b_{i},\widehat{x_{i,0}}\left(\theta,b_{i}\right)}(t)\right)^{T}\otimes\overline{u}_{\theta,b_{i},\widehat{x_{i,0}}\left(\theta,b_{i}\right)}(t)dt\\
 & + & 2\int_{0}^{T}\left(\frac{\partial}{\partial x_{i,0}}\overline{u}_{\theta,b_{i},\widehat{x_{i,0}}\left(\theta,b_{i}\right)}(t)\right)^{T}\frac{\partial}{\partial x_{i,0}}\overline{u}_{\theta,b_{i},\widehat{x_{i,0}}\left(\theta,b_{i}\right)}(t)dt
\end{array}
\]
and
\[
\begin{array}{lll}
\frac{\partial}{\partial\left(\theta,b_{i}\right)}\nabla_{x_{i,0}}\overline{\mathcal{C}_{i}}(\theta,b_{i},\widehat{x_{i,0}}\left(\theta,b_{i}\right)) & = & \sum_{j}\left(C\frac{\partial^{2}}{\partial^{2}\left(\theta,b_{i}\right)\partial x_{i,0}}\overline{X}_{\theta,b_{i},\widehat{x_{i,0}}\left(\theta,b_{i}\right)}(t_{ij})\right)^{T}\otimes\left(C\overline{X}_{\theta,b_{i},\widehat{x_{i,0}}\left(\theta,b_{i}\right)}-y_{ij}\right)\\
 & + & 2\sum_{j}\left(C\frac{\partial}{\partial\left(\theta,b_{i}\right)}\overline{X}_{\theta,b_{i},\widehat{x_{i,0}}\left(\theta,b_{i}\right)}(t_{ij})\right)^{T}C\frac{\partial}{\partial x_{i,0}}\overline{X}_{\theta,b_{i},\widehat{x_{i,0}}\left(\theta,b_{i}\right)}(t_{ij})\\
 & + & 2\int_{0}^{T}\left(\frac{\partial^{2}}{\partial\left(\theta,b_{i}\right)\partial x_{i,0}}\overline{u}_{\theta,b_{i},\widehat{x_{i,0}}\left(\theta,b_{i}\right)}(t)\right)^{T}\otimes\overline{u}_{\theta,b_{i},\widehat{x_{i,0}}\left(\theta,b_{i}\right)}(t)dt\\
 & + & 2\int_{0}^{T}\left(\frac{\partial}{\partial\left(\theta,b_{i}\right)}\overline{u}_{\theta,b_{i},\widehat{x_{i,0}}\left(\theta,b_{i}\right)}(t)\right)^{T}\frac{\partial}{\partial x_{i,0}}\overline{u}_{\theta,b_{i},\widehat{x_{i,0}}\left(\theta,b_{i}\right)}(t)dt
\end{array}
\]
the behavior of $\frac{\partial}{\partial\left(\theta,b_{i}\right)}\widehat{x_{i,0}}\left(\theta,b_{i}\right)$
is controlled by the partial derivatives of $\overline{X}_{\theta,b_{i},\widehat{x_{i,0}}\left(\theta,b_{i}\right)}$
and $\overline{u}_{\theta,b_{i},\widehat{x_{i,0}}\left(\theta,b_{i}\right)}$,
in particular if their moment of order two are bounded the same holds
for $\frac{\partial}{\partial\left(\theta,b_{i}\right)}\widehat{x_{i,0}}\left(\theta,b_{i}\right).$ 

Now, for the third part of the lemma, we establish the partial derivative
of $\overline{X}_{\theta,b_{i},x_{0}}^{(j)}$ and $p_{\theta,b_{i},x_{0}}^{(j)}$
with respect to $\left(\theta,b_{i},x_{0}\right)$ have finite moment
of order 2. This in turn will be sufficient to conclude for the partial
derivative of $\overline{u}_{\theta,b_{i},x_{0}}$ and $\widehat{x_{i,0}}\left(\theta,b_{i}\right)$
and to conclude the proof of lemma \ref{lem:dh_d2h_derivative_boundedness}.
Now we differentiate the conjugate equation with respect to $\left(b_{i},\theta\right)$:
\[
\begin{array}{lll}
\frac{d}{dt}\frac{\partial}{\partial\left(\theta,b_{i}\right)}\overline{X}_{\theta,b_{i},x_{0}}^{(j)} & = & (t_{j+1}-t_{j})\frac{\partial}{\partial\left(\theta,b_{i}\right)}f_{\theta,b_{i}}(t_{j}+t(t_{j+1}-t_{j}),\overline{X}_{\theta,b_{i},x_{0}}^{(j)})\\
 & + & (t_{j+1}-t_{j})\frac{\partial}{\partial x}f_{\theta,b_{i}}(t_{j}+t(t_{j+1}-t_{j})),\overline{X}_{\theta,b_{i},x_{0}}^{(j)})\frac{\partial}{\partial\left(b_{i},\theta\right)}\overline{X}_{\theta,b_{i},x_{0}}^{(j)}\\
 & - & \frac{BU^{-1}B^{T}}{2p^{0}}\frac{\partial}{\partial\left(b_{i},\theta\right)}p_{\theta,b_{i},x_{0}}^{(j)}\\
\frac{d}{dt}\frac{\partial}{\partial x_{0}}\overline{X}_{\theta,b_{i},x_{0}}^{(j)} & = & (t_{j+1}-t_{j})\frac{\partial}{\partial x}f_{\theta,b_{i}}(t_{j}+t(t_{j+1}-t_{j})),\overline{X}_{\theta,b_{i},x_{0}}^{(j)})\frac{\partial}{\partial x_{0}}\overline{X}_{\theta,b_{i},x_{0}}^{(j)}\\
 & - & \frac{BU^{-1}B^{T}}{2p^{0}}\frac{\partial}{\partial x_{0}}p_{\theta,b_{i},x_{0}}^{(j)}\\
\frac{d}{dt}\frac{\partial}{\partial\left(\theta,b_{i}\right)}p_{\theta,b_{i},x_{0}}^{(j)} & = & -(t_{j+1}-t_{j})\frac{\partial^{2}}{\partial\left(\theta,b_{i}\right)\partial x}f_{\theta,b_{i}}(t_{j}+t(t_{j+1}-t_{j}),\overline{X}_{\theta,b_{i},x_{0}}^{(j)})^{T}p_{\theta,b_{i},x_{0}}^{(j)}\\
 & - & (t_{j+1}-t_{j})\left(\frac{\partial^{2}}{\partial^{2}x}f_{\theta,b_{i}}(t_{j}+t(t_{j+1}-t_{j}),\overline{X}_{\theta,b_{i},x_{0}}^{(j)})\frac{\partial}{\partial\left(\theta,b_{i}\right)}\overline{X}_{\theta,b_{i},x_{0}}^{(j)}\right){}^{T}p_{\theta,b_{i},x_{0}}^{(j)}\\
 & - & (t_{j+1}-t_{j})\frac{\partial}{\partial x}f_{\theta,b_{i}}(t_{j}+t(t_{j+1}-t_{j}),\overline{X}_{\theta,b_{i},x_{0}}^{(j)})^{T}\frac{\partial}{\partial\left(\theta,b_{i}\right)}p_{\theta,b_{i},x_{0}}^{(j)}\\
\frac{d}{dt}\frac{\partial}{\partial x_{0}}p_{\theta,b_{i},x_{0}}^{(j)} & = & -(t_{j+1}-t_{j})\left(\frac{\partial^{2}}{\partial^{2}x}f_{\theta,b_{i}}(t_{j}+t(t_{j+1}-t_{j}),\overline{X}_{\theta,b_{i},x_{0}}^{(j)})\frac{\partial}{\partial x_{0}}\overline{X}_{\theta,b_{i},x_{0}}^{(j)}\right){}^{T}p_{\theta,b_{i},x_{0}}^{(j)}\\
 & - & (t_{j+1}-t_{j})\frac{\partial}{\partial x}f_{\theta,b_{i}}(t_{j}+t(t_{j+1}-t_{j}),\overline{X}_{\theta,b_{i},x_{0}}^{(j)})^{T}\frac{\partial}{\partial x_{0}}p_{\theta,b_{i},x_{0}}^{(j)}\\
\frac{\partial}{\partial\left(\theta,b_{i},x_{0}\right)}\overline{X}_{\theta,b_{i},x_{0}}^{(j)}(0) & = & \frac{\partial}{\partial\left(\theta,b_{i},x_{0}\right)}\overline{X}_{\theta,b_{i},x_{0}}^{(j-1)}(1),\,\frac{\partial}{\partial\left(\theta,b_{i}\right)}\overline{X}_{\theta,b_{i},x_{0}}^{(0)}(0)=0,\,\frac{\partial}{\partial x_{0}}\overline{X}_{\theta,b_{i},x_{0}}^{(0)}(0)=I_{d}\\
\frac{\partial}{\partial\left(\theta,b_{i},x_{0}\right)}p_{\theta,b_{i},x_{0}}^{(j)}(1) & = & 2p^{0}C^{T}C\frac{\partial}{\partial\left(\theta,b_{i},x_{0}\right)}\overline{X}_{\theta,b_{i},x_{0}}^{(j)}(1).
\end{array}
\]
We get the inequalities:
\[
\begin{array}{lll}
\left\Vert \frac{\partial}{\partial\left(b_{i},\theta\right)}\overline{X}_{\theta,b_{i},x_{0}}^{(j)}(t)\right\Vert _{2}^{2} & \leq & 4T\overline{\partial_{(b_{i},\theta)}f}+4T\overline{\partial_{x}f}\int_{0}^{t}\left\Vert \frac{\partial}{\partial\left(b_{i},\theta\right)}\overline{X}_{\theta,b_{i},x_{0}}^{(j)}(s)\right\Vert _{2}^{2}ds\\
 & + & \left\Vert \frac{BU^{-1}B}{p^{0}}\right\Vert _{2}^{2}\left\Vert \frac{\partial}{\partial\left(b_{i},\theta\right)}p_{\theta,b_{i},x_{0}}^{(j)}\right\Vert _{L^{2}}+\left\Vert \frac{\partial}{\partial\left(b_{i},\theta\right)}\overline{X}_{\theta,b_{i},x_{0}}^{(j)}(0)\right\Vert _{2}^{2}\\
\left\Vert \frac{\partial}{\partial x_{0}}\overline{X}_{\theta,b_{i},x_{0}}^{(j)}(t)\right\Vert _{2}^{2} & \leq & \left\Vert \frac{\partial}{\partial x_{0}}\overline{X}_{\theta,b_{i},x_{0}}^{(j)}(0)\right\Vert _{2}^{2}+2T\overline{\partial_{x}f}\int_{0}^{t}\left\Vert \frac{\partial}{\partial x_{0}}\overline{X}_{\theta,b_{i},x_{0}}^{(j)}(s)\right\Vert _{2}^{2}ds\\
 & + & \left\Vert \frac{BU^{-1}B}{p^{0}}\right\Vert _{2}^{2}\left\Vert \frac{\partial}{\partial x_{0}}p_{\theta,b_{i},x_{0}}^{(j)}\right\Vert _{L^{2}}
\end{array}
\]
\[
\begin{array}{lll}
\left\Vert \frac{\partial}{\partial\left(b_{i},\theta\right)}p_{\theta,b_{i},x_{0}}^{(j)}(t)\right\Vert _{2}^{2} & \leq & \left\Vert 2p^{0}C^{T}C\frac{\partial}{\partial\left(b_{i},\theta\right)}\overline{X}_{\theta,b_{i},x_{0}}^{(j)}(1)\right\Vert _{2}^{2}+4T\overline{\partial_{(b_{i},\theta)x}f}\int_{t}^{1}\left\Vert p_{\theta,b_{i},x_{0}}^{(j)}(s)\right\Vert _{2}^{2}ds\\
 & + & 4T\overline{\partial_{xx}f}\int_{t}^{1}\left\Vert \frac{\partial}{\partial\left(b_{i},\theta\right)}\overline{X}_{\theta,b_{i},x_{0}}^{(j)}(s)\right\Vert \left\Vert p_{\theta,b_{i},x_{0}}^{(j)}(s)\right\Vert _{2}^{2}ds+2T\overline{\partial_{x}f}\int_{t}^{1}\left\Vert \frac{\partial}{\partial\left(b_{i},\theta\right)}p_{\theta,b_{i},x_{0}}^{(j)}\right\Vert _{2}^{2}ds\\
 & \leq & 8T\left\Vert p^{0}C^{T}C\right\Vert _{2}^{2}\left(\overline{\partial_{(b_{i},\theta)}f}+\overline{\partial_{x}f}\left\Vert \frac{\partial}{\partial\left(b_{i},\theta\right)}\overline{X}_{\theta,b_{i},x_{0}}^{(j)}\right\Vert _{L^{2}}^{2}\right)\\
 & + & \left\Vert 2p^{0}C^{T}C\right\Vert _{2}^{2}\left(\left\Vert \frac{BU^{-1}B}{p^{0}}\right\Vert _{2}^{2}\left\Vert \frac{\partial}{\partial\left(b_{i},\theta\right)}p_{\theta,b_{i},x_{0}}^{(j)}\right\Vert _{L^{2}}+\left\Vert \frac{\partial}{\partial\left(b_{i},\theta\right)}\overline{X}_{\theta,b_{i},x_{0}}^{(j)}(0)\right\Vert _{2}^{2}\right)\\
 & + & 4T\left(\overline{\partial_{xx}f}\left\Vert \frac{\partial}{\partial\left(b_{i},\theta\right)}\overline{X}_{\theta,b_{i},x_{0}}^{(j)}\right\Vert _{L^{2}}^{2}+\overline{\partial_{(b_{i},\theta)x}f}\right)\left\Vert p_{\theta,b_{i},x_{0}}^{(j)}\right\Vert _{L^{2}}\\
 & + & 2T\overline{\partial_{x}f}\int_{t}^{1}\left\Vert \frac{\partial}{\partial\left(b_{i},\theta\right)}p_{\theta,b_{i},x_{0}}^{(j)}\right\Vert _{2}^{2}ds\\
\left\Vert \frac{\partial}{\partial x_{0}}p_{\theta,b_{i},x_{0}}^{(j)}(t)\right\Vert _{2}^{2} & \leq & \left\Vert 2p^{0}C^{T}C\frac{\partial}{\partial x}\overline{X}_{\theta,b_{i},x_{0}}^{(j)}(1)\right\Vert _{2}^{2}\\
 & + & 2T\left(\overline{\partial_{xx}f}\int_{t}^{1}\left\Vert p_{\theta,b_{i},x_{0}}^{(j)}(s)\right\Vert _{2}^{2}ds+\overline{\partial_{x}f}\int_{t}^{1}\left\Vert \frac{\partial}{\partial\left(b_{i},\theta\right)}p_{\theta,b_{i},x_{0}}^{(j)}\right\Vert _{2}^{2}ds\right)\\
 & \leq & \left\Vert 2p^{0}C^{T}C\right\Vert _{2}^{2}\left(2T\overline{\partial_{x}f}\int_{0}^{t}\left\Vert \frac{\partial}{\partial x_{0}}\overline{X}_{\theta,b_{i},x_{0}}^{(j)}(s)\right\Vert _{2}^{2}ds+\left\Vert \frac{BU^{-1}B}{p^{0}}\right\Vert _{2}^{2}\left\Vert \frac{\partial}{\partial x_{0}}p_{\theta,b_{i},x_{0}}^{(j)}\right\Vert _{L^{2}}\right)\\
 & + & \left\Vert 2p^{0}C^{T}C\right\Vert _{2}^{2}\left\Vert \frac{\partial}{\partial x_{0}}\overline{X}_{\theta,b_{i},x_{0}}^{(j)}(0)\right\Vert _{2}^{2}\\
 & + & 2T\left(\overline{\partial_{xx}f}\int_{t}^{1}\left\Vert p_{\theta,b_{i},x_{0}}^{(j)}(s)\right\Vert _{2}^{2}ds+\overline{\partial_{x}f}\int_{t}^{1}\left\Vert \frac{\partial}{\partial\left(b_{i},\theta\right)}p_{\theta,b_{i},x_{0}}^{(j)}\right\Vert _{2}^{2}ds\right)
\end{array}
\]
and from this and the Gronwall lemma, we end up with 
\[
\begin{array}{lll}
\left\Vert \frac{\partial}{\partial\left(b_{i},\theta\right)}p_{\theta,b_{i},x_{0}}^{(j)}(t)\right\Vert _{2}^{2} & \leq & 8T\left\Vert p^{0}C^{T}C\right\Vert _{2}^{2}\left(\overline{\partial_{(b_{i},\theta)}f}+\overline{\partial_{x}f}\left\Vert \frac{\partial}{\partial\left(b_{i},\theta\right)}\overline{X}_{\theta,b_{i},x_{0}}^{(j)}\right\Vert _{L^{2}}^{2}\right)e^{2T\overline{\partial_{x}f}}\\
 & + & \left\Vert 2p^{0}C^{T}C\right\Vert _{2}^{2}\left(\left\Vert \frac{BU^{-1}B}{p^{0}}\right\Vert _{2}^{2}\left\Vert \frac{\partial}{\partial\left(b_{i},\theta\right)}p_{\theta,b_{i},x_{0}}^{(j)}\right\Vert _{L^{2}}+\left\Vert \frac{\partial}{\partial\left(b_{i},\theta\right)}\overline{X}_{\theta,b_{i},x_{0}}^{(j)}(0)\right\Vert _{2}^{2}\right)e^{2T\overline{\partial_{x}f}}\\
 & + & 4T\left(\overline{\partial_{xx}f}\left\Vert \frac{\partial}{\partial\left(b_{i},\theta\right)}\overline{X}_{\theta,b_{i},x_{0}}^{(j)}\right\Vert _{L^{2}}^{2}+\overline{\partial_{(b_{i},\theta)x}f}\right)\left\Vert p_{\theta,b_{i},x_{0}}^{(j)}\right\Vert _{L^{2}}e^{2T\overline{\partial_{x}f}}\\
\left\Vert \frac{\partial}{\partial x_{0}}p_{\theta,b_{i},x_{0}}^{(j)}(t)\right\Vert _{2}^{2} & \leq & \left\Vert 2p^{0}C^{T}C\right\Vert _{2}^{2}\left(2T\overline{\partial_{x}f}\int_{0}^{t}\left\Vert \frac{\partial}{\partial x_{0}}\overline{X}_{\theta,b_{i},x_{0}}^{(j)}(s)\right\Vert _{2}^{2}ds+\left\Vert \frac{BU^{-1}B}{p^{0}}\right\Vert _{2}^{2}\left\Vert \frac{\partial}{\partial x_{0}}p_{\theta,b_{i},x_{0}}^{(j)}\right\Vert _{L^{2}}\right)e^{2T\overline{\partial_{x}f}}\\
 & + & 2T\overline{\partial_{xx}f}\int_{t}^{T}\left\Vert p_{\theta,b_{i},x_{0}}^{(j)}(s)\right\Vert _{2}^{2}dse^{2T\overline{\partial_{x}f}}+\left\Vert 2p^{0}C^{T}C\right\Vert _{2}^{2}\left\Vert \frac{\partial}{\partial x_{0}}\overline{X}_{\theta,b_{i},x_{0}}^{(j)}(0)\right\Vert _{2}^{2}e^{2T\overline{\partial_{x}f}}
\end{array}
\]
so if we choose $K_{3}(U)=2\left\Vert C^{T}C\right\Vert _{2}^{2}\left\Vert BU^{-1}B\right\Vert _{2}^{2}e^{2T\overline{\partial_{x}f}}<1$,
we obtain:
\[
\begin{array}{lll}
\left\Vert \frac{\partial}{\partial\left(b_{i},\theta\right)}p_{\theta,b_{i},x_{0}}^{(j)}\right\Vert _{L^{2}}^{2} & \leq & \frac{8T\left\Vert p^{0}C^{T}C\right\Vert _{2}^{2}\left(\overline{\partial_{(b_{i},\theta)}f}+\overline{\partial_{x}f}\left\Vert \frac{\partial}{\partial\left(b_{i},\theta\right)}\overline{X}_{\theta,b_{i},x_{0}}^{(j)}\right\Vert _{L^{2}}^{2}\right)e^{2T\overline{\partial_{x}f}}}{1-K_{3}(U)}\\
 & + & \frac{\left\Vert 2p^{0}C^{T}C\right\Vert _{2}^{2}\left\Vert \frac{\partial}{\partial\left(b_{i},\theta\right)}\overline{X}_{\theta,b_{i},x_{0}}^{(j)}(0)\right\Vert _{2}^{2}e^{2T^{2}\overline{\partial_{x}f}}}{1-K_{3}(U)}\\
 & + & \frac{4T\left(\overline{\partial_{xx}f}\left\Vert \frac{\partial}{\partial\left(b_{i},\theta\right)}\overline{X}_{\theta,b_{i},x_{0}}^{(j)}\right\Vert _{L^{2}}^{2}+\overline{\partial_{(b_{i},\theta)x}f}\right)\left\Vert p_{\theta,b_{i},x_{0}}^{(j)}\right\Vert _{L^{2}}e^{2T\overline{\partial_{x}f}}}{1-K_{3}(U)}\\
\left\Vert \frac{\partial}{\partial x_{0}}p_{\theta,b_{i},x_{0}}^{(j)}(t)\right\Vert _{L^{2}} & \leq & \frac{2T\left(2\left\Vert p^{0}C^{T}C\right\Vert _{2}^{2}\overline{\partial_{x}f}\left\Vert \frac{\partial}{\partial x_{0}}\overline{X}_{\theta,b_{i},x_{0}}^{(j)}\right\Vert _{L^{2}}^{2}+\overline{\partial_{xx}f}\left\Vert p_{\theta,b_{i},x_{0}}^{(j)}\right\Vert _{L^{2}}\right)e^{2T\overline{\partial_{x}f}}}{1-K_{3}(U)}\\
 & + & \frac{\left\Vert 2p^{0}C^{T}C\right\Vert _{2}^{2}\left\Vert \frac{\partial}{\partial x_{0}}\overline{X}_{\theta,b_{i},x_{0}}^{(j)}(0)\right\Vert _{2}^{2}e^{2T\overline{\partial_{x}f}}}{1-K_{3}(U)}
\end{array}
\]
and by re-injecting in the inequality ruling $\left\Vert \frac{\partial}{\partial\left(b_{i},\theta\right)}\overline{X}_{\theta,b_{i},x_{0}}^{(j)}(t)\right\Vert _{2}^{2}$
and $\left\Vert \frac{\partial}{\partial x_{0}}\overline{X}_{\theta,b_{i},x_{0}}^{(j)}(t)\right\Vert _{2}^{2}$
we get: 
\[
\begin{array}{lll}
\left\Vert \frac{\partial}{\partial\left(b_{i},\theta\right)}\overline{X}_{\theta,b_{i},x_{0}}^{(j)}(t)\right\Vert _{2}^{2} & \leq & 4T\overline{\partial_{(b_{i},\theta)}f}+4T\overline{\partial_{x}f}\int_{0}^{t}\left\Vert \frac{\partial}{\partial\left(b_{i},\theta\right)}\overline{X}_{\theta,b_{i},x_{0}}^{(j)}(s)\right\Vert _{2}^{2}ds+\left\Vert \frac{\partial}{\partial\left(b_{i},\theta\right)}\overline{X}_{\theta,b_{i},x_{0}}^{(j)}(0)\right\Vert _{2}^{2}\\
 & + & \left\Vert BU^{-1}B\right\Vert _{2}^{2}\frac{8T\left\Vert C^{T}C\right\Vert _{2}^{2}\left(\overline{\partial_{(b_{i},\theta)}f}+\overline{\partial_{x}f}\left\Vert \frac{\partial}{\partial\left(b_{i},\theta\right)}\overline{X}_{\theta,b_{i},x_{0}}^{(j)}\right\Vert _{L^{2}}^{2}\right)e^{2T\overline{\partial_{x}f}}}{1-K_{3}(U)}\\
 & + & \left\Vert BU^{-1}B\right\Vert _{2}^{2}\frac{\left\Vert 2C^{T}C\right\Vert _{2}^{2}\left\Vert \frac{\partial}{\partial\left(b_{i},\theta\right)}\overline{X}_{\theta,b_{i},x_{0}}^{(j)}(0)\right\Vert _{2}^{2}e^{2T\overline{\partial_{x}f}}}{1-K_{3}(U)}\\
 & + & \left\Vert \frac{BU^{-1}B}{p^{0}}\right\Vert _{2}^{2}\frac{4T\left(\overline{\partial_{xx}f}\left\Vert \frac{\partial}{\partial\left(b_{i},\theta\right)}\overline{X}_{\theta,b_{i},x_{0}}^{(j)}\right\Vert _{L^{2}}^{2}+\overline{\partial_{(b_{i},\theta)x}f}\right)\left\Vert p_{\theta,b_{i},x_{0}}^{(j)}\right\Vert _{L^{2}}e^{2T\overline{\partial_{x}f}}}{1-K_{3}(U)}\\
\left\Vert \frac{\partial}{\partial x_{0}}\overline{X}_{\theta,b_{i},x_{0}}^{(j)}(t)\right\Vert _{2}^{2} & \leq & \left\Vert \frac{\partial}{\partial x_{0}}\overline{X}_{\theta,b_{i},x_{0}}^{(j)}(0)\right\Vert _{2}^{2}\left(1+\left\Vert BU^{-1}B\right\Vert _{2}^{2}\frac{\left\Vert 2C^{T}C\right\Vert _{2}^{2}e^{2T\overline{\partial_{x}f}}}{1-K_{3}(U)}\right)\\
 & + & 2T\overline{\partial_{x}f}\int_{0}^{t}\left\Vert \frac{\partial}{\partial x_{0}}\overline{X}_{\theta,b_{i},x_{0}}^{(j)}(s)\right\Vert _{2}^{2}ds\\
 & + & \left\Vert \frac{BU^{-1}B}{p^{0}}\right\Vert _{2}^{2}\frac{2T\left(2\left\Vert p^{0}C^{T}C\right\Vert _{2}^{2}\overline{\partial_{x}f}\left\Vert \frac{\partial}{\partial x_{0}}\overline{X}_{\theta,b_{i},x_{0}}^{(j)}\right\Vert _{L^{2}}^{2}+\overline{\partial_{xx}f}\left\Vert p_{\theta,b_{i},x_{0}}^{(j)}\right\Vert _{L^{2}}\right)e^{2T\overline{\partial_{x}f}}}{1-K_{3}(U)}
\end{array}
\]
which leads to 
\[
\begin{array}{lll}
\left\Vert \frac{\partial}{\partial\left(b_{i},\theta\right)}\overline{X}_{\theta,b_{i},x_{0}}^{(j)}\right\Vert _{L^{2}}^{2} & \leq & \left(4T\overline{\partial_{(b_{i},\theta)}f}+\left\Vert \frac{\partial}{\partial\left(b_{i},\theta\right)}\overline{X}_{\theta,b_{i},x_{0}}^{(j)}(0)\right\Vert _{2}^{2}\right)e^{4T\overline{\partial_{x}f}}\\
 & + & \left\Vert BU^{-1}B\right\Vert _{2}^{2}\frac{8T\left\Vert C^{T}C\right\Vert _{2}^{2}\left(\overline{\partial_{(b_{i},\theta)}f}+\overline{\partial_{x}f}\left\Vert \frac{\partial}{\partial\left(b_{i},\theta\right)}\overline{X}_{\theta,b_{i},x_{0}}^{(j)}\right\Vert _{L^{2}}^{2}\right)e^{6T\overline{\partial_{x}f}}}{1-K_{3}(U)}\\
 & + & \left\Vert BU^{-1}B\right\Vert _{2}^{2}\frac{\left\Vert 2C^{T}C\right\Vert _{2}^{2}\left\Vert \frac{\partial}{\partial\left(b_{i},\theta\right)}\overline{X}_{\theta,b_{i},x_{0}}^{(j)}(0)\right\Vert _{2}^{2}e^{6T\overline{\partial_{x}f}}}{1-K_{3}(U)}\\
 & + & \left\Vert \frac{BU^{-1}B}{p^{0}}\right\Vert _{2}^{2}\frac{4T\left(\overline{\partial_{xx}f}\left\Vert \frac{\partial}{\partial\left(b_{i},\theta\right)}\overline{X}_{\theta,b_{i},x_{0}}^{(j)}\right\Vert _{L^{2}}^{2}+\overline{\partial_{(b_{i},\theta)x}f}\right)\left\Vert p_{\theta,b_{i},x_{0}}^{(j)}\right\Vert _{L^{2}}e^{6T\overline{\partial_{x}f}}}{1-K_{3}(U)}\\
\left\Vert \frac{\partial}{\partial x_{0}}\overline{X}_{\theta,b_{i},x_{0}}^{(j)}\right\Vert _{L^{2}}^{2} & \leq & \left\Vert \frac{BU^{-1}B}{p^{0}}\right\Vert _{2}^{2}\frac{2T\left(2\left\Vert p^{0}C^{T}C\right\Vert _{2}^{2}\overline{\partial_{x}f}\left\Vert \frac{\partial}{\partial x_{0}}\overline{X}_{\theta,b_{i},x_{0}}^{(j)}\right\Vert _{L^{2}}^{2}+\overline{\partial_{xx}f}\left\Vert p_{\theta,b_{i},x_{0}}^{(j)}\right\Vert _{L^{2}}\right)e^{6T\overline{\partial_{x}f}}}{1-K_{3}(U)}\\
 & + & \left\Vert \frac{\partial}{\partial x_{0}}\overline{X}_{\theta,b_{i},x_{0}}^{(j)}(0)\right\Vert _{2}^{2}\left(1+\left\Vert BU^{-1}B\right\Vert _{2}^{2}\frac{\left\Vert 2C^{T}C\right\Vert _{2}^{2}e^{2T\overline{\partial_{x}f}}}{1-K_{3}(U)}\right)e^{4T\overline{\partial_{x}f}}.
\end{array}
\]
So if $U$ is chosen such that $K_{4}(U)=\frac{4T\left\Vert BU^{-1}B\right\Vert _{2}^{2}e^{6T\overline{\partial_{x}f}}}{1-K_{3}(U)}\left(2\left\Vert C^{T}C\right\Vert _{2}^{2}\overline{\partial_{x}f}+\frac{\overline{\partial_{xx}f}}{\left|p^{0}\right|}\right)<1$
and $K_{5}(U)=\frac{4T}{1-K_{3}(U)}\left\Vert BU^{-1}B\right\Vert _{2}^{2}\left\Vert C^{T}C\right\Vert _{2}^{2}\overline{\partial_{x}f}e^{6T\overline{\partial_{x}f}}<1$
, then:
\[
\begin{array}{lll}
\left\Vert \frac{\partial}{\partial\left(b_{i},\theta\right)}\overline{X}_{\theta,b_{i},x_{0}}^{(j)}\right\Vert _{L^{2}}^{2} & \leq & \frac{\left(4T\overline{\partial_{(b_{i},\theta)}f}+\left\Vert \frac{\partial}{\partial\left(b_{i},\theta\right)}\overline{X}_{\theta,b_{i},x_{0}}^{(j)}(0)\right\Vert _{2}^{2}\right)}{1-K_{4}(U)}e^{4T\overline{\partial_{x}f}}\\
 & + & \left\Vert BU^{-1}B\right\Vert _{2}^{2}\left\Vert C^{T}C\right\Vert _{2}^{2}e^{6T\overline{\partial_{x}f}}\frac{\left(8T\left\Vert C^{T}C\right\Vert _{2}^{2}\overline{\partial_{(b_{i},\theta)}f}+2\left\Vert \frac{\partial}{\partial\left(b_{i},\theta\right)}\overline{X}_{\theta,b_{i},x_{0}}^{(j)}(0)\right\Vert _{2}^{2}\right)}{\left(1-K_{4}(U)\right)\left(1-K_{3}(U)\right)}\\
 & + & \left\Vert \frac{BU^{-1}B}{p^{0}}\right\Vert _{2}^{2}\frac{4T\overline{\partial_{(b_{i},\theta)x}f}\left\Vert p_{\theta,b_{i},x_{0}}^{(j)}\right\Vert _{L^{2}}e^{6T\overline{\partial_{x}f}}}{\left(1-K_{4}(U)\right)\left(1-K_{3}(U)\right)}\\
\left\Vert \frac{\partial}{\partial x_{0}}\overline{X}_{\theta,b_{i},x_{0}}^{(j)}\right\Vert _{L^{2}}^{2} & \leq & \left\Vert \frac{BU^{-1}B}{p^{0}}\right\Vert _{2}^{2}\frac{2T\overline{\partial_{xx}f}\left\Vert p_{\theta,b_{i},x_{0}}^{(j)}\right\Vert _{L^{2}}e^{6T\overline{\partial_{x}f}}}{\left(1-K_{5}(U)\right)\left(1-K_{3}(U)\right)}\\
 & + & \frac{\left\Vert \frac{\partial}{\partial x_{0}}\overline{X}_{\theta,b_{i},x_{0}}^{(j)}(0)\right\Vert _{2}^{2}}{\left(1-K_{5}(U)\right)}\left(1+\left\Vert BU^{-1}B\right\Vert _{2}^{2}\frac{\left\Vert 2C^{T}C\right\Vert _{2}^{2}e^{2T\overline{\partial_{x}f}}}{1-K_{3}(U)}\right)e^{4T\overline{\partial_{x}f}}.
\end{array}
\]
Now let us focus on the parameter value $\left(\theta,b_{i},x_{0}\right):=\left(\overline{\theta},\widehat{b_{i}}(\overline{\eta}),\widehat{x_{0}}(\overline{\theta},\widehat{b_{i}}(\overline{\eta})\right).$
We already show that $\left\Vert p_{\overline{\theta},\widehat{b_{i}}(\overline{\eta})}^{(j)}(t)\right\Vert _{2}^{2}\leq\sup_{\left(\theta,b,x_{0}\right)}\left\Vert p^{0}C^{T}\left(C\overline{X}_{\overline{\theta},\widehat{b_{i}}(\overline{\eta})}^{(j)}(1)-y_{j}\right)\right\Vert _{2}^{2}e^{T^{2}\overline{\partial_{x}f}}$
from which we derive $\left\Vert p_{\overline{\theta},\widehat{b_{i}}(\overline{\eta})}^{(j)}(t)\right\Vert _{2}^{2}$
has a finite moment of order 2, since we already establish the result
for $\left\Vert p^{0}C^{T}\left(C\overline{X}_{\overline{\theta},\widehat{b_{i}}(\overline{\eta})}^{(j)}(1)-y_{j}\right)\right\Vert _{2}^{2}$
in the previous lemma. Since $\left\Vert \frac{\partial}{\partial\left(b_{i},\theta\right)}\overline{X}_{\overline{\theta},\widehat{b_{i}}(\overline{\eta})}^{(0)}(0)\right\Vert _{2}^{2}=0$
and $\left\Vert \frac{\partial}{\partial x_{0}}\overline{X}_{\overline{\theta},\widehat{b_{i}}(\overline{\eta})}^{(0)}(0)\right\Vert _{2}^{2}=d$,
we can use the previous inequality to show that $\left\Vert \frac{\partial}{\partial\left(b_{i},\theta,x_{0}\right)}\overline{X}_{\overline{\theta},\widehat{b_{i}}(\overline{\eta}),}^{(0)}\right\Vert _{L^{2}}$
has a finite moment of order 2, from this it follows the same holds
for $\left\Vert \frac{\partial}{\partial\left(b_{i},\theta,x_{0}\right)}\overline{p}_{\overline{\theta},\widehat{b_{i}}(\overline{\eta})}^{(0)}\right\Vert _{L^{2}}$
and $\left\Vert \frac{\partial}{\partial\left(b_{i},\theta,x_{0}\right)}\overline{u}_{\overline{\theta},\widehat{b_{i}}(\overline{\eta})}^{(0)}\right\Vert _{L^{2}}$
this in turn leads to the boundedness of moment of order 2 for $\left\Vert \frac{\partial}{\partial\left(b_{i},\theta,x_{0}\right)}\overline{X}_{\overline{\theta},\widehat{b_{i}}(\overline{\eta})}^{(0)}(t)\right\Vert _{2}$
and $\left\Vert \frac{\partial}{\partial\left(b_{i},\theta,x_{0}\right)}\overline{u}_{\overline{\theta},\widehat{b_{i}}(\overline{\eta})}^{(0)}(t)\right\Vert _{2}.$
As in the previous part of the lemma, we can use a recursive argument
to show boundedness of moments of order two of $\left\Vert \frac{\partial}{\partial\left(b_{i},\theta,x_{0}\right)}\overline{X}_{\overline{\theta},\widehat{b_{i}}(\overline{\eta})}^{(j)}(t)\right\Vert _{2}$
and $\left\Vert \frac{\partial}{\partial\left(b_{i},\theta,x_{0}\right)}\overline{u}_{\overline{\theta},\widehat{b_{i}}(\overline{\eta})}^{(j)}(t)\right\Vert _{2}$
for $j>0.$ Since $\frac{\partial}{\partial\left(\theta,b_{i}\right)}\widehat{x_{i,0}}\left(\theta,b_{i}\right)=-\frac{\partial^{2}}{\partial^{2}x_{i,0}}\overline{\mathcal{C}_{i}}(\theta,b_{i},\widehat{x_{i,0}}\left(\theta,b_{i}\right))^{-1}\frac{\partial}{\partial\left(\theta,b_{i}\right)}\nabla_{x_{i,0}}\overline{\mathcal{C}_{i}}(\theta,b_{i},\widehat{x_{i,0}}\left(\theta,b_{i}\right))$,
the same results holds for $\left\Vert \frac{\partial}{\partial\left(\theta,b_{i}\right)}\widehat{x_{i,0}}\left(\overline{\theta},\widehat{b_{i}}(\overline{\eta})\right)\right\Vert _{2}.$
The proof is similar for the derivative of higher dimension.
\end{proof}

\bibliographystyle{rss}
\bibliography{biblio_OCA_iterative_approach}